\RequirePackage{fix-cm}
\documentclass[twocolumn]{svjour3}          % twocolumn
\smartqed  % flush right qed marks, e.g. at end of proof
\usepackage{graphicx}
\usepackage[numbers]{natbib}
\usepackage{xtab}
\usepackage{amsmath,amssymb,amsfonts}
\usepackage[]{multirow}
\usepackage{array}
\usepackage{booktabs}
\usepackage{subcaption}
\usepackage[belowskip=10pt,aboveskip=7pt]{caption}
\usepackage{lscape}
\newcolumntype{L}[1]{>{\raggedright\let\newline\\\arraybackslash\hspace{0pt}}m{#1}}
\newcolumntype{C}[1]{>{\centering\let\newline\\\arraybackslash\hspace{0pt}}m{#1}}
\newcolumntype{R}[1]{>{\raggedleft\let\newline\\\arraybackslash\hspace{0pt}}m{#1}}
\newcolumntype{v}{>{\raggedright}p}
\usepackage{placeins}

\usepackage[compact]{titlesec}
\titlespacing{\section}{1pt}{2ex}{1ex}
\titlespacing{\subsection}{1pt}{1ex}{0ex}
\titlespacing{\subsubsection}{1pt}{0.5ex}{0ex}
\journalname{Archives of Computational Methods in Engineering
}
\begin{document}

\title{A Survey and Analysis on Automated Glioma Brain Tumor Segmentation and Overall Patient Survival Prediction 

\thanks{The article has been published in Springer Journal, ``Archives of Computational Methods in Engineering", which is accessible using the link: https://link.springer.com/article/10.1007/s11831-021-09559-w. One can access the article with complementary access rights using the link:https://rdcu.be/cf2Zz.}
}


\author{Rupal R. Agravat         \and
        Mehul S. Raval  %etc.
}


\institute{Rupal R. Agravat \at
              School of Engineering and Applied Science, \\
              Ahmedabad University, \\ Ahmedabad, India. \\
              \email{rupal.agravat@iet.ahduni.edu.in}           %  \\
           \and
           Mehul S. Raval \at
              School of Engineering and Applied Science, \\
              Ahmedabad University, \\ Ahmedabad, India. \\
              \email{mehul.raval@ahduni.edu.in}
}

\date{Received: 3 May, 2020 / Accepted: 28 January, 2021}

\maketitle

\begin{abstract}
	
Glioma is the deadliest brain tumor with high mortality. Treatment planning by human experts depends on the proper diagnosis of physical symptoms along with Magnetic Resonance (MR) image analysis. Highly variability of a brain tumor in terms of size, shape, location, and a high volume of MR images make the analysis time-consuming. Automatic segmentation methods achieve a reduction in time with excellent reproducible results. The article aims to survey the advancement of automated methods for Glioma brain tumor segmentation. It is also essential to make an objective evaluation of various models based on the benchmark. Therefore, the 2012 - 2019 BraTS challenges evaluate the state-of-the-art methods. The complexity of the tasks facing this challenge has grown from segmentation (Task 1) to overall survival prediction (Task 2) to uncertainty prediction for classification (Task 3). The paper covers the complete gamut of brain tumor segmentation using handcrafted features to deep neural network models for Task 1. The aim is to showcase a complete change of trends in automated brain tumor models. The paper also covers end to end joint models involving brain tumor segmentation and overall survival prediction. All the methods are probed, and parameters that affect performance are tabulated and analyzed.	
	
\end{abstract}

\section{Introduction}
\label{intro}

From the days the medical images were captured by imaging devices and digitally preserved, researchers have started to build computerized automated and semi-automated analysis techniques to address a variety of problems like detection, segmentation, classification, and registration. Till 1990s, medical image analysis was done with pixel processing techniques to detect edge/line with filters, regions based on similarity of pixels or fit mathematical models to detect lines/elliptical shapes. Later, shape models, atlas models and probabilistic models became successful for medical image analysis, where the model learns to predict the unseen data based on observations (training data). This trend moved towards the machine learning models where features were extracted from the data and fed into the computer to make it learn the underlying class/pattern as per the input features and make predictions about unseen data in the future. This approach has changed the trend of human-dependent systems to machine dependent systems where the machine learns from the example data. Such algorithms work very well with the high dimensional feature space to find the optimal decision boundary. Here the only thing which is not done by the computer is feature extraction, which leads to the era of deep learning where the computer learns the optimal set of features for the problem at hand. Deep learning models transform the input data from image/audio/video/text to output data which is location/presence/spread and incrementally learn high dimensional features with the help of a set of intermediate layers between the input and output layers. Medical image analysis has reached the extent where such algorithms play a significant role in the early detection of a disease based on the initial symptoms leading to better treatments. Deadly diseases like cancer (Glioma-the cancerous brain tumor), if detected in the early stages, can increase the life expectancy of the patient.

\subsection{Brain Tumors and Their Types}
\label{sec:1.1}

When the natural cycle of the tissues in the brain breaks and growth becomes uncontrollable, it results in a brain tumor\cite{jhm:2020}. Brain tumors are of two types: (1) primary, and (2) secondary. A primary tumor starts with the brain tissues and grows within the brain, whereas a secondary tumor spreads in the brain and from the other cancerous organs\cite{rochester:2020}. More than 100 types of brain tumors are named based on the tissue and the brain part where it starts to grow. Out of all these tumors, Glioma is the most life-threatening brain tumor. It occurs in the glial cells of the brain. The severity grades of the Glioma tumors depend on \cite{amercr:2020}:

\begin{itemize}
\item the tumorous cell growth rate. 
\item blood supply to the tumorous cells.
\item presence of necrosis (dead cells at the center of the tumor).
\item location of the tumor within the brain.
\item confined area of the tumor.
\item its structural similarity with the healthy cells.
\end{itemize}

Grade I and II, are known as Low-Grade Glioma (LGG), which are benign tumors. Grade III and IV are known as High-Grade Glioma (HGG), which are malignant tumors. When symptoms like nausea, vomiting, fatigue, loss of sensation or movement, difficulty in balancing, persist for a longer duration, it is advisable to go through image screening to know the internal structural changes in the brain due to the tumor.

\subsection{Brain Imaging Modalities}
\label{sec:1.2}

Various imaging techniques are used for brain tumor screening which include positron emission tomography (PET), Computed Tomography(CT) and Magnetic Resonance Imaging(MRI). In PET, a radioactive tracer is injected in the body, which captures a high level of chemical activities of the disease infected part of the body. In the CT, an X-ray tube rotates around the patient’s body and emits a narrow X-ray beam, which is passed through the patient’s body to generate cross-sectional images of the brain. An MRI uses a strong magnetic field around the patient’s body, which aligns the protons in the body to the generated magnetic field, which is followed by the passing of radiofrequency signals to the body. When the current is off, the protons emit the energy and try to align with the magnetic field. The emitted energy from an image records the response of various tissues of the brain. There are two types of MRIs: 1)fMRI(Functional MRI): It measures the brain activities from the change in the blood flow, 2) sMRI(Structural MRI): It captures the anatomy and pathology of the brain. The proposed article uses the sMRI as the focus to deal with the pathology in the brain. Various modalities of sMRI capture responses of the tissues which lead to distinct biological information in the images. Various modalities of sMRI are:

\begin{itemize}
\item \textbf{Diffusion Weighted Image(DWI) MRI:} MR imaging technique measuring the diffusion of water molecules within tissue voxels. DWI is often used to visualize hyperintensities.
\item \textbf{FLAIR MRI:} an MRI pulse sequence which suppresses the fluid (mainly cerebrospinal fluid (CSF)) and enhances an edema.
\item \textbf{T1w MRI:}basic MRI pulse sequence that captures longitudinal relaxation time (time constant required for excited protons to return to equilibrium) differences of tissues.
\item \textbf{T1Gd MRI:} a contrast enhancing agent, Gadolinium is injected into the body and after that a T1 sequence is acquired. This contrast enhancing agent shortens the T1 time which results in bright appearance of blood vessels and pathologies like tumors.
\item \textbf{T2w MRI:} basic MRI pulse sequence that captures transverse relation time(T2) differences of tissues.
\end{itemize}

\begin{figure}
\centering
\begin{subfigure}{.25\textwidth}
  \centering
  \includegraphics[width=3cm,height=3cm]{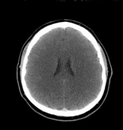}
  \caption{}
\end{subfigure}%
\begin{subfigure}{.25\textwidth}
  \centering
  \includegraphics[width=3cm,height=3cm]{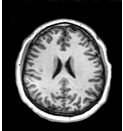}
  \caption{}
\end{subfigure}
\begin{subfigure}{.25\textwidth}
  \centering
  \includegraphics[width=3cm,height=3cm]{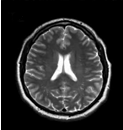}
  \caption{}
\end{subfigure}%
\begin{subfigure}{.25\textwidth}
  \centering
  \includegraphics[width=3cm,height=3cm]{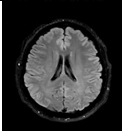}
  \caption{}
\end{subfigure}
\caption{Imaging modalities \cite{healthcare:2016} a) CT b) T1 MRI c) T2 MRI and d) DWI MRI.}
\label{fig:1}% Give a unique label
\end{figure}

In general, for brain tumors, CT and MRI are the commonly used techniques. Both the imaging techniques are essential, and the differences between CT and MRI are listed in Table \ref{tab:1}. Fig. \ref{fig:1} shows the imaging difference between a CT and an MRI, along with their various modalities. Healthy brain tissues are of three classes: 1) Gray Matter(GM), 2) White Matter (WM), and 3) Cerebrospinal Fluid(CSF). Such soft tissue detailing is captured clearly in an MRI as compared to a CT.

\begin{table}[htbp]
\caption{Difference between CT and MRI \cite{amzreg:2019}.}
\label{tab:1}
\renewcommand{\arraystretch}{1.2}
\begin{tabular}{@{}p{4 cm} p{4 cm}@{}}
\toprule
\multicolumn{2}{l}{\textbf{Advantages and Disadvantages}} \\
\textbf{CT} & \textbf{MRI} \\
\midrule
Non-invasive. & Non-invasive.\\
Fast processing. & Slow processing.\\
Cheap. & Expensive. \\
Less accurate. & More accurate. \\
Coarse tissue details. & Fine tissue details. \\
Single modality & Various modalities to record the reaction of different tissues differently (T1, T2, T1c, T2c, FLAIR, DWI). \\
Generates 3D images. & Generates 3D images. \\
Images can be in axial, coronal and sagittal views. & Images can be in axial, coronal and sagittal views. \\
\hline
\multicolumn{2}{l}{\textbf{Risks}} \\
\textbf{CT} & \textbf{MRI} \\
\hline
Harms unborn babies. & Reacts with the metals in the body due to magnetic field. \\
A small dosage of radiation. & The loud noise of the machine causes hearing issues. \\
Reacts to the use of dyes. &	Increases body temperature when exposed for a longer duration. \\
 & Imaging difficulties in case of claustrophobia. \\
 \bottomrule
\end{tabular}
\end{table}

\begin{figure}
  \centering
\begin{subfigure}{.25\textwidth}
  \centering
  \includegraphics[width=3cm,height=3cm]{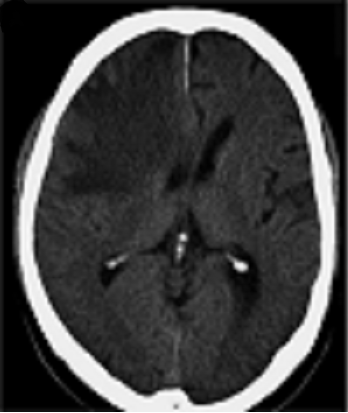}
  \caption{}
\end{subfigure}%
\begin{subfigure}{.25\textwidth}
  \centering
  \includegraphics[width=3cm,height=3cm]{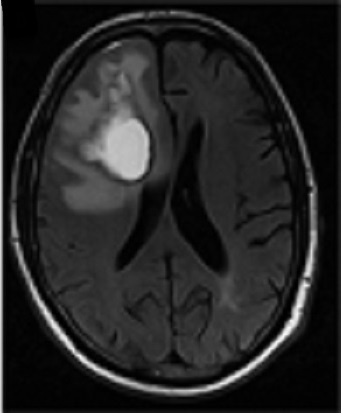}
  \caption{}
\end{subfigure}
\caption{Appearance of Tumor in a) CT and b) MRI \cite{janssen2012teaching}.}
\label{fig:2}       % Give a unique label
\end{figure}

There are other different MRI modalities that capture responses of various brain tissues differently. The tumor adds one more tissue class in the brain. Fig. \ref{fig:2} shows the difference in the appearance of a tumor in a CT and an MRI.

Table \ref{tab:2} shows the intensity variation of a tumor in different MRI modalities. The tumor class overlaps the normal tissue intensities, e.g., in T1 MR images, the GM, the CSF, and the tumor appear to be dark, whereas, in T2 the GM, the CSF and the tumor appear to be bright. It is desirable to use a combination of various modality MRI images for the purposes of analysis \cite{agravat2016brain}. The rest of the paper focuses on the methods based on MR images.

\begin{table}[htbp]
\caption{The appearance of normal brain tissues and Tumor in various MRI modalities.}
\label{tab:2}
\centering
\begin{tabular}{@{}lllll@{}}
\toprule
& \textbf{GM} & \textbf{WM} & \textbf{CSF} & \textbf{Tumor} \\
\midrule
\textbf{T1} & Dark & White & Dark & Dark \\
\textbf{T2} & Light & Dark & White & Bright \\
\textbf{T1c} & Dark & White & Dark & Bright \\
\textbf{FLAIR} & Light & Dark & Dark & Bright \\
\bottomrule
\end{tabular}
\end{table}

The availability of a benchmark dataset has boosted research in the area of computer-assisted analysis for brain tumor segmentation. Various types of methods for the segmentation task include semi-automated and automated methods. The semi-automated methods require user input to initiate the process. Circular automata and other random field methods require a seed point, a diameter, or rough boundary selection for further computation. Atlas-based methods try to fit the pathological image with a healthy image to locate the abnormal brain area. Pathological atlas creation is another approach to determine the abnormality in the brain. Expectation maximization methods iteratively refine the categories of the brain voxels from the input of Gaussian Mixture Models or atlas. The automated methods include machine learning methods, which use the features of the image for voxel classification. Later on, deep learning methods, specifically Convolution Neural Network (CNN) based methods, have shown success in the field of semantic segmentation, and such methods are adopted widely for brain tumor segmentation. CNN methods with various architectures as well as ensemble approaches have proven to be the best methods for the task of segmentation. In addition to the segmentation task, the survival prediction task predicts the survival days of patients. The contribution of the paper is as follows:

\begin{enumerate}
\item It is the most exhaustive review covering brain tumor imaging modalities, challenges in medical image analysis, evaluation metrics, BraTS dataset evolution, Pre-processing and post processing methods, segmentation methods, proposed architecture, hardware and software for implementation, overall survival predictions, and limitations.
\item It exhaustively traces the development in brain tumor segmentation by covering models based on handcrafted features to deep neural networks. It helps to understand state-of-the-art development more comprehensively.
\item A fair comparison among models is made by covering the BraTS benchmark dataset. The methods are classified, their parameters tabulated and analyzed for performance.
\item The paper also covers a survey on end-to-end methods for brain tumor segmentation and overall survival prediction. It helps to understand the impact of segmentation on overall survival prediction.

\end{enumerate}

The flow of the paper is as follows: section \ref{sec:2} covers the challenges for computer aided medical image analysis. Section \ref{sec:3} covers the problem statement, dataset, and evaluation framework. In contrast, section \ref{sec:4} includes segmentation methods using hand-crafted features with limitations, Section \ref{sec:5} covers segmentation and Overall Survival (OS) prediction using CNN methods, Section \ref{sec:6} covers the limitations of tumor segmentation and OS prediction methods followed by a conclusion and discussion in Section \ref{sec:7}.

\section{Challenges in Medical Image Analysis}
\label{sec:2}

Volumetric brain MRI images are analyzed and interpreted by human experts (neurologists, radiologists) to segment various brain tissues as well as to locate the tumor. This analysis is time-consuming. Besides, this type of segmentation is non-reproducible. Accuracy of brain tumor segmentation, which is desirable to plan proper treatment like medication or surgery, mainly depends on the human expert with utmost precision. Computer-aided analysis helps a human expert to locate the tumor in less time as well as it regenerates the analysis results. The intended analysis by computerized methods requires appropriate input with correct working methods. Input to the method may face the following challenges:
\vspace{-\topsep}

\begin{enumerate}
\item Low signal to noise ratio (SNR) and artifacts in raw MRI data are mainly due to electronic interferences in the receiver circuits, radiofrequency emissions due to thermal motion of the ions in the patient body, coils, and electronic circuits in MRI scanners. This random fluctuation reduces the image contrasts due to signal-dependent data bias \cite{goyal2018noise}.

\item Non-uniformity is an irrelevant additional intensity variation throughout the MRI signal. Possible causes of non-uniformity are radio-frequency coils, acquisition pulse sequence, the geometry and nature of the sample.

\item Unwanted information like skull, fat and skin acquired by MR machines along with brain images.

\item The intensity profile of MR images may vary due to the variety of MRI machine configurations.

\item Publicly available brain tumor images for computer-aided analysis are very few. The collection of MR images from various hospitals has privacy or confidentiality related issues. 

\item Class imbalance problem is another major issue in medical image analysis. The images for an abnormal class might be challenging to find because abnormal classes are rare compared to the normal classes.
\end{enumerate}
\vspace{-\topsep}

\section{Glioma Brain Tumor Segmentation}
\label{sec:3}

\begin{figure}[hbtp]
\centering
  \includegraphics[width=0.40\textwidth]{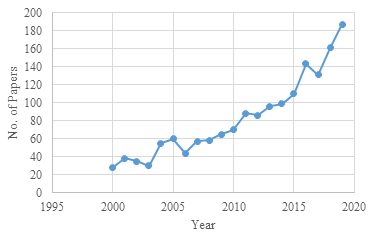}
\caption{Papers on PubMed \cite{pubmed:2020} with keywords `Brain Tumor Segmentation'.}
\label{fig:3}       % Give a unique label
\end{figure}

Focus on the methods to solve medical related issues has increased since the late 1990s, which is apparent by looking at the gradual increase in semi-automated or automated methods for tumor segmentation, as shown in Fig. \ref{fig:3}. By considering the same, the main focus of the article is on the Glioma brain tumor segmentation. An additional task is about the survival prediction techniques of the patients suffering from Glioma.

\subsection{Brain Tumor Segmentation(BraTS) Challenge Dataset}
\label{sec:3.1}

Literature proposes various methods for brain tumor segmentation. All the methods claim their own superiority and usefulness in some way. Initially, all such methods worked on the images taken from some hospitals or the radiology laboratories, which were private, and disclosure of those images to the other researchers was not allowed. This did not allow comparison of different methods. A publicly available dataset and evaluation framework  compares and evaluates methods on the same measure. The BraTS challenge dataset \cite{bakas2017segmentation,bakas2017segmentation1,bakas2017advancing} has the following characteristics:

\begin{itemize}
\item It contains multi-parametric MRI pre and post -operative scans in T1, T1Gd, T2, and T2-FLAIR volumes (The post-operative scans have been omitted since 2014);
\item The dataset contains images with different clinical protocols (2D or 3D) and various scanners from multiple institutions (1.5T or 3 T);
\item The dataset set includes images with pre-processing for their harmonization and standardization without affecting the apparent image information;
\item It has co-registration to the same anatomical template, interpolation to a uniform isotropic resolution ($1mm^{3}$), and skull-stripping.
\end{itemize}

Initially, clinical images in the dataset were very few and it was challenging to compare methods based on the results of such a small number of images. Comparison is possible with an increase in the number of sample images and accurate generation of the ground truth images. Ground truth is generated based on the evaluation by more than one expert to avoid inter-observer variability. The growth of the dataset from its inception is as shown in Table \ref{tab:3}.

\begin{table*}[htbp]
\caption{Growth of the BraTS dataset \cite{bakas2017segmentation,bakas2017segmentation1,bakas2017advancing}.}
\label{tab:3}
\centering
\begin{tabular}{@{}p{1 cm} p{1.5 cm} p{2 cm} p{1.5 cm} p{3 cm} p{2 cm} p{3 cm} @{}}
\toprule
\textbf{Year} & \textbf{Total Images} & \textbf{Training Images} & \textbf{Validation Images} & \textbf{Test Images} & \textbf{Tasks} & \textbf{Type of Data} \\
\midrule
2012 & Clinical:45, Synthetic:65 & Clinical data: 30(20HGG + 10LGG), Synthetic data:  50(25HGG + 25LGG) & N/A & Clinical data:15 Synthetic data:15 & Segmentation & Pre and post operative scans \\
\midrule
2013 & 65 & Clinical data from BraTS 2012 & N/A & Leaderboard: \newline Clinical:25 (15 from BraTS 2012 + 10 new) \newline Challenge: 10 & Segmentation & Pre and post operative scans \\
\midrule
2014 & 238 & 200 & N/A & 38 & Segmentation, Disease Progression & Pre-operative, Longitudinal \\
\midrule
2015 & 253 & 200 & N/A & 53 & Segmentation, Disease Progression & Pre-operative, Longitudinal \\
\midrule
2016 & 391 & 200 & N/A & 191 & Segmentation, Disease Progression & 	Pre-operative, Longitudinal \\
\midrule
2017 & 477 & 285 & 46 & 146 & Segmentation, Survival Prediction & Pre-operative, Longitudinal \\
\midrule
2018 & 542 & 285 & 66 & 191 & Segmentation, Survival Prediction & Pre-operative, Longitudinal\\
\midrule
2019 & 626 & 335 & 125 & 166 & Segmentation, Survival Prediction, Uncertainty Prediction & Pre-operative, Longitudinal \\
\bottomrule
\end{tabular}
\end{table*}

\begin{figure}
\centering
  \includegraphics[width=0.50\textwidth]{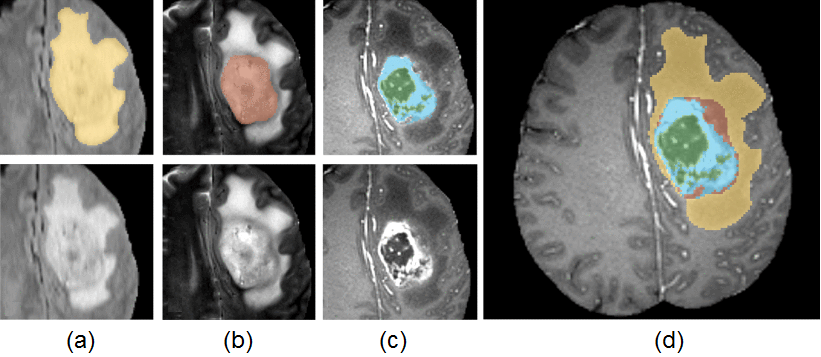}
\caption{Intra-tumoral structures appearance on three imaging modalities with manual annotations. (a) Top: whole tumor (yellow), Bottom: FLAIR, (b) Top: tumor core (red), Bottom: T2, (c) Top: enhancing tumor structures (light blue), surrounding the cystic/necrotic components of the core (green), bottom: T1c, (d) Fusion of the three labels \cite{menze2014multimodal}.}
\label{fig:4}       % Give a unique label
\end{figure}

Four different types of intra-tumoral structures are useful for ground truth generation: edema, enhancing core, non-enhancing(solid) core, and necrotic (or fluid-filled) core as shown in Fig. \ref{fig:4}. An annotation protocol was used by expert raters to annotate each case manually. Then the segmentation results from all raters were fused to obtain a single unanimous segmentation for each subject as the ground truth. The validation of the segmentation methods is based on: 1) Whole Tumor (WT): all intra-tumoral substructures, 2) Tumor Core (TC): enhancing tumor, necrosis, and non-enhancing tumor substructures and 3) Enhancing Tumor (ET): includes only enhancing substructure.

\subsubsection{Overall Survival Prediction}

On account of the availability of a sufficiently large dataset, the additional task of OS prediction has been introduced in BraTS challenge since 2017. This task focuses on the OS prediction of HGG patients. The dataset includes age, survival days, and resection status: Gross Total Resection (GTR) or Sub-Total Resection (STR) information for HGG patients in addition to the images. The task is to classify the patients into: long-survivors ($>$15 months), mid-survivors(between 10 to 15 months), and short-survivors($<$10 months)\cite{bakas2018identifying}. A detailed description of the OS prediction task is given in Table \ref{tab:4}. In 2019, an additional task included the quantification of uncertainty prediction in the segmentation. This task focuses on the uncertainty prediction in the context of glioma tumor segmentation.

\begin{table*}[htbp]
\caption{The distribution of BraTS dataset \cite{bakas2017segmentation,bakas2017segmentation1,bakas2017advancing} features in survival classes.}
\label{tab:4}
\centering
\begin{tabular}{@{} p{0.75 cm} p{1.5 cm} p{1 cm} p{0.75 cm} p{1.25 cm} p{1.25 cm} p{0.75 cm} p{1.25 cm} p{1.25 cm} p{0.75 cm} p{1.25 cm} p{1.25 cm}@{}}
\toprule
\multirow{2}{*}{\textbf{Year}} & \multirow{2}{*}{\textbf{\# records}} & \multirow{2}{*}{\textbf{Features}} & \multicolumn{3}{C{3.25 cm}}{\textbf{Short survivors (\textless 10 months)}} & \multicolumn{3}{C{3.25 cm}}{\textbf{Mid survivors (between 10 and 15 months)}} & \multicolumn{3}{C{3.25 cm}}{\textbf{Long Survivors (\textgreater 15 months)}} \\
\cmidrule{4-12}
 & & & Count & Age ($\mu\pm\sigma$) & OS days ($\mu\pm\sigma$) & Count & Age ($\mu\pm\sigma$) & OS days ($\mu\pm\sigma$) & Count & Age ($\mu\pm\sigma$) & OS days ($\mu\pm\sigma$) \\
 \midrule
2017 and 2018 & 163 & Age & 65 & 65.44 $\pm$ 10.68 & 147.44 $\pm$ 83.08	& 50 & 58.70 $\pm$ 11.26 & 394 $\pm$ 49.32 & 48 & 55.11 $\pm$ 12.19 & 826.23 $\pm$ 370.91 \\
 \midrule
2019 & 212 & Age, Resection status & 82 & 66.66 $\pm$ 11.42 & 150.21 $\pm$ 84.72 & 54 & 59.14 $\pm$ 10.98 & 377.43 $\pm$ 40.44 & 76 & 57.16 $\pm$ 11.84 & 796.38 $\pm$ 354.32 \\
\bottomrule
\end{tabular}
\end{table*}

\subsection{Evaluation Metrics for Brain Tumour Segmentation and OS prediction}
\label{sec:3.2}

The standard evaluation framework for tumor segmentation and OS prediction includes the following metrics.

\begin{enumerate}
	
\item Dice Similarity Coefficient (DSC) (or F1 measure): It is the overlap of two objects divided by the total size of both the objects. True Positive (TP) is the outcome where the model correctly predicts the positive class. In contrast, False Positive (FP) is the outcome where the model incorrectly predicts the negative class to be positive. False Negative (FN) is the outcome where the model incorrectly predicts the positive class to be negative. This has been shown in equation \ref{equation:1}.	
	
\begin{equation}
\label{equation:1}
DSC = \frac{2TP}{2TP+FP+FN}
\end{equation}

\item Jaccard Similarity Coefficient: It is known as the intersection over the union of two different sets. This has been shown in equation \ref{equation:2}.

\begin{equation}
\label{equation:2}
Jaccard = \frac{TP}{TP+FP+FN}
\end{equation}

\item Sensitivity: It is a measure that correctly identifies tumorous voxels. This has been shown in Equation \ref{equation:3}.

\begin{equation}
\label{equation:3}
Sensitivity = \frac{TP}{TP+FN}
\end{equation}

\item Hausdorff distance: It measures how far two subsets of a metric space are from each other. If $x$ and $y$ are two non-empty subsets of a metric space $(M,d)$, then their Hausdorff distance $d_{H}(x, y)$ can be defined by:

\begin{equation}
d_H(x,y) = max \{ \underset{x \in X} {sup} \, \underset{y \in Y} {inf} \, d(x,y), \underset{y \in Y} {sup} \, \underset{x \in X} {inf} \, d(x,y) \}
\end{equation}

where, \textit{sup} represents the supremum and \textit{inf} the infimum.

\item The OS prediction is measurable with accuracy, which is defined to be the quality of being precise.

\end{enumerate}

As DSC is the most commonly used evaluation metric, this article compares all the methods based on DSC unless specified explicitly.

\subsection{Image Pre-processing}
\label{sec:3.3}

Medical image pre-processing plays a significant role in the appropriate input to computer-assisted analysis techniques. The pre-processed images help to get accurate outputs as such images show proper voxel relationships. As mentioned in section \ref{sec:3.1}, the dataset has images with different clinical protocols and scanners; the variability has to be standardized , and it is necessary to put all the scans on a single scale. In addition to image registration, uniform isotropic resolution, and skull-stripping, the following types of pre-processing further improve the image input:

\begin{itemize}
	
\item Bias Field Correction(BFC): Bias field is a multiplicative field added in the image due to the magnetic field and radio signal in the MR machine. Authors in \cite{tustison2010n4itk} have suggested a bias field correction technique.

\item Intensity Normalization (IN): Different modality images have a separate intensity scale. They must map to the same range. Standardization of all the scans considers zero mean and unit variance.

\item 	Histogram Matching (HM): Due to the different configuration of MR machines, the intensity profile of the acquired images may vary. Intensity profiles are to be brought to the same scale using the histogram matching process.

\item Noise Removal (NR): Noise in MR image at the time of acquisition is due to the radio signal, or the magnetic field. Various noise filtering techniques are useful for the removal of noise.

\end{itemize}

\subsection{Image Post-processing}
\label{sec:3.4}

Segmentation output generated by computer-assisted methods may contain false segmentation in the image due to improper or incorrect feature selection. Segmentation improves by applying post-processing techniques like:

\begin{enumerate}
\item Connected Component Analysis (CCA): CCA groups the voxels based on connectivity depending on similar voxel intensity values. The connected components which are very small are excluded from the result as such components are considered to be false positives due to spurious segmentation results.

\item Conditional Random Field (CRF): The classifier predicts the voxel class based on features related to that voxel, which does not depend on the relationship of that voxel with other nearby voxels. CRF takes this relationship into consideration and builds a graphical model to implement dependencies between predictions.

\item Morphological Operations: Such operations are applied to adjust the voxel value based on the value of the other voxels in its neighbourhood according to the size and shape.
\end{enumerate}

\section{Segmentation Methods using Handcrafted Features}
\label{sec:4}

The methods in this section are divided into two categories: interactive and noninteractive. Interactive methods require user input in the form of tumor diameter, boundary or seed point selection. Random field-based methods belong to this category. Non-interactive methods do not require user input. The first group of this category detects abnormality. In atlas-based methods, the abnormal image is non-linearly mapped to the normal/input specific atlas to identify the abnormality which in general is the area of the image where atlas mapping fails. The other approach is Expectation Maximisation (EM), where the intensity distribution of the normal and abnormal voxels is learnt with Gaussian Mixture Models (GMM) or probabilistic atlases. The second group of this category is machine learning approaches. The clustering approach groups the voxels in numbers of clusters such that one of these clusters will result in a tumorous voxels group. In random forest (RF) and neural network (NN) approaches, high dimensional features of the images are given for training. The trained model later classifies the unseen voxels. A detailed description of all these approaches is covered in the following sub-sections.

\subsection{Random Field Based Methods}
\label{sec:4.1}

Authors in \cite{hamamci2011tumor} took user input for the largest possible tumor diameter from HGG images to find Volume of Interest (VoI) for a tumor and background from T1C MRI images, followed by Cellular Automata (CA) to obtain the probability maps for both the regions. A level set surface was further applied to those probability maps to get the final probability maps. They further extended their approach in \cite{hamamci2012multimodal}, which considered multimodal images (T1C and FLAIR images) to segment tumors, edemas as well as LGG images.

A semi-automatic method in \cite{guo2013semi} took the rough tumor region boundary as user input and fine-tuned it with a global and local active contour-based model. The tumor region was broken into sub-regions with adaptive thresholding based on a statistical analysis of the intensities of various tumor regions to separate an edema from an active tumor core. The process was repeated for all the slices of the MRI of a patient. In \cite{corso2008efficient}, the main contribution was to incorporate soft model assignments into the calculations of model-free affinities, which were then integrated with model-aware affinities for multilevel segmentation by a weighted aggregation algorithm.

In \cite{xiao2012hierarchical}, Random Walk (RW) based interactive as well as an iterative method was applied to fine-tune the tumor boundary. RW was applied as an edge-weighted graph in the discrete feature space based on the variation of the distribution density of the voxels in the feature space. The user made an initial tumor seed selection for tumors as well as edema. Next, RW was applied to the feature space as well as on the image. If the user did not approve the results, then the segmentation process was reinitiated.

In \cite{doyle2013fully}, the Hidden Markov Random Field (HMRF) based model was used with a modified Pott’s Model to panelize the neighboring pixels belonging to different classes. In \cite{taylor2013map}, various modality intensity images along with their neighbourhood voxel intensities fed into the map-reduced Hidden Markov Model (HMM), and the model was corrected iteratively based on the class labels.

In \cite{subbanna2012probabilistic}, Gabor filter bank-based Bayesian classification was followed by MRF classification. Initially, each voxel was divided into its constituent class by applying the Gabor filter bank to the input vector (made up of intensities of four modalities at a voxel) to classify the voxel in five different classes (GM, WM, CSF, tumor, and edema). Next, an MRF based classifier was applied to the tumor as well as edema classes. It uses voxel intensity and spatial intensity differences over the neighbouring voxels. Authors in \cite{dera2016interactive} used the Non-Negative Matrix to find voxel clusters, which showed the tumor and level set methods to fine-tune the region boundary.

\subsection{Atlas Based Methods}
\label{sec:4.2}

Authors in \cite{prastawa2004brain}, initially identified abnormal brain tissues by registering a tumorous brain image with a healthy tissue brain atlas. This step was followed by identifying the presence of an edema using T2 images, and finally, geometric and spatial constraints were applied to detect the tumor and edema regions. Authors in \cite{cuadra2004atlas} applied an affine transformation to the atlas image to globally match the patient. The lesion was segmented using the Adaptive Template Moderated Spatially Varying statistical Classification (ATM SVC) algorithm. This atlas was then manually seeded by an expert with a single voxel placed on the estimated origin of the patient’s lesion, which was followed by the nonlinear demons registration algorithm with the model of the lesion growth to deform the seeded atlas to match the patient. Four volumes of the contrast-enhanced agent with meningioma implemented the model.

The paper in \cite{kwon2014multimodal} applied a semi-automatic method, which required user input to give the seed point for a tumor, the radius for each tumor, and the seed point for each regular tissue class. The random walk generated tumor priors using initial tumor seeds. The patient-specific atlas was modified to accommodate tumor classes, using a tumor growth model. The Empirical Bayes model used the EM framework to update the posterior of the tumor, growth model parameters as well as a patient-specific atlas. This work was then extended in \cite{bakas2015glistrboost}, where instead of a single seed point for various labels, multiple seed points were taken into consideration to find the intensity mean and variance of a specific label. This work focused on preoperative MRI scans. Whereas, it was further extended in \cite{zeng2016segmentation} to include post-operative scans along with additional features to the GMM. Here the need for manual selection of the seed point was also omitted.

\subsection{Expectation Maximisation Based Methods}
\label{sec:4.3}

The authors in \cite{menze2012segmenting} used spatially varying probability prior (atlas) to labelling healthy tissues in the brain. The latent probabilistic atlas ‘alpha’ finds the probability of having a tumor at that voxel. Gaussian distribution for healthy tissue classes, as well as tumorous tissue class, was taken into consideration to identify the tumorous tissues in the image by expectation maximization. For spatial tumor regularization, latent atlas alpha used MRF using channel-specific regularization parameters. In contrast, authors in \cite{raviv2012multi} used probability priors to predict voxel probability using the EM algorithm, followed by a level set framework with a gradient descent approach for parameter estimation.

The authors in \cite{zhao2012brain} applied the Gaussian Mixture Model with EM to divide the data into five probable classes. Afterwards, the snake method was used to find the subtle boundary between the tumor and the edema. In \cite{tomas2012automatic}, the Bayesian intensity tissue model used the expectation-maximization problem using a trimmed likelihood estimator (TLE), which was robust against the outliers. This step was followed by a graph cut algorithm to differentiate between tumorous tissues as well as false positives.

Authors in \cite{zhao2013automatic} selected super voxels using Simple Linear Iterative Clustering (SLIC) 3D based on the colour and proximity of the voxels. A graph cut on MRF was implemented for initial segmentation, followed by histogram construction, histogram matching, and likelihood estimation applied to predict the probability of voxels. In \cite{piedra2016brain} the posterior probability using the Bayes theorem was calculated, which was followed by Super voxel partitioning using the SLIC algorithm.

\subsection{Clustering Based Methods}
\label{sec:4.4}

In \cite{cordier2013patch} from all the images, D patches of size 3x3x3 were selected in the database. A single patch of this small cube used all the four modalities. Once the database was ready, from the test dataset, the same size patches were extracted and mapped with the database patches with $k$-nearest neighbors where $k = 5$. These five nearest neighbors generated the test patch label.

The authors in\cite{clark1998automatic} used knowledge-based multispectral analysis on top of the unsupervised clustering technique. The rule-based expert system then extracted the intracranial region. The training was performed on three T1, Proton Density(PD), and T2 images, whereas the testing was done on thirteen such volumes. In \cite{saha2016brain} authors used rough-set approximation to fine-tune the prediction done by k-means clustering.

In \cite{rajendran2012brain} initially, the enhanced probabilistic fuzzy C-means clustering was applied to get a rough estimation of the tumor region. This estimation and cluster centroid was given to the gradient-vector-flow snake model to refine the tumor boundary. \cite{shin2012hybrid} learnt a sparse dictionary of size 4x4 for different tissue types using four image modalities followed by logistic regression for tissue classification. This initial stage classified the image voxels in various classes. This step was followed by k-means clustering, which used a very high dimensional feature vector as input to find the classification of the tumor as well as the edema region as an overlap of the output of the previous step.

\subsection{Random Forest Based methods}
\label{sec:4.5}

Meier et al. \cite{meier2013hybrid} implemented a generative- discriminative model. The generative model estimated tissue probability using density forest (similar to GMM). The discriminative model implemented classification forest, which took 51 features (gradient information first-order texture features and symmetry-based features, prior tissue probabilities based on the density forest) as input to generate the probability of the tissue. This probability was then supplied to the CRF to fine-tune the result.

Authors in \cite{zikic2012decision} used context-aware spatial features, along with the tissue appearance probability generated by the Gaussian Mixture Model, to train a decision forest. The authors in \cite{bauer2012segmentation} worked on Random forest classification with CRF regularization to predict the probability of tissue in multiple classes, i.e., GM, WM, CSF, Edema, Necrotic core and enhancing tumor. 28-dimensional feature vector including the intensity of each modality along with first-order statistics like mean, variance, skewness, kurtosis, energy, and entropy were computed from local patches around each voxel in each modality.

In \cite{geremia2012spatial} each voxel was characterized by signal modality as well as spatial prior. Averaging across a 3x3x3 cube removed noise. The random forest trained using local features (intensity or priors), as well as context-specific features (region-based features or symmetry-based features). The forest was made up of 30 trees with a depth of 20 for each tree and trained on synthetic data.

Zikic et al. \cite{zikic2012context} used a discriminative multiclass classification forest. It used spatially non-local context-sensitive high dimensional features along with the prior probability for the tissue to be classified. Prior probability was available to the GMM. It classified in three classes, i.e., background, edema, and tumor. Three types of features train 40 trees with a depth of 20 each, and such features were intensity difference, intensity mean difference, and intensity range on a 3D line to check for structural changes. A 2000 combination is selected to design the decision trees. Festa et al. \cite{festa2013automatic} trained a random forest of 50 trees each of depth 25 on 120000 samples with 324 features which comprised of intensity, neighbourhood information, context information, and texture information.

In \cite{reza2013multi}, a random forest classifier used intensities, the difference of intensities (T1 - T2, T1- Flair, T1 – T1c), and the texture-based features such as fractal and texton to train a RF using three-fold cross-validation. The authors further extended their work in \cite{reza2014improved} for post-processing, where the connected component analysis removed tiny regions in the 3D volume, and holes in between the various parts of the segmented tumor according to the neighbouring region.

In \cite{goetz2014extremely} Extremely Randomized Trees (ExtraTrees) were used, which introduced more randomness at the time of training. The classifier was trained on 208 features extracted from all four modalities, which included intensity values, local histograms, first-order statistics, second-order statistics, and basic histogram-based segmentation. In the paper, the ExtraTrees trained with the best threshold rather than the threshold derived from individual features. In \cite{kleesiek2014ilastik}, pixel classification was done with ten random forests with ten trees each, which were trained in parallel to reduce the training time and finally merged into a single forest with Gini impurity. One thousand samples for tumorous class and 1000 samples for the non-tumorous class trained the RF. A Classification forest in \cite{meier2014appearance} used 237 features which included appearance specific features (image intensities, first-order texture features, and gradient features), context sensitive features (ray feature, symmetry intensity difference features). Authors in \cite{ellwaa2016brain}, \cite{le2016lifted}, \cite{lefkovits2016brain}, \cite{maier2015image}, \cite{meier2015parameter}, \cite{meier2016crf}, \cite{phophalia2017multimodal}, \cite{song2016anatomy} used a random forest classifier with a combination of various intensity based features, gradient based features, texture based features and rotation invariant features.

In \cite{bharath2017tumor} tensor features were extracted along with mean, entropy and standard deviation features. Authors in \cite{serrano2019brain} extracted features for super voxels from multi scale images and created sparse feature vectors to segment the whole tumor. Subregions of the tumor were then separated using CRF.

\subsection{Neural Network Based methods}
\label{sec:4.6}

In \cite{buendia2013grouping} Grouping Artificial Immune Network (GAIN) took voxel intensity from 2D as well as 3D slices as input and statistical features and texture features for training as well as segmentation of brain MRI images. Information was in the form of bits from various image modalities. In \cite{agn2015brain} Convolutional Restricted Boltzmann Machine trained GMM and spatial tissue priors.

Table \ref{tab:5} summarizes the methods for the type of pre-processing, the dataset, the number of images used as well as the DSC achieved. The DSC is shown for various tumor sub-regions which include WT, ET, TC, Edema(ED) and Necrosis(NC) for training, validation and test datasets.

\begin{figure*}
\centering
  \includegraphics[width=0.80\textwidth]{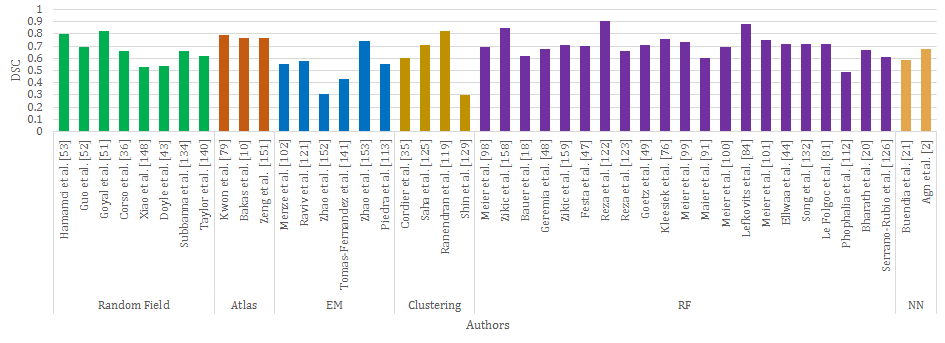}
\caption{Comparison of segmentation methods using hand-crafted features 
(DSC for validation/Test for TC of all images/HGG). }
\label{fig:5}       % Give a unique label
\end{figure*}

The DSC comparison of different methods for tumor segmentation is shown in Fig. \ref{fig:5}. The comparison is based on the TC region of validation or test set of either all the images or only the HGG images. The atlas and RF-based methods performed well compared to all other approaches.

Random field methods, atlas-based methods, expectation maximization methods, and clustering methods do not use any post-processing techniques. Random forest-based methods were used by authors in \cite{ellwaa2016brain}, \cite{festa2013automatic}, \cite{kleesiek2014ilastik}, \cite{maier2015image}, \cite{meier2014appearance}, \cite{meier2015parameter}, \cite{reza2014improved} and \cite{song2016anatomy} for segmentation. However, \cite{festa2013automatic}, \cite{kleesiek2014ilastik} and \cite{reza2014improved} used connected component analysis, \cite{meier2014appearance}, \cite{meier2015parameter}, \cite{song2016anatomy} applied spatial regularization and authors in \cite{ellwaa2016brain} and \cite{maier2015image} applied morphological operations to refine the segmentation output.

\tablefirsthead{%
     \toprule  \\
       \textbf{Ref.} & \textbf{Pre-processing} & \textbf{Dataset} & \textbf{\# Images} & \textbf{DSC Mean} \\
     \midrule}
 
%This is the header for the remaining page(s) of the table...
\tablehead{
    
    \multicolumn{5}{c}
    {{\bfseries \tablename\ \thetable{} --
       continued \ldots}} \\
     \toprule
     \textbf{Ref.} & \textbf{Pre-processing} & \textbf{Dataset} & \textbf{\# Images} & \textbf{DSC Mean} \\
   }
 
%This is the footer for all pages except the last page of the table...
\tabletail{%
	\midrule 
}
\tablelasttail{\bottomrule}
\tablecaption{Summarization of segmentation methods using handcrafted features.}
\label{tab:5}

\begin{small}
\begin{xtabular}{@{}p{0.75 cm} p{1.5 cm} p{1.3 cm} p{1.1 cm} p{1.75 cm}@{}}
\multicolumn{5}{l}{\textbf{Random Field methods}} \\
\midrule
\cite{hamamci2011tumor} & - & Custom & 29 & Training TC:0.89 Test TC:0.80 \\
\midrule
\cite{hamamci2012multimodal} & - & BraTS 2012 & 30 & TC:0.69,ED:0.37 \\
\midrule
\cite{guo2013semi} & - & BraTS 2013 & 30 & TC:0.82 \\
\midrule
\cite{corso2008efficient} & IN & Custom & 20 & Training TC:0.70,ED:0.66 Test TC:0.66,ED:0.61 \\
\midrule
\cite{xiao2012hierarchical} & NR & BraTS 2012 & 30 & TC:0.53,ED:0.25 \\
\midrule
\cite{doyle2013fully} & - & BraTS 2013 & 30 & HGG WT:0.84,TC:0.54, ET:0.67 \newline LGG WT:0.81,TC:0.54, ET:0.11 \\
\midrule
\cite{subbanna2012probabilistic} & - & BraTS 2012 & 28 & TC:0.66,ED:0.56 \\
\midrule
\cite{taylor2013map} & BFC & BraTS 2013 & 30 & HG \newline TC:0.62,ED:0.59 \\
\midrule
\multicolumn{5}{l}{\textbf{Atlas Based Methods}}\\
\midrule
\cite{kwon2014multimodal} & BFC, IN & BraTS 2014 & 200 & Validation WT:0.86,TC:0.79, ET:0.59 Test WT:0.88,TC:0.83, ET:0.72 \\
\midrule
\cite{bakas2015glistrboost} & BFC, IN & BraTS 2015 & 186 & Training WT:0.88,TC:0.77, ET:0.68 \\
\midrule
\cite{zeng2016segmentation} & NR, HM & BraTS 2016 & 200 & WT:0.89,TC:0.77, ET:0.67\\
\midrule
\multicolumn{5}{l}{\textbf{Expectation Maximization Based Methods}}\\
\midrule
\cite{menze2012segmenting} & - & BraTS 2012 & 30 & HGG TC:0.55,ED:0.57 LGG TC:0.24,ED:0.42  \\
\midrule
\cite{raviv2012multi} & - & BraTS 2012 & 30 & HGG TC:0.58,ED:0.60 LGG TC:0.32,ED:0.36 \\
\midrule
\cite{zhao2012brain} & - & BraTS 2012 & 30 & TC:0.31,ED:0.35 \\
\midrule
\cite{tomas2012automatic} & - & BraTS 2012 & 30 & TC:0.43,ED:0.55\\
\midrule
\cite{zhao2013automatic} & IN & BraTS 2013 & 30 & HGG WT:0.83,TC:0.74, ET:0.68 \newline LGG WT:0.83,TC:0.58, ET:0.51\\
\midrule
\cite{piedra2016brain} & - & BraTS 2015 & 200 & Training WT:0.74,TC:0.55, ET:0.54 \\
\midrule
\multicolumn{5}{l}{\textbf{Clustering Based Methods}}\\
\midrule
\cite{cordier2013patch} & - & BraTS 2013 & 30 & HGG WT:0.79,TC:0.60, ET:0.59 \newline LGG WT:0.76,TC:0.64, ET:0.44\\
\midrule
\cite{saha2016brain} & - & BraTS 2013 BraTS 2015 & 200 & WT:0.82,TC:0.71, ET:0.72 \\
\midrule
\cite{rajendran2012brain} & - & Custom & 15 & TC:0.82 \\
\midrule
\cite{shin2012hybrid} & - & BraTS 2012 & 30 & TC:0.30,ED:0.39 \\
\midrule
\multicolumn{5}{l}{\textbf{Random Forest Based Methods}}\\
\midrule
\cite{meier2013hybrid} & - & BraTS 2013 & 30 & HGG WT:0.80,TC:0.69, ET:0.69 \newline LGG WT:0.76,TC:0.58, ET:0.20 \\
\midrule
\cite{zikic2012decision} & - & Custom & 40 & TC:0.85,NC:0.75, ET:0.80 \\
\midrule
\cite{bauer2012segmentation} & BFC, IN, HM & BraTS 2012 & 30 & HGG-real TC:0.62,ED:0.61 LGG-real TC:0.49,ED:0.35 \\
\midrule
\cite{geremia2012spatial} & HM & BraTS 2012 & 30 & HGG-real TC:0.68,ED:0.56 LGG-real TC:0.52,ED:0.29 \\
\midrule
\cite{zikic2012context} & BFC & BraTS 2012 & 30 & HGG-real TC:0.71,ED:0.70 LGG-real TC:0.62,ED:0.44 \\
\midrule
\cite{festa2013automatic} & BFC, HM & BraTS 2013 & 30 & HGG WT:0.83,TC:0.70, ET:0.75 \newline LGG WT:0.72,TC:0.47, ET:0.21 \\
\midrule
\cite{reza2013multi} & BFC, HM & BraTS 2013 & 30 & HGG WT:0.92,TC:0.91, ET:0.88 \newline LGG WT:0.92,TC:0.91, ET:0.88 \\
\midrule
\cite{reza2014improved} & - & BraTS 2014 & 200 & Training WT:0.81,TC:0.66, ET:0.71 \\
\midrule
\cite{goetz2014extremely} & BFC, HM & BraTS 2013 & 208 & WT:0.83,TC:0.71, ET:0.68 \\
\midrule
\cite{kleesiek2014ilastik} & HM, IN & BraTS 2014 & 200 & Training WT:0.84,TC:0.68, ET:0.72 Valid/Test WT:0.87,TC:0.76, ET:0.64 \\
\midrule
\cite{meier2014appearance} & NR, IN, BFC & BraTS 2013 & 30 & Training WT:0.83,TC:0.66, ET:0.58 Challenge WT:0.84,TC:0.73, ET:0.68 \\
\midrule
\cite{maier2015image} & BFC, IN & BraTS 2015 & 252 & WT:0.75,TC:0.60, ET:0.56 \\
\midrule
\cite{meier2015parameter} & - & BraTS 2015 & 65 & WT:0.83,TC:0.69, ET:0.63 \\
\midrule
\cite{lefkovits2016brain} & BFC, NR, IN & BraTS 2013 & 30 & WT:0.87,TC:0.88 \\
\midrule
\cite{meier2016crf} & - & BraTS 2013 & 30 & Valid WT:0.79,TC:0.75, ET:0.66 \newline Challenge WT:0.83,TC:0.76, ET:0.71 \\
\midrule
\cite{ellwaa2016brain} & BFC, HM & BraTS 2016 & 200 & WT:0.80,TC:0.72, ET:0.73 \\
\midrule
\cite{song2016anatomy} & IN & BraTS 2015 & 274 & Training WT:0.87,TC:0.72, ET:0.75 \\
\midrule
\cite{le2016lifted} & IN & BraTS 2015 & 274 & Training WT:0.84,TC:0.72, ET:0.71 \\
\midrule
\cite{phophalia2017multimodal} & BFC & BraTS 2017 & 285 & Training
WT:0.64,TC:0.49, ET:0.47 Test WT:0.63,TC:0.41, ET:0.42 \\
\midrule
\cite{bharath2017tumor} & IN, HM & BraTS 2017 & 285 & Validation
WT:0.79,TC:0.67, ET:0.61 Test WT:0.77,TC:0.61, ET:0.50 \\
\midrule
\cite{serrano2019brain} &  - & BraTS 2018 & 285 & Validation 
WT:0.80,TC:0.63, ET:0.57 Test WT:0.73,TC:0.58, ET:0.50 \\
\midrule
\multicolumn{5}{l}{\textbf{Neural Network Based Methods}}\\
\midrule
\cite{buendia2013grouping} & - & BraTS 2013 & 30 & WT:0.73,TC:0.59, ET:0.63 \\
\midrule
\cite{agn2015brain} & - & BraTS 2015 & 200 & Test WT:0.81, TC:0.68, ET:0.65 \\
\end{xtabular}
\end{small}
\vspace{2pt}
~

Limitations associated with methods working on handcrafted features are as follows:

\begin{enumerate}
\item \textbf{Identifying tissue probability classes:} tumor tissue intensities overlap with that of the healthy tissues, as mentioned in Table \ref{tab:2}; in such a case identifying the probable class for tumorous tissue is quite challenging.	
	
\item \textbf{Atlas matching (healthy or tumorous atlas):} Usually, the brain atlas contains the normal brain tissue distribution map. Due to the deformation of the healthy tissue by the tumor, the atlas matching of a tumorous brain may result in the wrong map.

\item \textbf{Manual seed point identification for the tumor or its subparts:} Almost all semiautomated methods require some initial selection for the tumorous voxel, its diameter, or its rough outer boundary. The selection depends on the expert. Its repetition over all the slices of the brain is a time-consuming task.

\item \textbf{Feature extraction from the images:} RF training depends on the features extracted from the brain images. All the MRI modalities contain different biological information. This variation in the information complicates the tasks of feature extraction as well as selection to training the RF.

\item \textbf{Discontinuity:} The results generated by such methods are spurious, which increases the chances of false positives. Proper post-processing techniques are required to fine-tune the generated results.

\end{enumerate} 

\section{Deep Neural Network}
\label{sec:5}

Deep Neural Network (DNN) is an artificial intelligence function which mimics the human brain when working for data processing and pattern creation in decision making. There are mainly four reasons contributing to its success:

\begin{enumerate}

\item The DNN models solve problems in an end-to-end manner. The models learn the features from the data automatically with the help of various functions. Feature learning improves from simple features at initial layers to complex features at deeper layers of the model. Automatic feature learning has eliminated the need for domain expertise.
\item Computational capabilities of the hardware in terms of GPU and efficient implementation of the model on a GPU with various open source libraries have made the training of the DNN 10 to 50 times faster than the CPU.
\item Efficient optimization techniques for robust learning contribute to the success of DNN for optimal network performance.
\item The availability of benchmark datasets allows training and testing of various deep learning models to be implemented successfully.
\end{enumerate}

\begin{figure}
\centering
  \includegraphics[width=0.50\textwidth]{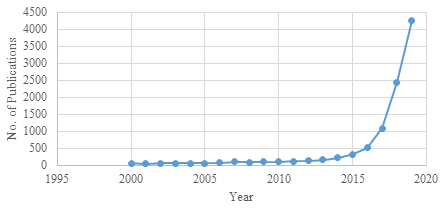}
\caption{The growth rate of research papers with the keyword `deep learning' on PubMed \cite{pubmed:2020}.}
\label{fig:6}       % Give a unique label
\end{figure}

\begin{figure}
\centering
  \includegraphics[width=0.50\textwidth]{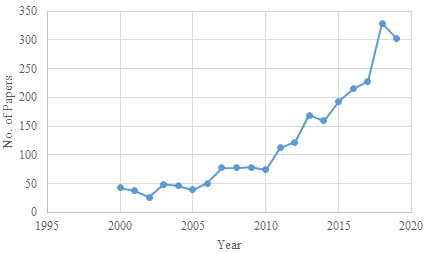}
\caption{The growth rate research papers with the keywords `deep learning' and  `brain tumor segmentation' on PubMed \cite{pubmed:2020}.}
\label{fig:7}       % Give a unique label
\end{figure}

The exponential growth of the usage of DNN techniques to solve a variety of problems is shown in Fig. \ref{fig:6}. A similar growth pattern is identified for solving the brain tumor segmentation problem, which is as shown in Fig. \ref{fig:7}.

\begin{figure*}
\centering
  \includegraphics[width=0.80\textwidth,height=6cm]{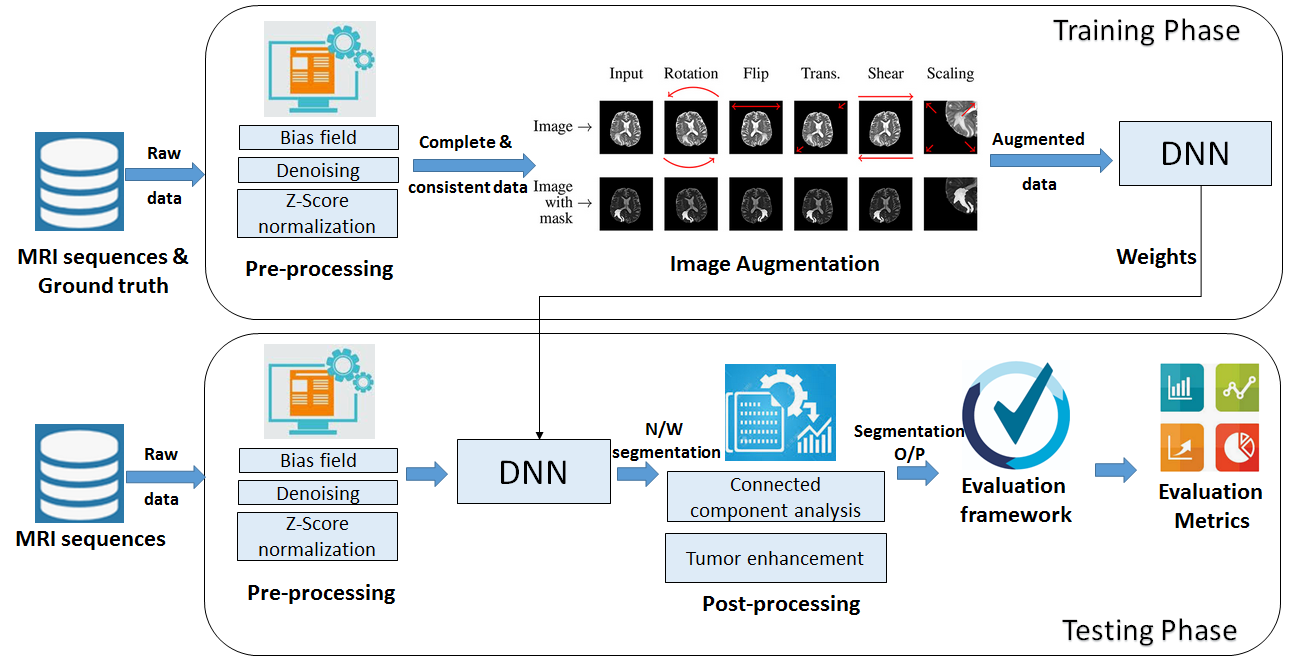}
\caption{Generalize DNN network.}
\label{fig:8}       % Give a unique label
\end{figure*}

The general block diagram of any deep learning technique is as shown in Fig. \ref{fig:8}. The crucial task is to get the labelled data set. After the availability of the dataset, it is divided into training and validation sets, followed by appropriate pre-processing techniques as per the task on hand. Actual DNN applies to the training data, which makes the network learn the network parameters. The output of DNN is spurious in some brain areas, and post-processing fine-tunes the segmentation result. And at last, the evaluation framework measures the performance of the network.

\subsection{Evolution of DNN}
\label{sec:5.1}

In medical image analysis, the semantic segmentation task is common, e.g., segmentation of an organ or a lesion. The Convolutional Neural Network (CNN), specifically the DNN architecture gained its popularity from 1990 with the architecture of LeNet\cite{lecun1998gradient}, which was two layers of architecture. After the availability of fast GPUs and other computing facilities, over fifteen years later, AlexNet was proposed by authors in \cite{krizhevsky2012imagenet} with five convolutional layers. The CNNs are designed with a variety of layers (like convolution layers, non-linearity layers, pooling layers, fully-connected layers), regularization, optimization and normalization, loss functions, as well as network parameter initializations. Authors in \cite{bernal2019deep,litjens2017survey} have nicely explained the architectural elements of CNN which are as follows:

\begin{description}
			
\item[\textbf{Convolution layer:}] extracts representative features from the input. It achieves: 1) weight-sharing mechanism, 2) exploits local connectivity of the input, and 3) provides shift invariance to some extent.

\item[\textbf{Non-linearity layer:}] provides sparse representation of input space which achieves data variability invariance and a computationally efficient representation. Types of non-linearity layers are Rectified Linear Unit (ReLU), Leaky ReLU(LReLU), Parametric ReLU (PReLU), S-shaped ReLU(SReLU), Maxout and its variants, Exponential Linear Unit(ELU) and its variants.

\item[\textbf{Pooling/subsampling layer:}] extracts prominent features from a non-overlapping neighbourhood. It is used to: 1) reduce the number of parameters, 2) reduce over-fitting, and 3) achieve translation invariance. Commonly used pooling techniques are max pooling and average pooling.

\item[\textbf{Fully connected layer:}] converts 2D features to a 1D feature vector. It helps to predict an input image class label.

\item[\textbf{Loss functions:}] improve the learning process by improving within class similarity and between class separability.

\item[\textbf{Regularizations:}] deal with over-fitting issues. Commonly used regularization techniques are L1 and L2 regularizations, dropout, early stopping and batch normalization.

\item[\textbf{Optimization:}] used for proper updates of network parameters during back propogation. Various techniques of optimization include Nesterov accelerated gradient descent, adaptive gradient algorithm (Adagrad), Root Means Square Propogation (RMSProp).

\item[\textbf{Weight initialization and normalization:}] boosts the learning process by helping the weight update with proper initial values.
\end{description} 

The convolution layers extract features from the input by applying kernels to it. The output feature map depends on the type of the kernel and its size. At the initial layers simple features are extracted from the input like edge or lines. The gradual increase of the network depth requires higher number of feature maps to extract complex shapes \cite{agravat2018deep}. The activation function is applied to the feature maps to learn the non-linear relationship within the data and allows the errors to backpropagate to the initial layers for accurate parameter updates. An increase in the network depth exponentially increases network parameters, which is computationally highly expensive. Pooling layers are introduced to down sample the input feature maps and reduce their spatial size by considering only the prominent features. It balances the growth of the network parameters. The fully connected layers at the end of the network flatten the result of the input layers before actual classification. The loss function at the classification layer calculates the error in the prediction. Based on this error, the network parameters are updated using the gradient descent methods by backpropagation. Commonly used loss functions are:

\begin{itemize}
\item Cross Entropy loss function:
\begin{equation}
J = - \frac{1}{N} \left (\sum_{voxels} y_{true} \cdot log \,\hat{y}_{pred} \right )
\end{equation}

\item Dice Loss function:
\begin{equation}
J = 1 - \frac{2 \sum_{voxels}y_{true} y_{pred} + \epsilon}{\sum_{voxels}y_{true}^{2} + \sum_{voxels} y_{pred}^{2}  + \epsilon}
\end{equation}
 \end{itemize}
Here, N = Number of voxels, $y_{true}$ = ground truth label, 
$y_{pred}$ = network predicted label, and $\epsilon$ is to avoid zero denominator.

The CNNs, with convolution layers followed by fully-connected layers, classify an entire image in a single category. GoogleNet(Inception)\cite{szegedy2015going} and InceptionV3\cite{szegedy2016rethinking} networks have introduced the inception module which implements the kernels of different sizes to reduce network parameters. ResNet\cite{targ2016resnet} has introduced the residual connection between the convolution layers such that it learns the identity function which allows the effective training of deeper networks. In DenseNet\cite{huang2017densely} the layers are very narrow and add very few number of feature maps to the network which again allows to design deeper architectures and training is efficient as each layer has direct access to the gradient of the loss function.

For semantic segmentation, the CNN can simply be used to classify each voxel of the image individually by presenting it with several patches extracted around the particular voxel. Each voxel of the image is classified with the same process, resulting in the segmentation of the entire image. This ‘sliding-window’ approach repeats the convolution operations for adjacent patches of neighbouring voxels. The improvement to this approach is the replacement of fully-connected layers with convolution layers, which generates the probability map of the entire input image rather than generating output for a single voxel. Such networks are known to be Fully Convolutional Neural Networks(FCNN). FCN\cite{long2015fully} is a type of FCNN where skip connections are introduced to reconstruct a high resolution image. U-Net\cite{ronneberger2015u}, a very well-known highly adapted network architecture for tumor segmentation has taken the encoder - decoder approach where every encoding layer is connected with its peer decoding layer with a skip connection to reconstruct the dimension as well to get the particular spatial information from the encoding layer. SegNet\cite{badrinarayanan2017segnet} and DeepLab\cite{chen2017deeplab} are other types of FCNN architectures adopted to solve the problem of brain tumor segmentation.

\subsection{Handling Class Imbalance Problem}
\label{sec:5.2}

In medical image analysis, finding the number of abnormal images as compared to the normal images is difficult as an abnormality like a tumor is rare. This problem is called `class imbalance'. All images in this article are of the tumor; the class imbalance issue persists because, in a single brain volume, the proportion of the tumor is less compared to the average brain volume. Even on a single brain slice, the area of the tumor is small compared to the brain part, which leads to a class imbalance problem. The proportion of the brain volume (BV) and background volume (BGV) with respect to the tumor volume across all the slices for BraTS 2019 dataset is as shown in Table \ref{tab:6}.

\begin{table}[htbp]
\caption{The proportion of non tumor brain volume (NTBV) and non tumor background volume (NTBGV) vs. tumorous volume(TV).}
\label{tab:6}
\begin{tabular}{@{}p{1 cm} p {1 cm} c c p{1.5 cm}@{}}
\toprule
 & \% NTBV & \% Necrosis  & \% Edema  & \% ET \\ 
\midrule
BV & 75.75 & 0.70 & 17.78 & 5.77 \\
\midrule
BGV & 99.11 & 0.03 & 00.65 & 0.21 \\
\bottomrule
\end{tabular}
\end{table}

The following approaches address data imbalance problem.
\begin{itemize}

\item Patch sampling: The patch sampling-based methods can mitigate the imbalanced data problem. The sampling process includes equiprobable patches from all the tumorous regions as well as the non-tumorous regions.
\item Improvement in loss functions: Some of the loss functions, when used in their raw form, may not suit the tumor segmentation task, as they consider balanced datasets. These functions adopt an imbalanced dataset with modifications as:
\begin{enumerate}
\item The weighted cross-entropy loss function: Voxel-wise class prediction averaged for all voxels may lead to error if the class is imbalanced in the image. Even if the weighted cross-entropy loss function weighs each voxel individually, the issue of class imbalance will not be addressed. Since the background regions dominate the training set, it is reasonable to incorporate the weights of multiple classes into the cross-entropy such that more weight is given to the voxels of the positive class. This has been shown in equation \ref{equation:5}.
\begin{equation}
\label{equation:5}
Loss_{WCE} = \sum_{voxels} \sum_{classes}  y_{true} \cdot log \hat{y}_{pred}
\end{equation}
\item Generalized Dice Loss Function: Authors in \cite{sudre2017generalised} propose using class rebalancing properties of the generalized dice. It provides a robust and accurate deep-learning loss function for unbalanced tasks. This has been shown in equation \ref{equation:6}.
\begin{equation}
	\label{equation:6}
Loss_{GDL} = 1 - 2 \frac{\sum_{classes}w \sum_{voxels}y_{true}\, y_{pred}}{\sum_{classes}w \sum_{voxels}y_{true} + y_{pred}} 
\end{equation}
\item Focal Loss Function: The detection task uses focal loss. It encourages the model to down-weight easy examples and focuses training on hard negatives. Formally, the focal loss defines a modulating factor to the cross-entropy loss and a parameter for class balancing \cite{lin2017focal}. It has been shown in equation \ref{equation:7}.

\begin{equation}
\label{equation:7}
\begin{aligned}
Loss_{FL}(p_{t}) = - \alpha _{t} (1- p_{t})^{\gamma} log(p_{t}) \\
p_{t}  = \begin{cases} 
p_{t} & if \, y_{i}=1 \\
1-p_{t} & otherwise
\end{cases}
\end{aligned}
\end{equation}

where $y \epsilon \{1, -1\} $ is the ground-truth class, and $p_{t} \epsilon [0,1] $  is the estimated probability for the class with label $y = 1$. The weighting parameter $ \alpha $ deals with the imbalanced dataset. The focusing parameter $ \gamma $ smoothly adjusts the rate at which easy examples are down-weighted. Setting $ \gamma > 0 $ can reduce the relative loss for well-classified examples. It places the focus on hard and misclassified examples; the focal loss is equal to the original cross-entropy loss when $ \gamma = 0 $.

\end{enumerate}
\item Augmentation techniques: Most of the time, a large number of labels for training are not available for several reasons. Labelling the dataset requires an expert in this field, which is expensive and time-consuming. Training large neural networks from limited training data causes an over-fitting problem. Data augmentation is a way to reduce over-fitting and increase the amount of training data. It creates new images by transforming (rotated, translated, scaled, flipped, distorted, and adding some noise such as Gaussian noise) the ones in the training dataset. Both the original image and the created images are input to the neural network. For example, a large variety of data augmentation techniques include random rotations, random scaling, random elastic deformations, gamma correction augmentation, and mirroring on the fly during training.

\end{itemize}

\subsection{CNN Methods Classification for Tumor Segmentation}
\label{sec:5.3}

\begin{figure*}
\centering
  \includegraphics[width=0.75\textwidth,height=7cm]{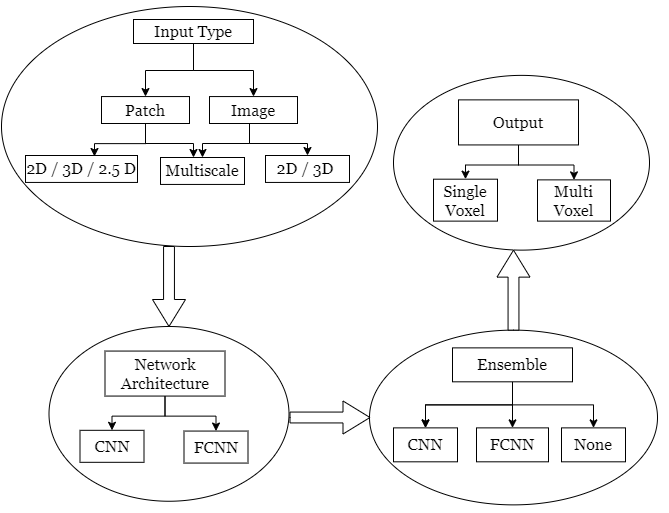}
\caption{Design aspects of CNN architectures.}
\label{fig:9}       % Give a unique label
\end{figure*}

The classification of CNNs for tumor segmentation uses the combination of design aspects as shown in Fig. \ref{fig:9}.

\begin{description}
\item[\textbf{Input type:}] The network may take 2D/3D input in the form of patches or images. The CNNs with fully-connected layers classify the centre voxel of the patch whereas FCNN predicts multiple or all voxels of the patch/image. The network may take multi-scale patches to extract coarse and fine details of the input.

\item[\textbf{Output Type:}] The output of the network depends on the problem to be solved. It predicts a single output for the classification problem and multiple voxel outputs for the semantic segmentation problem.

\item[\textbf{Type of network:}] The CNN approach indicates a convolution network with fully-connected layers at the end whereas FCN indicates a network with all convolution layers.

\item[\textbf{Ensemble Approach:}] Ensemble approaches can be classified into serial and parallel approaches. In the serial approach multiple networks combine in a series to fine tune the end output. The input of one network depends on the output of the other. In the parallel approach multiple networks work in parallel and take the same/different input to gather comprehensive details from the input. The final output of the network is decided based on the majority voting or averaging of all the network output.

\end{description}
 
\begin{figure*}
\centering
  \includegraphics[width=0.70\textwidth]{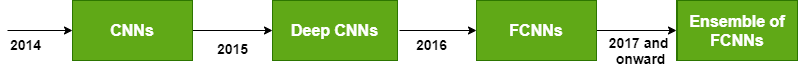}
\caption{Variational growth of CNN for tumor segmentation.}
\label{fig:10}       % Give a unique label
\end{figure*}

\begin{figure*}
\centering
  \includegraphics[width=0.70\textwidth]{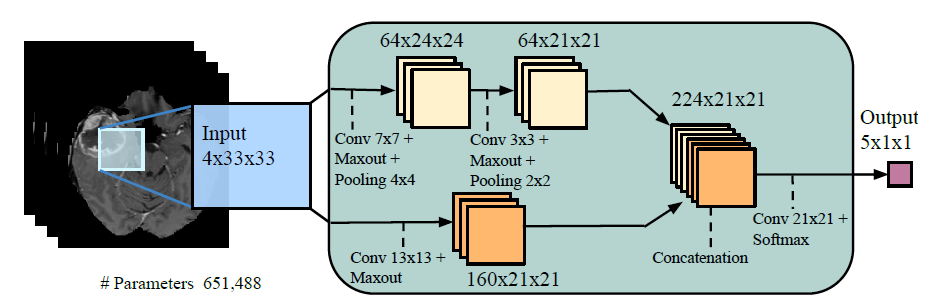}
\caption{TwopathwayCascadedCNN Architecture\cite{havaei2017brain}.}
\label{fig:11}       % Give a unique label
\end{figure*}

\begin{figure*}
\centering
  \includegraphics[width=0.80\textwidth,height=4cm]{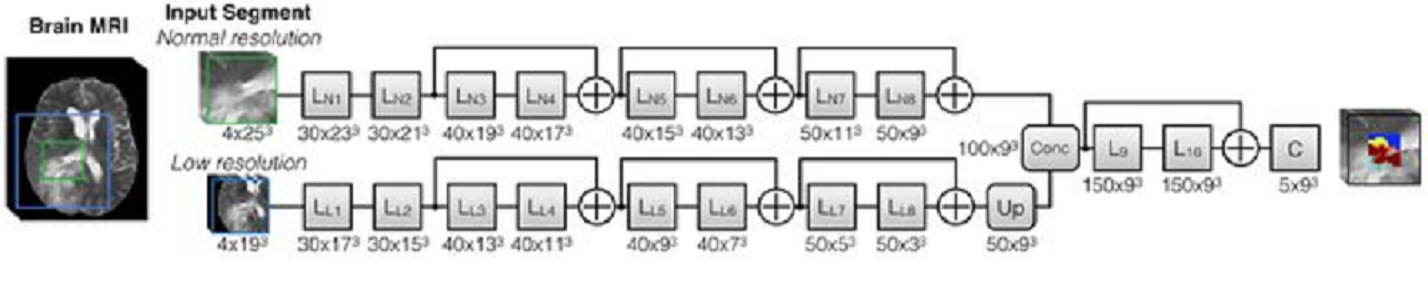}
\caption{Deep Medic Architecture \cite{kamnitsas2016deepmedic}.}
\label{fig:12}       % Give a unique label
\end{figure*}

\begin{figure*}
\centering
  \includegraphics[width=0.70\textwidth,height=7cm]{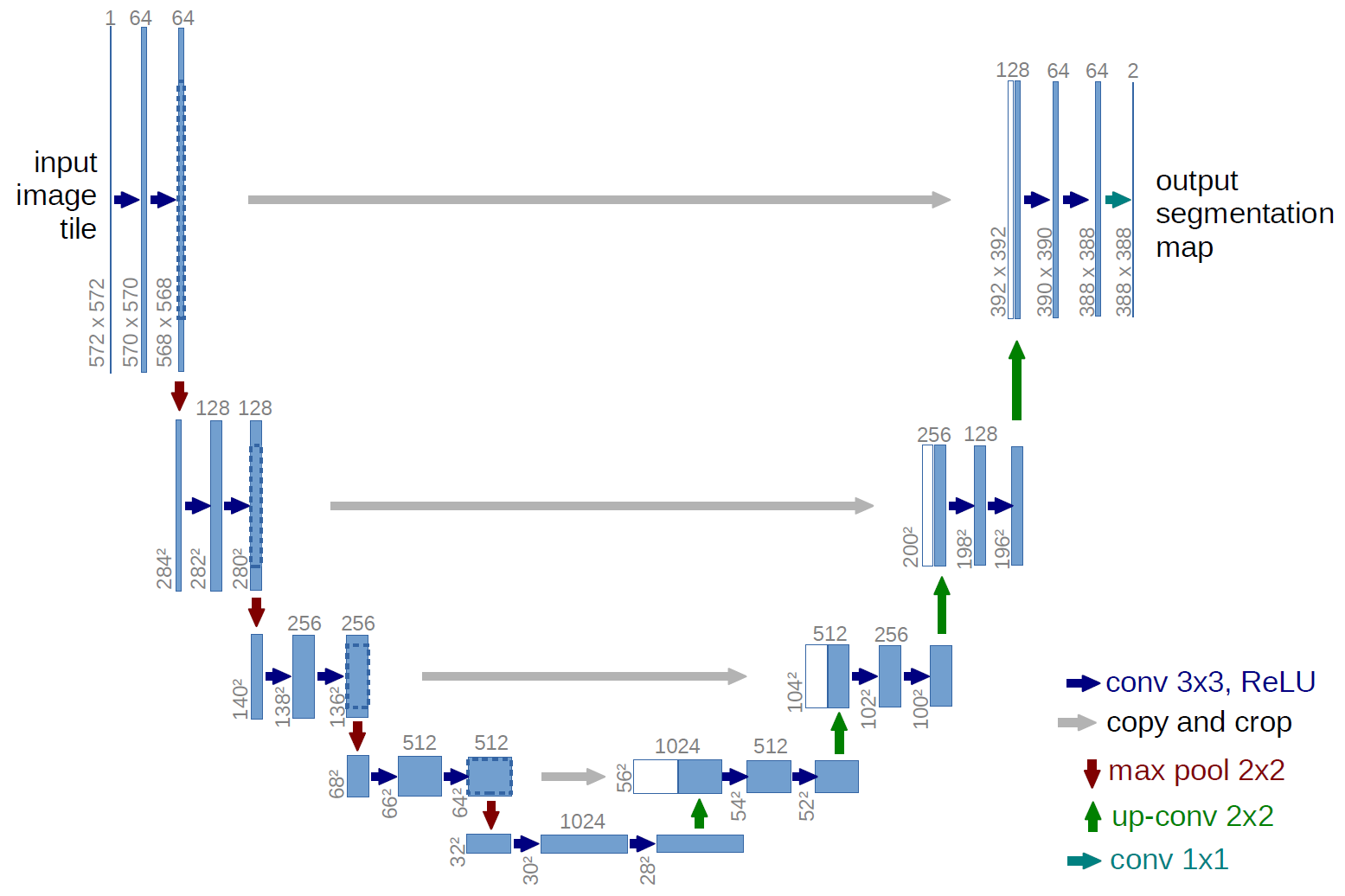}
\caption{U-net architecture\cite{ronneberger2015u}.}
\label{fig:13}       % Give a unique label
\end{figure*}

\begin{figure*}
\centering
  \includegraphics[width=0.70\textwidth]{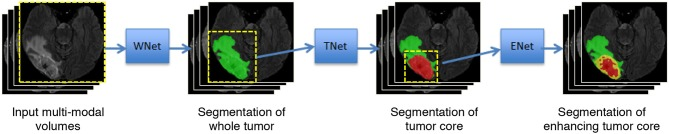}
\caption{Anisotropic Cascaded Architecture\cite{wang2017automatic}.}
\label{fig:14}       % Give a unique label
\end{figure*}

The evolution of CNN based methods for tumor segmentation is as shown in Fig. \ref{fig:10}. Some of well-known CNN architectures for brain tumor segmentation are represented in Fig. \ref{fig:11}, Fig. \ref{fig:12}, Fig. \ref{fig:13} and Fig. \ref{fig:14}.

The architecture of \cite{havaei2017brain} was a two pathway CNN which took 2D multi-resolution input patches, applied the convolution operations and concatenated the output of both the pathways. The deepmedic \cite{kamnitsas2016deepmedic} also followed two pathways with 3D multiresolution input patches and incorporated the residual connections and predicted the output for multiple voxels at a time.

U-net \cite{ronneberger2015u}  is an encoder-decoder architecture with skip-connections between the peer layers of analysis and synthesis path. This architecture has gained the most popularity. Anisotropic \cite{wang2017automatic} architecture followed the serial ensemble approach. The first network segments the whole tumor, the second segments the tumor core and considers the output of first network and finally the third network segments the enhancing tumor with the help of the output of the second network.

Initially, shallow CNN performed voxel-based image segmentation. The authors in \cite{havaei2017brain} proposed voxel-wise classification using a CNN multipath way architecture. One pathway used 2D patches of size 32x32, and the other used a fully connected input of 5x5 patch size having a center pixel the same as the 32x32 patch. Patch selection was made such that the labels were equiprobable. L1 and L2 regularizations were used, and are even now being used, to overcome overfitting. In \cite{urban2014multi}, a voxel-wise class probability prediction used separate 3D CNNs for HGG and LGG images. The final probability classified the voxel into the six classes. In \cite{zikic2014segmentation}, a five layer deep 2D CNN architecture performed voxel-wise classification.

Gradually the depth of CNN has increased to accommodate more layers in the network. In \cite{pereira2015deep}, a 2D deep CNN with fully connected output layers separated HGG and LGG. This approach was further extended by \cite{randhawa2016improving} with a two-phase process along with a weighted loss function;  initially, the network trained using equiprobable patches that follow actual patch training without the class imbalance problem. In \cite{havaei2015convolutional}, the authors designed a 2D Input Cascaded CNN, which took the output of a Two Path CNN to train other 2D CNNs with the input images. After the successful implementation of FCN by the authors in \cite{long2015fully}, the authors in \cite{kamnitsas2016deepmedic} proposed a two-pathway architecture, where both pathways included residual connections and trained on the different input patch sizes. As the network was fully convolutional, multiple voxels of the input patch label at a time.

\cite{zhao2016brain} used a 2D FCNN approach along with CRF. The FCNN trained on patches and the CRF on slices. In \cite{chang2016fully}, cascaded encoder-decoders like FCNNs along with residual connections were used for segmentation. The first FCNN segmented the whole tumor, followed by the internal tumor regions by the second FCNN. Authors in \cite{mckinley2016nabla} proposed an encoder-decoder FCNN based architecture to segment various tumor sub-regions. In \cite{casamitjana20163d}, the authors proposed three different FCNN architectures and showed that the architectures with multi-resolution features performed better compared to a single-resolution architecture. Authors in \cite{lopez2017dilated} implemented a Dilated Residual Network for patch-based training where equiprobable patches were supplied to the network for training.

Authors in \cite{dong2017automatic} adopted a U-net architecture for brain tumor segmentation. Authors in \cite{feng2017patch} modified the U-net, which took 3D input, and the depth of the network was reduced to three. Authors in [64], optimized the training of the network proposed in their previous work \cite{isensee2017brain}. In \cite{chen2018drinet}, authors proposed a novel encoder-decoder architecture that worked well on multiple biomedical image segmentation problems. Various other CNN based approaches where a single CNN was used for segmentation task were \cite{casamitjana2017cascaded}, \cite{castillo2017brain}, \cite{hu20173d}, \cite{islam2017multi}, \cite{jesson2017brain}, \cite{kim2017brain}, \cite{marcinkiewicz2018segmenting}, \cite{mehta20183d}, \cite{nuechterlein20193d},  \cite{pawar2017residual},  \cite{pourreza2017brain}, \cite{shaikh2017brain}. 

An ensemble of CNNs performs better compared to a single CNN, as in \cite{kamnitsas2017ensembles}, where the authors implemented an ensemble of seven networks using DeepMedic, FCN, and U-net along with variations of those three networks. They also tried three different approaches for pre-processing on all these networks. The output of individual networks with all the pre-processing generates the final label. Authors in \cite{mckinley2017pooling} extended their work proposed in \cite{mckinley2016nabla} where the dense module and the dilated module were introduced in the encoder decoder cascaded architecture of two networks. The pooling layers were replaced with dilated convolution layers.

The authors in \cite{myronenko20183d} implemented an ensemble of ten encoder-decoder based architectures, which included an auto-encoder stream to reconstruct the original image for additional information and regularization purposes. Authors in \cite{mckinley2018ensembles} extended their approach proposed in \cite{jungo2017towards}. They used a combination of Unet and Densenet with U-net like architecture containing dense blocks of dilated convolution. The network of \cite{zhao2016brain} was extended in \cite{zhao20173d} to create an ensemble of three networks which were trained on three image views. A two network cascaded path was used in \cite{colmeiro2017multimodal}, where a Coarse Segmentation Network segmented the WT and a Fine Segmentation Network segmented the sub-regions of the network. Both the networks used 3D U-net with a four level deep architecture.

Authors in \cite{wang2017automatic} used three networks (WNet, TNet, and ENet) to prepare the cascaded path. The multi-scale prediction averaged for the private network. These networks trained on three views of images (axial, coronal, sagittal), and used the results from the averages to generate the final segmentation output. In \cite{zhou2018learning}, a cascaded network was proposed, which initially segmented the whole tumor, followed by the tumor core, and enhancing tumor segmentation refinement. In \cite{xu2018multi} the authors proposed a cascaded UNet with three networks. The network process downsampled the input and the generated output was passed to the next network in a cascaded sequence. Authors in \cite{xu2018multi} proposed multi-scale mask 3D U-nets with atrous spatial pyramid pooling layers, where the WT segmentation generated by the first network was passed to the second for TC generation, which in sequence passed to the final network to generate an ET output. Other ensemble based CNN approaches were explained in \cite{albiol2018extending}, \cite{chandra2019context}, \cite{choudhury2018segmentation}, \cite{hua2018multimodal}, \cite{kermi2018deep}, \cite{ma2018automatic},  \cite{wang2018automatic}, \cite{yao2019automatic}.

The comparison of the methods summarized in Table \ref{tab:7}  is based on the pre-processing techniques, DNN architecture, activation function, loss function, postprocessing, and the DSC achieved. The pre-processing techniques considered for the comparison are:

\begin{enumerate}
\item Intensity clipping: 1 \% of highest and lowest frequencies are clipped.
\item Bias field correction.
\item Z-score normalization: $ Z = (x - \mu) / \sigma $.
\item Histogram matching: Histogram of all the images is match with the reference histogram.
\item Image normalization: Min-max normalization.
\item Intensity standardization with Nyul approach \cite{nyul1999standardizing}.
\item Image denoising: applies noise filtering e.g, Gaussian noise filtering. 
\item Intensity rescaling : rescaling the intensity range between some specific limits.
\end{enumerate}
The post-processing techniques used for segmentation result improvement are:
\begin{enumerate}
\item Connected component analysis: Analyse the connected components and removes the component with the volume below some threshold.
\item Conditional random field.
\item Morphological operators to remove false positives and fill the holes
\item relabelling the output label: Enhancing tumor labels below some threshold are relabelled as necrosis.  
\end{enumerate}

\begin{figure*}[!h]
\centering
  \includegraphics[width=0.80\textwidth]{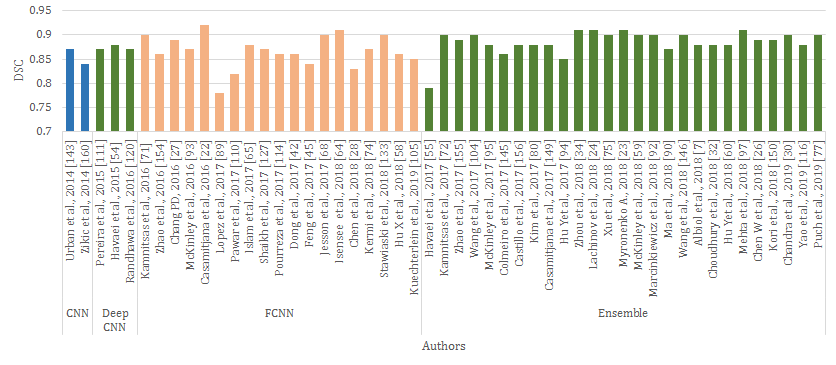}
\caption{Validation set DSC comparison for whole tumor region of CNN methods.}
\label{fig:15}       % Give a unique label
\end{figure*}

Fig. \ref{fig:15} shows a pictorial representation of the DSC of various CNN methods for the whole tumor region of the validation set. The average DSC of a CNN is 0.86, for a deep CNN and FCN it is 0.87, and for an ensemble approach it is 0.89. The ensemble of the CNNs/FCNNs learns robust features from the input.

\subsection{Hardware and Software for DNN}
\label{sec:5.4}
\textbf{Hardware}: DNN methods have gained in popularity after the availability of the Graphical Processing Units (GPU). Nowadays efficient parallel processing for manipulation of large amounts of data is possible with the help of General Purpose GPU(GPGPU). Computing libraries like CUDA and OpenGL allow the efficient implementation of the processing code on a GPU. The performance of the GPU mainly depends on GPU computing cores (CUDA cores), Tensor processing cores, Thermal Design Power (TDP), and on-board GPU memory. Various types of GPUs used for implementation of the CNN methods for segmentation tasks are as shown in Table \ref{tab:8}. As the computing capacity of the GPU increases, it allows more complex networks with a higher number of network parameters that can be trained with less time.

\begin{small}
\begin{table}[!h]
\caption{GPU specifications\cite{gpuspec:2019}.}
\label{tab:8}
\begin{tabular}{@{}p{1 cm} p{1.5 cm} p{1 cm} p{1 cm} p{1 cm} p{1 cm}@{}}
\toprule
\textbf{Year} &  \textbf{GPU Type} & \textbf{CUDA Cores} & \textbf{Tensor Cores} & \textbf{TDP (Watts)} & \textbf{RAM (GB)} \\ \midrule
2016 & Tesla K80 &  2496 & N/A & 300 & 12 \\
2016 & GeForce GTX 980 Ti & 2816 & N/A & 165 & 6 \\
2016 - 2017 & GeForce GTX 1080 Ti & 3584 & N/A  & 250 & 11 \\
2017 & GTX Titan X & 3072 &  N/A & 240 & 12 \\
2018 & Quadro P4000 & 1792 & N/A & 105 & 8 \\
2018 & Quadro P5000 & 2560 & N/A & 180 & 16 \\
2018 & Titan Xp & 3840 & N/A & 250 & 12 \\
2018 & Tesla P100 & 3584 & N/A & 250 & 16 \\
2018 & Tesla V100 & 5120 & 640 & 300 & 16/32 \\
2018 & DGX-1 & 40960 (8xV100) & 5120 & 3500 & 128(16/ GPU) \\
2020 & RTX 2080 Ti & 4352 & 544 & 250 & 11 \\ \bottomrule
\end{tabular}
\end{table}
\end{small}

\textbf{Software}: Open software library packages provide implementation of various CNN operations like convolution. Most popular python library packages for CNN implementations are: Caffe\cite{jia2014caffe}, Tensorflow\cite{abadi2016tensorflow}, Theano \cite{bastien2012theano}, and PyTorch\cite{paszke2019pytorch}. Some of the third-party packages which work on the top of these networks are Keras\cite{chollet2018keras}, Lasagne\cite{lasagne}, and TensorLayer\cite{dong2017tensorlayer}.

\onecolumn

\tablefirsthead{%
     \toprule  \\
     \textbf{Ref.} & \textbf{Pre-processing} & \textbf{Input Modality + Augmentation}  & \textbf{Patch/ Image} & \textbf{Input view} & \textbf{Network Architecture}  & \textbf{\# Networks} & \textbf{Ensemble Type} & \textbf{Loss Function} & \textbf{Post-processing} & \textbf{Dataset} & \textbf{DSC Mean} \\      
       }
 
%This is the header for the remaining page(s) of the table...
\tablehead{%
      \multicolumn{12}{c}
    {{\bfseries \tablename\ \thetable{} --
       continued \ldots}} \\
     \toprule
     \textbf{Ref.} & \textbf{Pre-processing} & \textbf{Input Modality + Augmentation}  & \textbf{Patch/ Image} & \textbf{Input view} & \textbf{Network Architecture}  & \textbf{\# Networks} & \textbf{Ensemble Type} & \textbf{Loss Function} & \textbf{Post-processing} & \textbf{Dataset} & \textbf{DSC Mean} \\  }
 
%This is the footer for all pages except the last page of the table...
\tabletail{%
	\midrule
}
\tablelasttail{\bottomrule}
\tablecaption{Summarization of Segmentation methods using CNN architectures.}
\label{tab:7}

\begin{landscape}
\begin{small}
\begin{xtabular}{@{\extracolsep{\fill}}p{0.8 cm} p{2 cm} p{2 cm} p{1.5 cm} p{1.5 cm} p{1.5 cm} p{1.5 cm} p{1.2 cm} p{1 cm} p{1 cm} p{1 cm} p{2.5 cm}@{\extracolsep{0pt}} }
\midrule
\multicolumn{12}{l}{\textbf{CNN}} \\
\midrule
\cite{urban2014multi} & 5 & T1, T2, T1c, FLAIR  & 3D patches & axial
 & 3D CNN & 1 & - & softmax	& 1 & BraTS 2013 & Test set: WT:0.87,TC:0.77,ET:0.73 \\
\midrule
\cite{zikic2014segmentation} & 2,downsampling & T1, T2, T1c, FLAIR  & 2D patches &	axial &	2D CNN & 1 & - & softmax & - & BraTS 2013 & Training set: WT:0.84,TC:0.74,ET:0.69 \\
\midrule
\multicolumn{12}{l}{\textbf{Deep CNN}} \\
\midrule
\cite{pereira2015deep} & 2, 6 & T1, T2, T1c, FLAIR, Rotation at certain angles  & 2D patches	& axial	& 2D CNN & 2 & - & categorical cross entropy & 3 & BraTS 2015 & Training set: WT:0.87,TC:0.73,ET:0.68 \\
\midrule
\cite{havaei2015convolutional} & 1,2,3 & T1, T2, T1c, FLAIR & 2D multiscale patches &	axial &	2D CNN & 4 & cascaded, parallel & softmax & 1 & BraTS 2013 & Test Set: WT:0.88,TC:0.79,ET:0.73 \\
\midrule
\cite{randhawa2016improving} & 1,2,3 & T1, T2, T1c, FLAIR, horizontal and vertical flipping & 2D patches & axial & 2D CNN & 2 & - & weighted cross entropy with L1 and L2 regularization & 1 & BraTS 2016 & Test Set: WT:0.87,TC:0.75,ET:0.71 \\
\midrule
\multicolumn{12}{l}{\textbf{FCNN}} \\
\midrule
\cite{kamnitsas2016deepmedic} & 3 & T1, T2, T1c, FLAIR + with flipping around a mid-sagittal plane & 3D multiscale patches & axial & 3D FCNN & 2 & parallel & - & 2
 & BraTS 2015  & Without residual connections: WT:0.90,TC:0.75,ET:0.72 With residual connection: WT:0.90,TC:0.76,ET:0.72 \\
\midrule
\cite{zhao2016brain} & 1, 5 & T1c, T2, FLAIR & 2D multiscale patches & axial & 2D FCNN	& 2 & cascaded & - & 1 & BraTS 2013 &  Validation set: WT:0.86,TC:0.73,ET:0.62 Test set: WT:0.87,TC:0.83,ET:0.76 \\
\midrule
\cite{chang2016fully} & 5 & T1, T2, T1c, FLAIR,
Rotation at 45 degree, bias field corrected images, additional images & 2D patches &	axial & 2D FCNN & 2 & cascaded & softmax with L2 regularization & - & BraTS 2016 & 	Test set:  WT:0.89,TC:0.83,ET:0.78 \\
\midrule
\cite{mckinley2016nabla} & - & T1, T2, T1c, FLAIR & 2D patches & axial & 2D FCNN  & 1	 & - & sigmoid & - & BraTS 2012 & Training set: WT:0.87,TC:0.69,ET:0.56 \\
\midrule
\cite{casamitjana20163d} & - & T1, T2, T1c, FLAIR & 3D multiscale patches & axial & 3D FCNN & 2 & parallel & softmax & - & BraTS 2015 & Training set:\newline 3Dnet1 WT:0.90,TC:0.77,ET:0.63 \newline 3DNet2 WT:0.92,TC:0.70,ET:0.74 \newline 3DNet3 WT:0.92,TC:0.84,ET:0.77 \\
\midrule
\cite{lopez2017dilated} & 5 & T1, T2, T1c, FLAIR & 2D patches & axial & 2D FCNN & 1 & - & dice & - & BraTS 2017 & Validation set: WT:0.78,TC:0.69,ET:0.57 Test set: WT:0.69,TC:0.61,ET:0.51 \\
\midrule
\cite{pawar2017residual} & 4 & T1, T2, T1c, FLAIR & 2D images & axial & 2D FCNN & 1 &	- & softmax & - & BraTS 2017 & Validation set: WT:0.82,TC:0.63,ET:0.58 Test set:  WT:0.78,TC:0.58,ET:0.50 \\
\midrule
\cite{islam2017multi} & - & T2, T1c, FLAIR & 2D images & axial & 2D FCNN &	1 &	- &	- &	-  & BraTS 2017 & Validation set: WT:0.88,TC:0.76,ET:0.69 Test set: WT:0.86,TC:0.70,ET:0.62 \\
\midrule
\cite{shaikh2017brain} & 3 & T1, T2, T1c, FLAIR & 2D images & axial & 2D FCNN & 1 & - & dice + cross entropy & 1,2 &	BraTS 2017 &	Validation set: WT:0.87,TC:0.68,ET:0.65 Test set: WT:0.83,TC:0.65,ET:0.65 \\
\midrule
\cite{pourreza2017brain} & 2,4 & T1, T2, T1c, FLAIR & 2D images & axial & 2D FCNN & 1 &	- &	-  & 3,1 &	BraTS 2017 & Validation set: WT:0.86,TC:0.60,ET:0.69 Test set: WT:0.80,TC:0.55,ET:0.55 \\
\midrule
\cite{dong2017automatic} & 3 & T1, T2, T1c, FLAIR & 2D images & axial & 2D FCNN & 1 &	- &	dice & - & BraTS 2015 & Training set: WT:0.86,TC:0.86,ET:0.65 \\
\midrule
\cite{feng2017patch} & 1,2,3 & T1, T2, T1c, FLAIR & 3D patches & axial & 3D FCNN & 1	& - &	- & -	& BraTS 2017 & Validation set: WT:0.84,TC:0.75,ET:0.66  \\
\midrule
\cite{jesson2017brain} & 3 &	T1, T2, T1c, FLAIR & 3D images & axial & 3D FCNN & 1 &	- &	multi scale weighted  & - &	BraTS 2017 & Validation set: WT:0.90,TC:0.75,ET:0.71 Test set: WT:0.86,TC:0.78,ET:0.71 \\
\midrule
\cite{isensee2018no} & 3 & T1, T2, T1c, FLAIR & 3D patches & axial & 3D FCNN & 1 & -  & 	weighted dice +  cross entropy & 4 & BraTS 2018 & Validation set: WT:0.91,TC:0.86,ET:0.81 Test set: WT:0.88,TC:0.81,ET:0.78 \\
\midrule
\cite{chen2018drinet} &  -  & T1, T2, T1c, FLAIR & 2D images & axial & 2D FCNN & 1 & -  &  cross entropy & - & BraTS 2017 & Training set: WT:0.83,TC:0.73,ET:0.65  \\
\midrule
\cite{kermi2018deep} &	1,3 &	T1, T2, T1c, FLAIR & 2D images & axial & 2D FCNN & 1 &	- &	weighted cross entropy + generalized dice loss & -	&	BraTS 2018 &	Validation set: WT:0.87,TC:0.81,ET:0.78 Test set: WT:0.81,TC:0.73,ET:0.65\\
\midrule
\cite{stawiaski2018pretrained} &	- &	T1, T2, T1c, FLAIR & 2D images & axial & 2D FCNN & 1 & - &	dice &	- &	BraTS 2018 &	Validation set: WT:0.90,TC:0.85,ET:0.79 Test set: WT:0.88,TC:0.79,ET:0.78 \\
\midrule
\cite{hu2018hierarchical} &	- &	T1, T2, T1c, FLAIR &  3D images & axial & 3D FCNN & 1 & - & cross entropy & 	-	 & BraTS 2018 &	Validation set: WT:0.86,TC:0.77,ET:0.72 \\
\midrule
\cite{nuechterlein20193d}	& 5 & 	T1, T2, T1c, FLAIR & 3D images & axial & 3D FCNN & 1 & - &	softmax & 	- & BraTS 2018 & Test set: WT:0.85,TC:0.68,ET:0.67 \\
\midrule
\multicolumn{12}{l}{\textbf{Ensemble of CNNs}} \\
\midrule
\cite{havaei2017brain} & 1, 2 on (T1,T1c), 3 & T1, T2, T1c, FLAIR & 2D multiscale patches & axial & 2D CNN & 3 & cascaded, parallel & softmax with L1 and L2 regularization & - & BraTS 2013 & Training set-HGG: WT:0.79,TC:0.68,ET:0.57 Training set-LGG: WT:0.81,TC:0.75,ET:0.54 \\
\midrule
\cite{kamnitsas2017ensembles} & 3,2,5 and its different combination & T1, T2, T1c, FLAIR & 3D multiscale patches & axial & 3D FCNN &	7 & parallel & 	- & 1 & BraTS 2017 & Validation set: WT:0.90,TC:0.80,ET:0.74 Test set: WT:0.87,TC:0.79,ET:0.73 \\
\midrule
\cite{zhao20173d} &	1, 5 &	T1c, T2, FLAIR & 2D images & axial, coronal, sagittal & 2D FCNN & 3 & parallel & - & 1	& BraTS 2017 & 	Validation set: WT:0.89,TC:0.79,ET:0.75 Test set: WT:0.88,TC:0.75,ET:0.76 \\
\midrule
\cite{myronenko20183d} & 3, intensity scale, intensity shift, flip & T1, T2, T1c, FLAIR & 3D images	& axial & 3D FCNN & 1 &	- & weighted average of L2, Dice,KL & - & BraTS 2018 & Validation Set: WT:0.91,TC:0.87,ET:0.82 Test set: WT:0.88,TC:0.82,ET:0.77 \\
\midrule
\cite{mckinley2018ensembles} & 3 on the individual volume & T1, T2, T1c, FLAIR & 2D images & axial, coronal, sagittal & 2D FCNN & 3	& parallel & - & - & BraTS 2018 & Validation set: WT:0.90,TC:0.85,ET:0.80 Test set: WT:0.89,TC:0.80,ET:0.73 \\
\midrule
\cite{wang2017automatic} & 3 & T1, T2, T1c, FLAIR & 3D patches & axial, coronal, sagittal & 2D FCNN & 3 & parallel of cascades & dice & - & BraTS 2017 & Validation set: WT:0.90,TC:0.83,ET:0.78 Test set: WT:0.87,TC:0.77,ET:0.78 \\
\midrule
\cite{zhou2018learning} & - & - & 3D multi scale patches & axial & - & 7 & parallel & - & 1, 4 & BraTS 2018  & Validation set: WT:0.91,TC:0.87,ET:0.81 Test set: WT:0.88,TC:0.80,ET:0.78 \\
\midrule
\cite{lachinov2018glioma} & 3,1 & T1, T2, T1c, FLAIR & 2D images & axial & 2D FCNN & 3 & cascaded & mean dice & - & BraTS 2018 & Validation set: WT:0.91,TC:0.84,ET:0.77 \\
\midrule
\cite{xu2018multi} & 2,3 & T1, T2, T1c, FLAIR & 3D patches & axial & 3D FCNN & 3 &	cascaded & cross entropy &  - & BraTS 2018 & Validation set: WT:0.90,TC:0.83,ET:0.80 \\
\midrule
\cite{mckinley2017pooling} &	3 &	T1, T2, T1c, FLAIR & 2D images & axial & 2D FCNN &	2 & cascaded & - & 1 &	BraTS 2017	 & Validation set: WT:0.88,TC:0.76,ET:0.71 Test set: WT:0.86,TC:0.71,ET:0.71 \\
\midrule
\cite{colmeiro2017multimodal} &	5,3 &	T1, T2, T1c, FLAIR & 3D patches & axial & 3D FCNN & 2 & cascaded &	dice  &	- &	BraTS 2017 & Validation set: WT:0.86,TC:0.69,ET:0.66 Test set: WT:0.82,TC:0.67,ET:0.60 \\
\midrule
\cite{castillo2017brain} & 5 &	T1, T2, T1c, FLAIR & 2D multiscale patches & axial & 3D CNN & 4 & parallel & softmax & - &	BraTS 2017 &	Validation set: WT:0.88,TC:0.68,ET:0.71 Test set: WT:0.86,TC:0.67,ET:0.65 \\
\midrule
\cite{kim2017brain} &	-	& T1, T2, T1c, FLAIR &	2D images & axial, coronal, sagittal & 2D FCNN & 3 & parallel & dice &	- &	BraTS 2017 & Validation set: WT:0.88,TC:0.73,ET:0.75 Test set: WT:0.86,TC:0.73,ET:0.72 \\
\midrule
\cite{casamitjana2017cascaded}	& - & 	T1, T2, T1c, FLAIR & 3D images & axial & 3D FCNN & 2 & cascaded &	weighted dice + cross entropy &	- &	BraTS 2017 & Validation set: WT:0.88,TC:0.64,ET:0.71 \\
\midrule
\cite{hu20173d}	& 5 & 	T1, T2, T1c, FLAIR & 2D patches & axial & 2D FCNN & 4 & parallel to cascaded &	dice  &	-	 & BraTS 2017 &	Validation set: WT:0.85,TC:0.70,ET:0.65 Test set: WT:0.81,TC:0.69,ET:0.55 \\
\midrule
\cite{marcinkiewicz2018segmenting} &	-	& T2, T1c, FLAIR &	2D images & axial & 2D FCNN & 2 & cascaded &	dice &	- &	BraTS 2018 &	Validation set: WT:0.90,TC:0.81,ET:0.75 Test set: WT:0.86,TC:0.73,ET:0.65 \\
\midrule
\cite{ma2018automatic}	& 3,4,5 & 	T2, T1c, FLAIR & 3D images & axial & 3D FCNN & 3 & parallel to cascaded & 	dice &	- &	BraTS 2018 &	Validation set: WT:0.87,TC:0.77,ET:0.74 Test set: WT:0.81,TC:0.73,ET:0.65 \\
\midrule
\cite{wang2018automatic} &	3	 & T1, T2, T1c, FLAIR + transformation and noise blurring &	3D images & axial & 3D FCNN(Test time augmentation) & 3 & cascaded &	dice  & -	& BraTS 2018 &	Validation set: WT:0.90,TC:0.86,ET:0.80 Test set: WT:0.88,TC:0.80,ET:0.75 \\
\midrule
\cite{albiol2018extending} &	3	& T1, T2, T1c, FLAIR &	3D patches & axial & 3D FCNN & 4 & parallel &	dice &	-	& BraTS 2018 & Validation set: WT:0.88,TC:0.78,ET:0.77 Test set: WT:0.85,TC:0.74,ET:0.72 \\
\midrule
\cite{choudhury2018segmentation} &	8 &	T1, T2, T1c, FLAIR & 2D images & axial, coronal, sagittal & 2D FCNN & 18 & parallel  &	- &	- &	BraTS 2018 &	Validation set: WT:0.88,TC:0.79,ET:0.71 \\
\midrule
\cite{hu2018brain} &	5 &	T1, T2, T1c, FLAIR &	2D images & axial, coronal, sagittal & 2D FCNN & 3 & parallel  &	- &	- &	BraTS 2018 &	Validation set: WT:0.88,TC:0.74,ET:0.69 Test set: WT:0.85,TC:0.72,ET:0.66 \\
\midrule
\cite{mehta20183d} &	3, 8 &	T1, T2, T1c, FLAIR & 3D images & axial & 3D FCNN & 5 & parallel &	weighted categorical cross entropy &	- & 	BraTS 2018 &	Validation set: WT:0.91,TC:0.83,ET:0.79 Test set: WT:0.87,TC:0.77,ET:0.71 \\
\midrule
\cite{chandra2019context} & 	4	& T1, T2, T1c, FLAIR &	3D images & axial & 3D FCNN & 3 & parallel & generalized dice & 2 &	BraTS 2018 & Validation set: WT:0.90,TC:0.81,ET:0.77 Test set: WT:0.83,TC:0.73,ET:0.62 \\
\midrule
\cite{yao2019automatic} & - & T1, T2, T1c, FLAIR &	3D images & axial, coronal, sagittal & 3D FCNN & 3 & parallel & dice & - & BraTS 2018 & Validation set: WT:0.88,TC:0.81,ET:0.78 Test set: WT:0.86,TC:0.77,ET:0.72 \\
\midrule
\cite{chen2018s3d} & 	2,4,8 &	T1, T2, T1c, FLAIR	& 3D images	& axial, coronal, sagittal & 3D FCNN & 3 & parallel &	multi class dice &	- &	BraTS 2018 & Validation set: WT:0.89,TC:0.83,ET:0.75 Test set: WT:0.84,TC:0.78,ET:0.69 \\
\midrule
\cite{puch2019global} &  &  T1, T2, T1c, FLAIR & 3D images & axial & 3D FCNN & 2 & parallel &	categorical cross entropy &	- &	BraTS 2018 & Validation set: WT:0.90,TC:0.77,ET:0.76 Test set: WT:0.86,TC:0.75,ET:0.69 \\
\midrule
\cite{kori2018ensemble} & 	3	& T1, T2, T1c, FLAIR & 2D images + 3D multi resolution patches & axial & 3D FCNN & 3 & parallel  &	- & 2,1 &	BraTS 2018 &	Validation set: WT:0.89,TC:0.76,ET:0.76 Test set: WT:0.83,TC:0.72,ET:0.69 \\
\end{xtabular}
\end{small}
\end{landscape}
\twocolumn 

\subsection{Proposed Architecture for Tumor Segmentation}
\label{sec:5.5}

\begin{figure*}
\centering
  \includegraphics[width=0.80\textwidth]{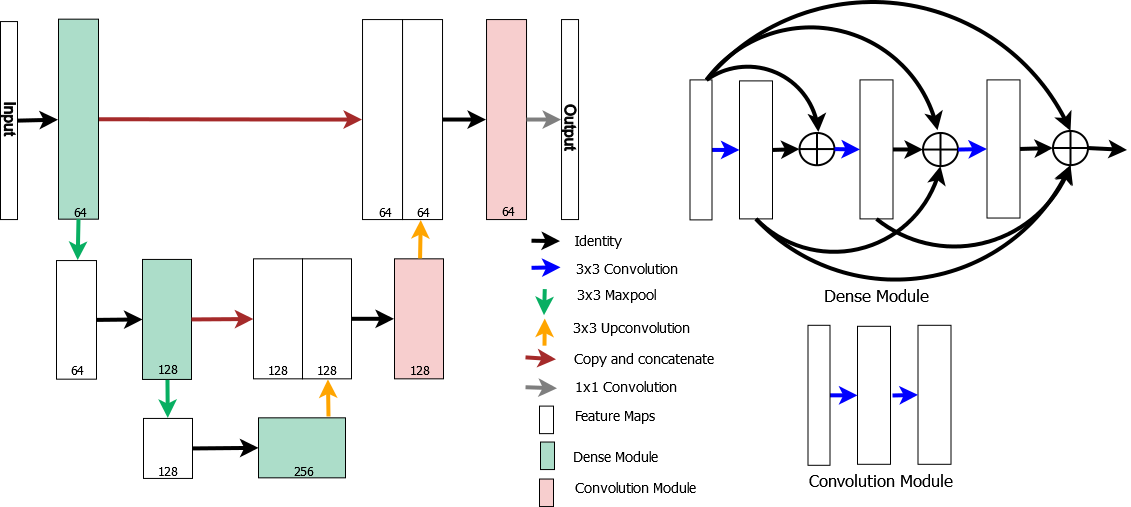}
\caption{Three stage U-net architecture \cite{agravat2019brain}.}
\label{fig:16}       % Give a unique label
\end{figure*}

The authors of the article have adopted a 2D U-net architecture with three layers\cite{agravat2019brain}. Each encoder layer is replaced with a dense module as shown in Fig. \ref{fig:16}. In the first phase, the network is trained for the whole tumor for 50 epochs with a dice loss function. Further the network parameters are initialized with whole tumor weights to train the network for necrosis, enhancing and edema subregions. The inductive transfer learning approach \cite{pan2009survey} for parameter initialization has reduced false positives and boosted the network training convergence. In the second phase, these subregion weights are used as initial parameters to train networks again for the same subregions using focal loss function. Network training continues for 50 epochs for each subregion. From the BraTS 2019 training dataset, 85\% of the images are given for training and 15\% are kept for validation. The input to the network is 2D slices of size 240x240 from all the four modalities. Blank slices of the dataset are removed as they do not contain any meaningful information. The following are the updates applied to the network:

\begin{itemize}
\item The bounding box is applied around the brain to reduce the input which contains zero intensity (blank) voxels.
\item The convolution layer is added in the dense module .
\end{itemize}

The results for the network are in Table \ref{tab:9}, which includes minor variations in the network, different input image sizes, different loss functions, and the number of images in training. The network is trained on 85\% of the total images and validated on the remaining 15\% images of the BraTS 2019 dataset. An online tool provided by the organizer generates the evaluation metric for the training set as well as a separate validation set in addition to the training set.

\begin{small}
\begin{table*}
\caption{Comparison of model variations of \cite{agravat2019brain}.}
\label{tab:9}
\begin{tabular}{@{}p{1.5 cm} p{2.25 cm} p{1.75 cm} p{1.75 cm} p{1.75 cm} p{1.75 cm} p{1.75 cm} p{1.75 cm}@{}}
\toprule
\textbf{Model}  & \textbf{Architectural change} & \textbf{Input Modalities} & \textbf{\# Feature maps} & \textbf{Input image size} & \textbf{Loss function} & \textbf{DSC Training Set} & \textbf{DSC Validation Set} \\
\midrule
1 & Original & T1, T1c, T2, FLAIR & 32, 64, 128, 64, 32 & 240 x 240 & Dice Loss & WT:0.90 TC:0.86 ET:0.75 &  WT:0.72 TC:0.63 ET:0.55 \\
\midrule
2 & Original & T1, T1c, T2, FLAIR & 32, 64, 128, 64, 32 & 156 x 200 & Dice Loss & WT:0.90 TC:0.86 ET:0.75 & WT:0.72 TC:0.63 ET:0.55\\
\midrule
3 & dense module + convolution layer & T1, T1c, T2, FLAIR & 32, 64, 128, 64, 32 & 156 x 200 & Dice Loss & WT:0.89 TC:0.84 ET:0.71 & WT:0.71 TC:0.61 ET:0.56 \\
\midrule
4 & dense module + convolution layer & T1, T1c, T2, FLAIR & 32, 64, 128, 64, 32 & 156 x 200 & Focal Loss & WT:0.93 TC:0.90 ET:0.83 & WT:0.75 TC:0.65 ET:0.60\\
\midrule
5 & dense module + convolution layer & T1c, T2, FLAIR & 32, 64, 128, 64, 32 & 156 x 200 & Dice Loss & WT:0.87 TC:0.81 ET:0.67 &	WT:0.71 TC:0.58 ET:0.51\\
\midrule
6 & dense module + convolution layer & T1c, T2, FLAIR & 32, 64, 128, 64, 32 & 156 x 200 & Focal Loss & WT:0.92 TC:0.89 ET:0.79 &	WT:0.76 TC:0.66 ET:0.60\\
\bottomrule
\end{tabular}
\end{table*}
\end{small}

\begin{figure}
\centering
\begin{subfigure}{.50\textwidth}
  \centering
  \includegraphics[width=6cm,height=3.25cm]{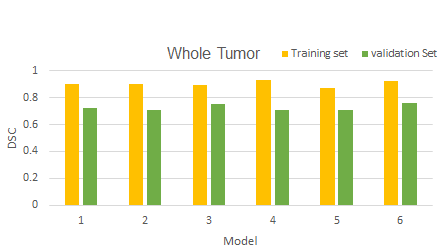}
  \caption{}
\end{subfigure}
\begin{subfigure}{.50\textwidth}
  \centering
  \includegraphics[width=6cm,height=3.25cm]{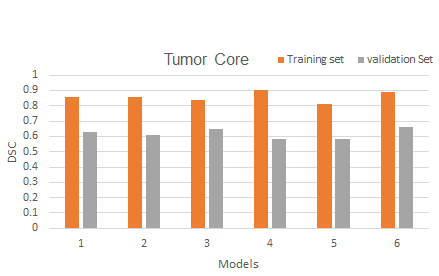}
  \caption{}
\end{subfigure}
\begin{subfigure}{.50\textwidth}
  \centering
  \includegraphics[width=6cm,height=3.25cm]{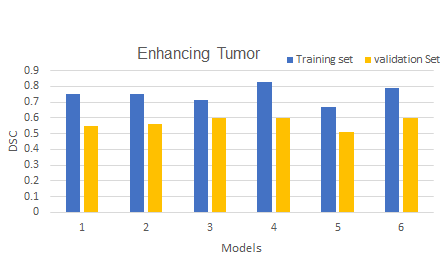}
  \caption{}
\end{subfigure}
\caption{DSC comparison for training and validation set a) whole tumor b)tumor core c) enhancing tumor.}
\label{fig:17}       % Give a unique label
\end{figure}

\begin{figure}
\centering
\begin{subfigure}{.25\textwidth}
  \centering
  \includegraphics[width=3cm,height=3cm]{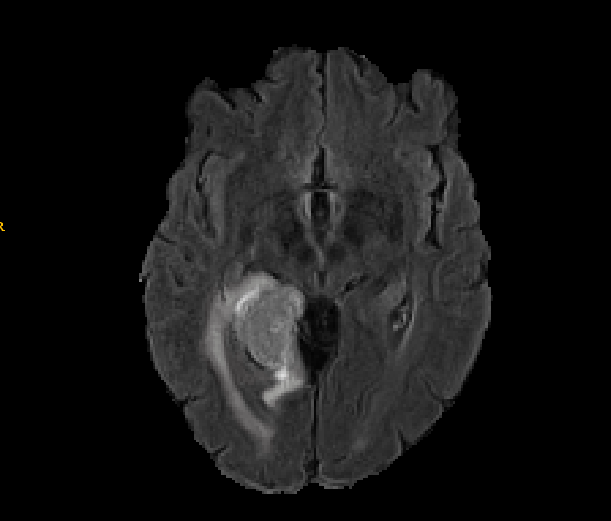}
  \caption{}
\end{subfigure}%
\begin{subfigure}{.25\textwidth}
  \centering
  \includegraphics[width=3cm,height=3cm]{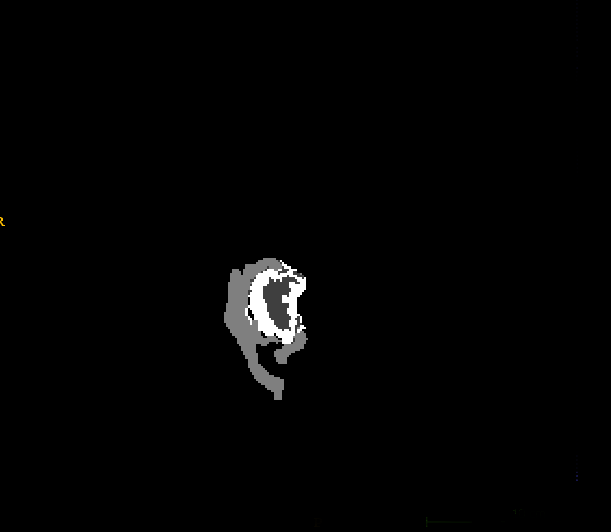}
  \caption{}
\end{subfigure}
\begin{subfigure}{.25\textwidth}
  \centering
  \includegraphics[width=3cm,height=3cm]{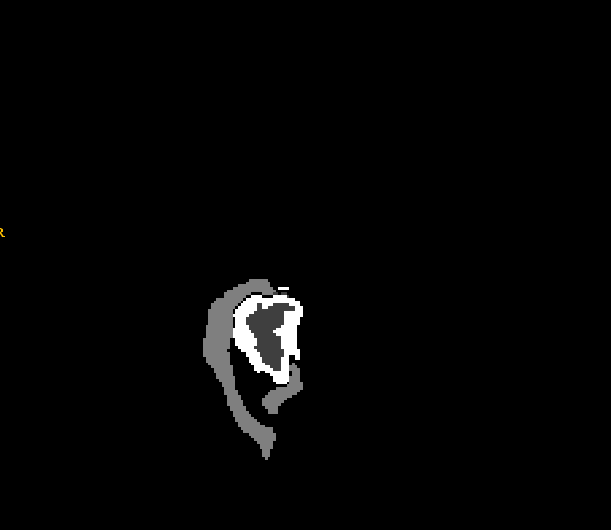}
  \caption{}
\end{subfigure}%
\begin{subfigure}{.25\textwidth}
  \centering
  \includegraphics[width=3cm,height=3cm]{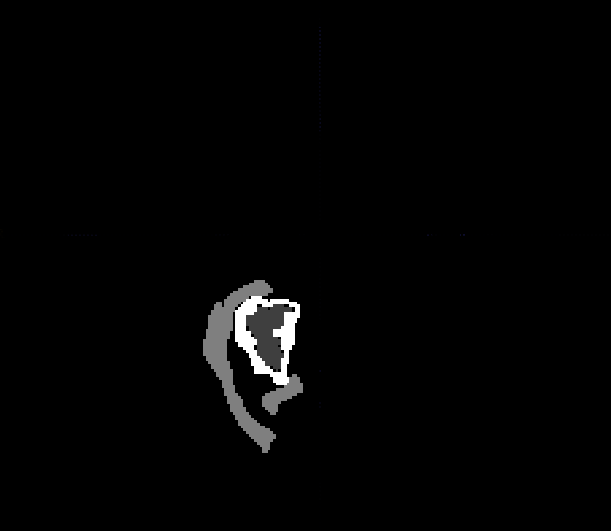}
  \caption{}
\end{subfigure}
\begin{subfigure}{.25\textwidth}
  \centering
  \includegraphics[width=3cm,height=3cm]{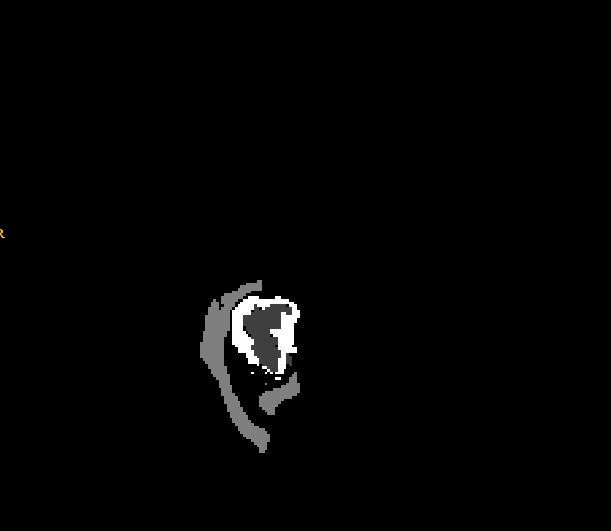}
  \caption{}
\end{subfigure}%
\begin{subfigure}{.25\textwidth}
  \centering
  \includegraphics[width=3cm,height=3cm]{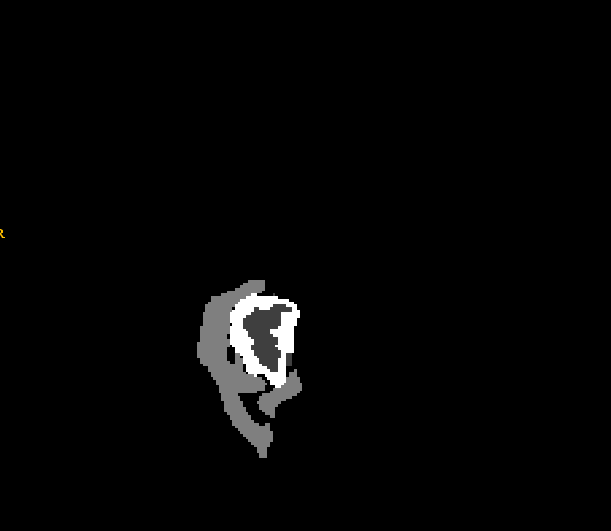}
  \caption{}
\end{subfigure}
\begin{subfigure}{.25\textwidth}
  \centering
  \includegraphics[width=3cm,height=3cm]{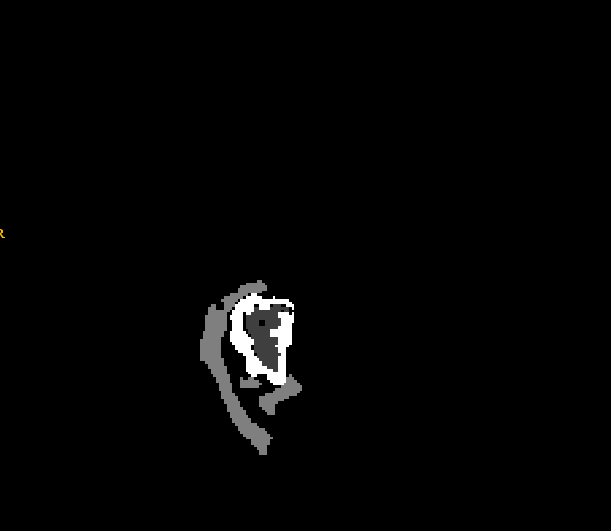}
  \caption{}
\end{subfigure}%
\begin{subfigure}{.25\textwidth}
  \centering
  \includegraphics[width=3cm,height=3cm]{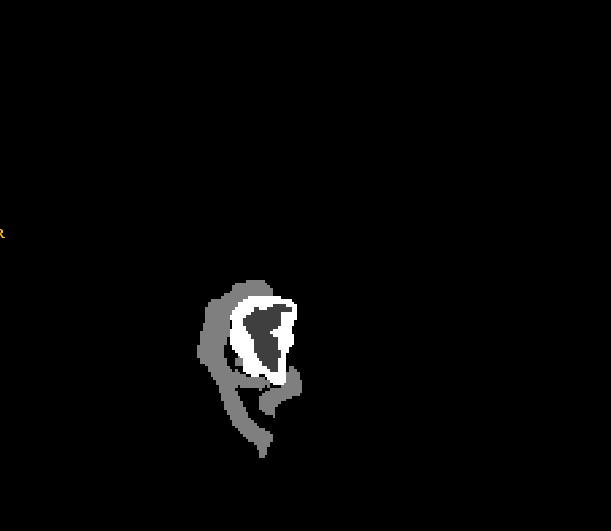}
  \caption{}
\end{subfigure}
\caption{Training set image, correct segmentation a) Original FLAIR image b) ground truth segmentation c) model 1 d) model 2 e) model 3 f) model 4 g) model 5 h) model 6.}
\label{fig:18}% Give a unique label
\end{figure}

\begin{figure}
\centering
\begin{subfigure}{.25\textwidth}
  \centering
  \includegraphics[width=3cm,height=3cm]{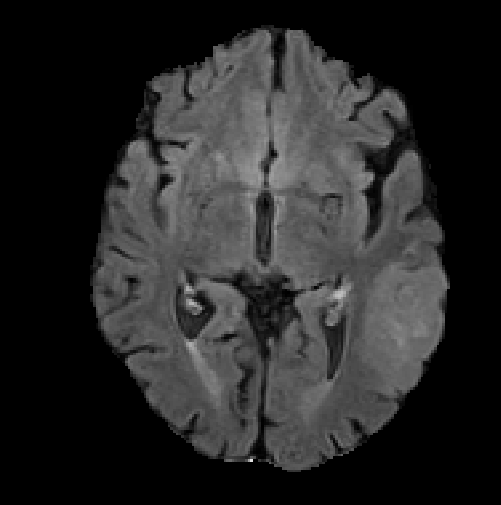}
  \caption{}
\end{subfigure}%
\begin{subfigure}{.25\textwidth}
  \centering
  \includegraphics[width=3cm,height=3cm]{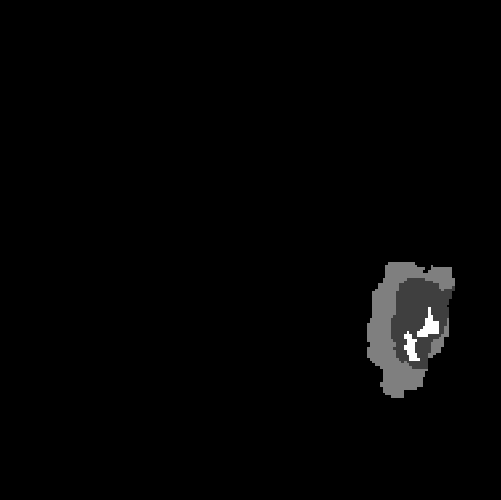}
  \caption{}
\end{subfigure}
\begin{subfigure}{.25\textwidth}
  \centering
  \includegraphics[width=3cm,height=3cm]{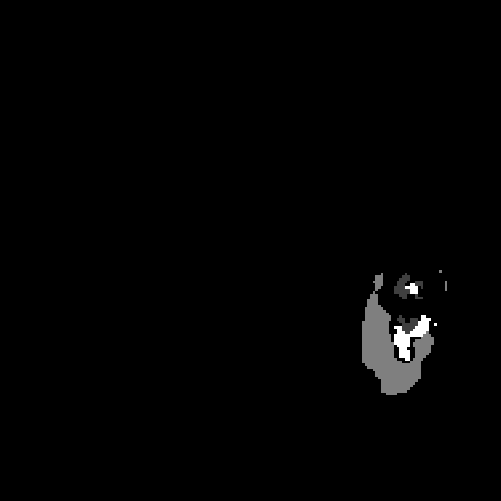}
  \caption{}
\end{subfigure}%
\begin{subfigure}{.25\textwidth}
  \centering
  \includegraphics[width=3cm,height=3cm]{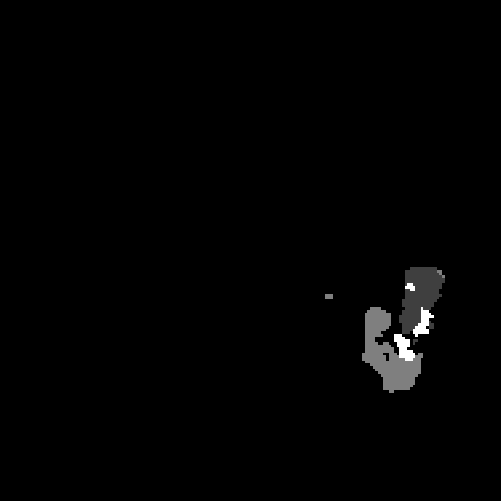}
  \caption{}
\end{subfigure}
\begin{subfigure}{.25\textwidth}
  \centering
  \includegraphics[width=3cm,height=3cm]{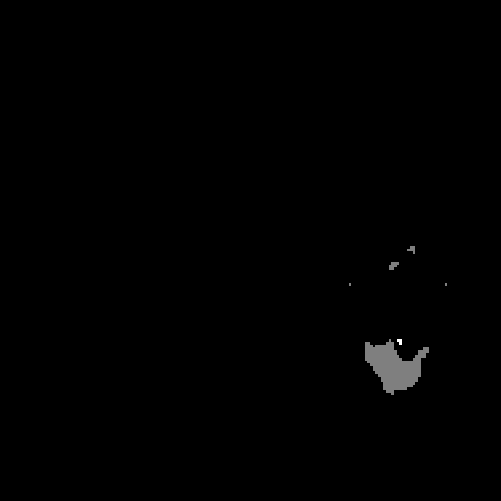}
  \caption{}
\end{subfigure}%
\begin{subfigure}{.25\textwidth}
  \centering
  \includegraphics[width=3cm,height=3cm]{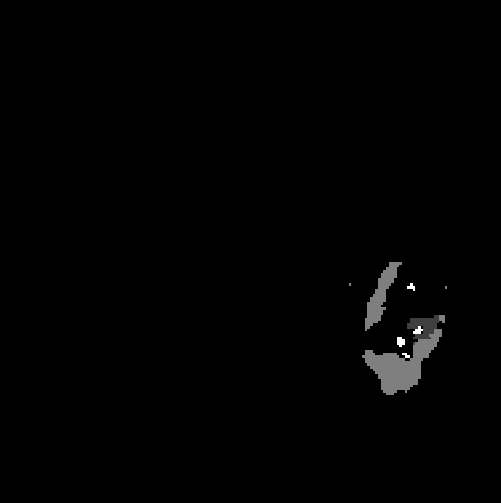}
  \caption{}
\end{subfigure}
\begin{subfigure}{.25\textwidth}
  \centering
  \includegraphics[width=3cm,height=3cm]{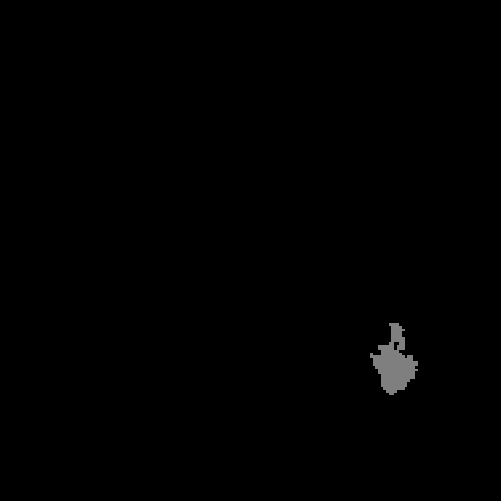}
  \caption{}
\end{subfigure}%
\begin{subfigure}{.25\textwidth}
  \centering
  \includegraphics[width=3cm,height=3cm]{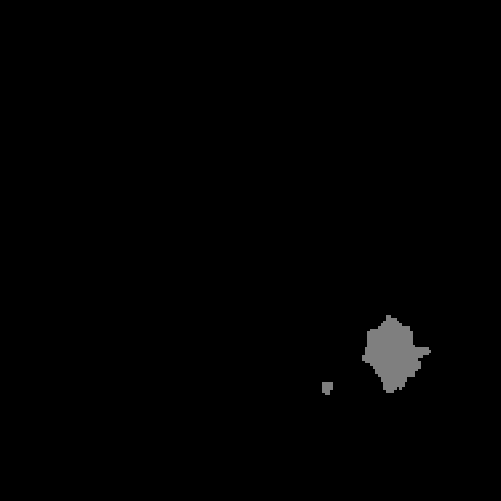}
  \caption{}
\end{subfigure}\caption{Training set image, incorrect segmentation a) Original FLAIR image b) ground truth segmentation c) model 1 d) model 2 e) model 3 f) model  g) model 5 h) model 6.}
\label{fig:19}% Give a unique label
\end{figure}

A DSC comparison of the training and validation sets for these variations is as shown in Fig. \ref{fig:17}. The segmentation results show that the method overfits the training data and does not generate good results for the unknown validation set. Segmentation results of the sample images from the training set for correct as well as incorrect segmentation are shown in Fig. \ref{fig:18} and Fig. \ref{fig:19}. The network does not distinguish between the subregions where the tumor appearance is homogeneous for all its subregions and the intensity is the same as normal brain tissues.

\subsection{End-to-End methods for tumor segmentation and OS prediction}
\label{sec:5.6}

Since 2017, the second task of survival prediction has been introduced in the challenge. Some of the methods participated in the end-to-end solutions in the challenge, i.e., segmentation followed by survival prediction.

In \cite{shboul2017glioblastoma}, an ensemble of RF and CNN segments the tumor and the random forest regressor (RFR) was used to predict the overall survival days using 240 features out of 1366 different features (Kaplan-Meier was used to find relevant and useful features). Authors in \cite{jungo2017towards} modified U-net with Full Resolution Residual Network (FRRN) and Residual Unit (RU) units along with weight scaling dropout. The survival prediction ANN worked with a linear activation function on four selected features.

A variant of U-net was used in \cite{isensee2017brain}, which took 3D input and included a context module and a localization module in each level of the architecture. The segmentation result was generated based on an element-wise summation of the output from the decoder layers. Survival prediction was the average of RFR and multilayer perceptron (MLP). RFR trained on 517 features extracted from three tumor sub-regions using the radiomics package \cite{van2017computational}. The output of RFR and MLP averaged 15 MLPs designs with three hidden layers, each with 64 neurons.

Authors in \cite{baid2018deep} implemented 3D Unet with three stages of encoder-decoder architecture. A egression model based radiomics features selection trains the MLP for OS prediction. Whereas in \cite{weninger2018segmentation}, two 3D U-nets used a four-stage encoder decoder architecture; the first network segmented the whole tumor, and the second one segmented the tumor subregion. In addition to four conventional modalities, they used an additional image as an input, which was the T1c-T1 subtracted image. This image provided additional information for the tumor core region. They used only the age feature to predict the OS using a linear regressor. The approach presented in \cite{li2017deep} used an FCNN named FCRN-101, which derived from pre-trained SegNet and U-net architecture. A three path network combined the result of three views, i.e., axial, coronal, and sagittal. The OS prediction used SPNet, a fully connected CNN, which took four modalities and the network segmentation result as input to predict the probability of OS prediction.

The authors in \cite{feng2018brain} used an ensemble of six 3D U-net type networks with variation in the input size, number of encoding/decoding blocks, and feature maps at every layer. The OS prediction used linear regression with ground truth image volume, surface area, age, and resection status. Features were input to the network after z-score normalization. Authors in \cite{puybareau2018segmentation} implemented FCN and generated results for three axes. They used majority voting to generate the final segmentation results. For OS prediction, ten features (focusing on necrosis and active tumor) from the segmentation results were generated, and the mean PCA and standard deviation PCA was used to train RF on the GTR images.

In \cite{sun2018tumor} an ensemble of three networks (U-net\cite{ronneberger2015u}, DFKZNet\cite{isensee2017brain}, and CA-CNN\cite{wang2017automatic}) was used, and majority voting applied for final segmentation. The OS prediction used RF with 14 radiomics features selected from various modality images, Laplacian of Gaussian Images and wavelet decomposed images. Authors in \cite{agravat2019prediction} implemented a 2D U-net architecture with three stages for tumor segmentation and age, volumetric, and shape features of the whole tumor were used to predict the OS.

All the approaches did not use location-based information of the tumor and its sub-regions. In contrast, \cite{kao2018brain}, used twenty-one brain parcellation regions as input along with four MR modalities. It emphasized the number of tumor regions in those specific parcellation areas. Those twenty-five input channels were given as input to the ensemble of 3D U-net as well as the ensemble of DeepMedic architectures with different kernel and input patch sizes. Tractrographic features from network segmented regions trained SVM classifiers with a linear kernel to predict the OS. Authors in \cite{agravat2019brain} implemented a 2D Unet of three stages with dense blocks at every encoder level, and the feature set of \cite{agravat2019prediction} of the necrosis tumor sub-region for the OS prediction.

The authors in \cite{soltaninejad2017mri} combined the features from VGG16 based FCN and texton maps to generate features and supply them to the RF classifier to generate the segmentation result. RF was also used for OS prediction using volumetric as well as the age feature of the patient.

Authors in \cite{agravat20203d}, \cite{dai2018automatic}, \cite{hua2018multimodal}, \cite{islam2018glioma}, \cite{raval2021glioblastoma},  \cite{zhou2017tpcnn} attempted an end-to-end segmentation approach. Table \ref{tab:10} compares the segmentation results of end-to-end methods. The approaches for pre-processing and post-processing techniques are as specified in Section \ref{sec:5.3} and Table \ref{tab:11} provides details related to survival prediction.

 \onecolumn
\begin{small}
\tablefirsthead{%
     \toprule  \\
     \textbf{Ref.} & \textbf{Pre-processing} & \textbf{Input Modality + Augmentation}  & \textbf{Patch/ Image} & \textbf{Input view} & \textbf{Network Architecture}  & \textbf{\# Networks} & \textbf{Ensemble Type} & \textbf{Loss Function} & \textbf{Post-processing} & \textbf{Dataset} & \textbf{DSC Mean} \\     
       }
 %This is the header for the remaining page(s) of the table...
\tablehead{%
      \multicolumn{12}{c}
    {{\bfseries \tablename\ \thetable{ } --
       continued \ldots}} \\
     \toprule
     \textbf{Ref.} & \textbf{Pre-processing} & \textbf{Input Modality + Augmentation}  & \textbf{Patch/ Image} & \textbf{Input view} & \textbf{Network Architecture}  & \textbf{\# Networks} & \textbf{Ensemble Type} & \textbf{Loss Function} & \textbf{Post-processing} & \textbf{Dataset} & \textbf{DSC Mean} \\ }
 
%This is the footer for all pages except the last page of the table...
\tabletail{%
	\midrule
}
\tablelasttail{\bottomrule}
\tablecaption{End-to-end Methods, Task 1: Brain Tumor Segmentation.}
\label{tab:10}
\begin{landscape}
\begin{xtabular}{@{\extracolsep{\fill}}p{0.8 cm} p{2 cm} p{2 cm} p{1.5 cm} p{1.5 cm} p{1.5 cm} p{1.5 cm} p{1.2 cm} p{1 cm} p{1 cm} p{1 cm} p{2.5 cm}@{\extracolsep{0pt}} }
\midrule
\cite{li2017deep} & 6 & T1, T2, T1c, FLAIR & 2D images & axial & 2D FCNN & 3 & 	Parallel & weighted focal & - & BraTS 2017 & Validation set: ET:0.75,WT:0.88,CT:0.71 \newline Test set: ET:0.69,WT:0.88,TC:0.71  \\
\midrule
\cite{isensee2017brain} & 3 followed by clipping and scaling & T1, T2, T1c, FLAIR & 3D patches	& axial & 3D FCNN & 	1 & -  & weighted dice & - & Brats 2017 & 
Validation set: WT:0.90,TC:0.80,ET:0.73 Test set: WT:0.86,TC:0.78,ET:0.65 \\
\midrule
\cite{shboul2017glioblastoma} & 2,3 & T1, T2, T1c, FLAIR & 2D patches & axial & 2D CNN  + RF & 1 & -  & softmax & - & BraTS 2017 & HGG Test set: ET:0.73,WT:0.83,TC:0.72 \\
\midrule	
\cite{jungo2017towards}	& - & T1, T2, T1c, FLAIR & 2D images & axial & 2D FCNN & 1 & -  & - & - & BraTS 2017 & Validation set: WT:0.90,TC:0.79,ET:0.75 Test set: WT:0.87,TC:0.74,ET:0.67 \\
\midrule
\cite{soltaninejad2017mri} & 1,3,4 &	T1, T2, T1c, FLAIR &	2D images & axial & 2D FCNN & 1 & - & 	- &	-	& BraTS 2017 & Validation set: WT:0.86,TC:0.78,ET:0.66 Test set: WT:0.85,TC:0.69,ET:0.67 \\
\midrule
\cite{zhou2017tpcnn} &	4,5 &	T1, T2, T1c, FLAIR & 2D + 3D patches & axial & 2D FCNN & 5 & cascaded to parallel & - & 1,4  &	BraTS 2017 & Training set: WT:0.88,TC:0.72,ET:0.73 \\
\midrule
\cite{feng2018brain} & 2,7 & T1, T2, T1c, FLAIR, Random Flip & 3D patches & axial &	3D FCNN & 6 & parallel  & weighted uniform & - & BraTS 2018 & Validation set: WT:0.91,TC:0.84,ET:0.79 \\
\midrule
\cite{puybareau2018segmentation} & 8 & T1, T2, T1c & 2D images & axial & 2D FCNN &	1 & -  & softmax & 3 & BraTS 2018 & Training set: WT:0.82 \\
\midrule
\cite{sun2018tumor} & 3, random flipping, gaussian noise & T1, T2, T1c, FLAIR & 3D images	& axial & 3D FCNN & 3 & parallel & - & - &  BraTS 2018 & Validation set: WT:0.91,TC:0.85,ET:0.81 Test set: WT:0.88,TC:0.80,ET:0.72 \\
\midrule
\cite{baid2018deep} & 2,3 & T1, T2, T1c, FLAIR & 3D patches & axial	& 3D FCNN &	1 &	-
 & - & 1,3, false positives removal & BraTS 2018 & Validation set: WT:0.88,TC:0.83,ET:0.75 Test set: WT:0.85,TC:0.77,ET:0.67 \\
\midrule
\cite{weninger2018segmentation} & 3 & T1, T2, T1c, FLAIR, (T1c-T1) & 3D patches	& axial & 3D FCNN & 2 & cascaded & dice & - & BraTS 2018 & Validation set: WT:0.89,TC:0.76,ET:0.71 Test set: WT:0.84,TC:0.73,ET:0.62 \\
\midrule
\cite{kao2018brain} & - & T1, T2, T1c, FLAIR, 21 binary brain parcellation images & 3D patches & axial & 3D FCNN	 & 26 & parallel & - & - & BraTS 2018 & Validation set: WT:0.91,TC:0.81,ET:0.79 Test set: WT:0.88,TC:0.79,ET:0.75 \\
\midrule
\cite{hua2018multimodal} &	2,4,8 &	T1, T2, T1c, FLAIR & 3d images & axial & 3D FCNN &	5 & parallel to cascaded &	focal & 1 & BraTS 2018 &	Validation set: WT:0.90,TC:0.84,ET:0.77 Test set: WT:0.87,TC:0.80,ET:0.74\\
\midrule
\cite{banerjee2018multi} &	- &	T1, T2, T1c, FLAIR & 2D images & axial, coronal, sagittal & 2D FCNN & 3 & parallel &	weighted cross entropy + generalized dice &	- &	BraTS 2018 & Validation set: WT:0.88,TC:0.80,ET:0.77 Test set: WT:0.87,TC:0.80,ET:0.74\\
\midrule
\cite{islam2018glioma} &	3 &	T2, T1c, FLAIR & 2D images & axial & 2D FCNN & 1 & - &	- &	- &	BraTS 2018 &	Validation set: WT:0.90,TC:0.83,ET:0.77 Test set: WT:0.87,TC:0.77,ET:0.70\\
\midrule
\cite{dai2018automatic}	 & 1,3 & T1, T2, T1c, FLAIR  & 3D images & axial & 2D FCNN & 	9 &	parallel &	confusion + multi class dice &	1,3	& BraTS 2018 &	Validation set: WT:0.91,TC:0.85,ET:0.81 Test set: WT:0.87,TC:0.79,ET:0.74 \\
\midrule
\cite{agravat2019prediction} & 3 & T1, T2, T1c, FLAIR & 2D images & axial & 2D FCNN & 1 & - & dice & - & BraTS 2018 & -  \\
\midrule
\cite{agravat2019brain} & 3 & T1, T2, T1c, FLAIR & 2D images & axial & 2D FCNN & 1 & - & focal & - & BraTS 2019 & Validation set: WT:0.73,TC:0.63,ET:0.59  Test set: WT:0.72,TC:0.66,ET:0.64 \\
\end{xtabular}
\end{landscape}
\end{small}
\twocolumn

\begin{table}
\caption{End-to-end Methods, Task 2: OS Prediction.}
\label{tab:11}
\begin{tabular}{@{}p{1 cm} p{2 cm} p{1.5 cm} p{1 cm} p{1 cm}@{}}
\toprule
\textbf{Ref.} & \textbf{Method}	& \textbf{\# features/ type of network} & \textbf{Dataset} & \textbf{Accuracy (\%)} \\
\midrule
\cite{li2017deep} & 2D CNN & - & BraTS 2017 & Valid:55 Test:45 \\
\midrule
\cite{isensee2017brain} & RFR and MLP & 66 & BraTS 2017 & Valid:52.6 \\
\midrule
\cite{shboul2017glioblastoma} & RFR & 240 & BraTS 2017 & Valid:66.7 Test:57.9 \\
\midrule
\cite{jungo2017towards} & ANN & 4 & BraTS 2017 & Valid:42.4 Test:56.8  \\
\midrule
\cite{feng2018brain} & Linear Regression Model & 9 & BraTS 2018 & Valid:32.1 \\
\midrule
\cite{puybareau2018segmentation} & PCA  + RF & 10 & BraTS 2018 & Test:61 \\
\midrule
\cite{sun2018tumor} & RF & 14 & BraTS 2018 & Valid:46.4 Test:61 \\
\midrule
\cite{baid2018deep} & MLP & 468 & BraTS 2018 & Valid:57.1 Test:55.8 \\
\midrule
\cite{weninger2018segmentation} & Linear Regressor & 1 & BraTS 2018 & Valid:50 Test:55.8 \\
\midrule
\cite{kao2018brain} & SVM with linear kernel & - & BraTS 2018 & Valid:35.7 Test:41.6 \\
\midrule
\cite{agravat2019prediction} & RF & 13 & BraTS 2018 & 59 \\
\midrule
\cite{agravat2019brain} & RF & 13 & BraTS 2019 & Valid:58.6 Test:57.9 \\
\midrule
\cite{soltaninejad2017mri} &	RF & 4 & BraTS 2017	& Valid:48.5 Test:41.1 \\
\midrule
\cite{zhou2017tpcnn} &	CNN+XBoost & 183 + CNN features & BraTS 2017 &	Training:63.3\\
\midrule 
\cite{hua2018multimodal} &	ensemble of Xboost, SVM, MLP, DT, RF, LDA &	900 & BraTS 2018 &	Test:51.9 \\
\midrule
\cite{banerjee2018multi} &	MLP &	83 &	BraTS 2018 & Valid:54 \\
\midrule
\cite{islam2018glioma} &	ANN &	50 &	BraTS 2018 & Valid:67.9 Test:46.8 \\
\midrule
\cite{dai2018automatic} &	Xboost &	195 & BraTS 2017 &	Valid:50 \\

\bottomrule
\end{tabular}
\end{table}
\FloatBarrier
\section{Limitations of tumor segmentation and OS prediction}
\label{sec:6}

Tumor segmentation result depends on the architectural design from shallow CNN to the ensemble/cascaded CNN, the amount of training data, input pre-processing, type of input(2D/3D), network optimization as well as post-processing of the generated output. Still, the DL methods have certain limitations, which include:

\begin{itemize}
\item Over-fitting: The common problem of the DNN based approach is over-fitting. It may occur due to the unavailability of an adequate amount of labelled training data for brain tumor segmentation, which refers to a model that has an excellent performance on the training dataset but does not perform well on new data. The over-fitting problem can be handled either by reducing the network complexity (in terms of network layers and parameters) or by generating an ample amount of training data using image augmentation techniques. The augmentation techniques produce new images by performing data transformations and the corresponding ground truth that includes operations of scaling, rotation, translation, brightness variation, elastic deformations, horizontal flipping, and mirroring.
\item Class imbalance: Class imbalance is another issue in tumor segmentation where the background class dominates the foreground class(tumor). Class imbalance can be handled by proper training data sampling, improved loss functions, and augmentation techniques.
\end{itemize}

The highest reported accuracy for the survival prediction task does not exceed 63\% \cite{feng2018brain}. It is due to the dependency on the extraction of the features from the segmentation results. Incorrect segmentation results in wrong feature extraction for survival prediction. The biological importance of the extracted features also plays an important role. If the relevance of the features is not known correctly, the survival prediction cannot be accurate. Besides, the importance of tumor sub-regions plays a role in feature extraction. The dataset includes all the pre-operative scans and does not give any other information like the success of tumor removal, post-operative treatments to the patients, and the response of the patients to such treatments. Chances of tumor reoccurrence are high if the patient is exposed to a radiation environment. Features related to it may further improve the OS prediction.

\section{Conclusion and Future Research Directions}
\label{sec:7}

The availability of the benchmark dataset (BraTS) has grown the field of computer-assisted medical image analysis for brain tumor segmentation. The article covers a detailed literature survey for brain tumor segmentation techniques. Tumor segmentation is approached with various techniques like semi-automated and automated methods. The semi-automated methods work on the input provided by the user. Such methods suffer from limitations like manual seed point selection and atlas creation. Methods have gradually been improved to include machine learning techniques like clustering, RF  and ANN. The limitation of such methods is the selection of the features for training, which requires knowledge of biological information of the image. The need of domain knowledge is removed by CNNs - deep neural network. CNN extracts high level features at deeper layers successively from the low level features from the preceding layers. This feature learning has improved the performance of CNN for tumor segmentation. Improvements to CNN are its variants, FCNN and ensemble of CNN/FCNN. Details of all the methods along with pre-processing, post-processing, prominent highlights of the methods and evaluation measures are given in Table \ref{tab:5}, Table \ref{tab:7}, Table \ref{tab:10} and Table \ref{tab:11}.

The ensemble generates robust segmentation results as well as improvements in the accuracy of the network. All these methods generate spurious segmentation results, which improve with the help of post-processing techniques like connected component analysis, spatial regularization and morphological operations to fine-tune the output. Class imbalance is the primary concern in the training of CNNs for medical image analysis. Balanced input selection and loss function for a positive class will resolve the issue. Due to the unavailability of a large amount of data for training, it may be possible that the network overfits the training data. Regularization and dropout resolve such issues. Moreover, as Bayesian CNN \cite{shridhar2019comprehensive} handles epistemic and aleatoric uncertainties in the presence of limited data and knowledge, such type of a network is useful for semantic segmentation. Adaptive loss \cite{barron2019general}, approximates varieties of loss functions with a single latent variable. This function can further be thought of to solve the problem of semantic segmentation. Despite their popularity such methods have following limitations:

\begin{itemize}
\item computational efficiency
\item memory requirement: As the depth of the network increases, the number of network parameters increases. This increase in the parameters requires additional memory as well as time to tune those on each epoch.
\item Ample amount of annotated training data requirement: The annotated data generation is itself a challenge as this process is very time consuming and the annotation results may change depending on variability of the expert. In addition, specific annotation tools are required by the expert for proper delineation and annotation purposes. In place of voxel annotation, image labelling for the presence/ absence of a tumor is less time consuming, does not require much expertise and specialized annotation tools. Use of the mixed supervision \cite{lee2019ficklenet} of image labelling in addition to voxel labelling may help the network to learn relevant features for segmentation with less annotated data.

\end{itemize}

\begin{acknowledgements}
The authors would like to thank NVIDIA Corporation for donating the Quadro K5200 and Quadro P5000 GPU used for this research, Dr. Krutarth Agravat (Medical Officer, Essar Ltd) for clearing our doubts related to medical concepts, Po-yu Kao, and Ujjawal Baid for their continuous support and help, Dr. Spyros and his entire team for BraTS dataset. The authors acknowledge continuous support from Professor Sanjay Chaudhary, Professor N.  Padmanabhan, and Professor Manjunath Joshi for this work.\end{acknowledgements}

\section*{Conflict of interest}
 On behalf of all authors, the corresponding author states that there is no conflict of interest.

\bibliographystyle{spbasic}      % basic style, author-year citations

\FloatBarrier

\bibliography{bibtexref}

\begin{thebibliography}{163}
\providecommand{\natexlab}[1]{#1}
\providecommand{\url}[1]{{#1}}
\providecommand{\urlprefix}{URL }
\expandafter\ifx\csname urlstyle\endcsname\relax
  \providecommand{\doi}[1]{DOI~\discretionary{}{}{}#1}\else
  \providecommand{\doi}{DOI~\discretionary{}{}{}\begingroup
  \urlstyle{rm}\Url}\fi
\providecommand{\eprint}[2][]{\url{#2}}

\bibitem[{Abadi et~al.(2016)Abadi, Barham, Chen, Chen, Davis, Dean, Devin,
  Ghemawat, Irving, Isard et~al.}]{abadi2016tensorflow}
Abadi M, Barham P, Chen J, Chen Z, Davis A, Dean J, Devin M, Ghemawat S, Irving
  G, Isard M, et~al. (2016) Tensorflow: A system for large-scale machine
  learning. In: 12th $\{$USENIX$\}$ Symposium on Operating Systems Design and
  Implementation ($\{$OSDI$\}$ 16), pp 265--283

\bibitem[{Agn et~al.(2015)Agn, Puonti, af~Rosensch{\"o}ld, Law, and
  Van~Leemput}]{agn2015brain}
Agn M, Puonti O, af~Rosensch{\"o}ld PM, Law I, Van~Leemput K (2015) Brain tumor
  segmentation using a generative model with an rbm prior on tumor shape. In:
  BrainLes 2015, Springer, pp 168--180

\bibitem[{Agravat and Raval(2019{\natexlab{a}})}]{agravat2019brain}
Agravat R, Raval MS (2019{\natexlab{a}}) Brain tumor segmentation and survival
  prediction. arXiv preprint arXiv:190909399

\bibitem[{Agravat and Raval(2020)}]{agravat20203d}
Agravat R, Raval MS (2020) 3d semantic segmentation of brain tumor for overall
  survival prediction. arXiv preprint arXiv:200811576

\bibitem[{Agravat and Raval(2016)}]{agravat2016brain}
Agravat RR, Raval MS (2016) Brain tumor segmentation-towards a better life. CSI
  Communication 40:31--35

\bibitem[{Agravat and Raval(2018)}]{agravat2018deep}
Agravat RR, Raval MS (2018) Deep learning for automated brain tumor
  segmentation in mri images. In: Soft Computing Based Medical Image Analysis,
  Elsevier, pp 183--201

\bibitem[{Agravat and Raval(2019{\natexlab{b}})}]{agravat2019prediction}
Agravat RR, Raval MS (2019{\natexlab{b}}) Prediction of overall survival of
  brain tumor patients. In: TENCON 2019-2019 IEEE Region 10 Conference
  (TENCON), IEEE, pp 31--35

\bibitem[{Albiol et~al.(2018)Albiol, Albiol, and Albiol}]{albiol2018extending}
Albiol A, Albiol A, Albiol F (2018) Extending 2d deep learning architectures to
  3d image segmentation problems. In: International MICCAI Brainlesion
  Workshop, Springer, pp 73--82

\bibitem[{Badrinarayanan et~al.(2017)Badrinarayanan, Kendall, and
  Cipolla}]{badrinarayanan2017segnet}
Badrinarayanan V, Kendall A, Cipolla R (2017) Segnet: A deep convolutional
  encoder-decoder architecture for image segmentation. IEEE transactions on
  pattern analysis and machine intelligence 39(12):2481--2495

\bibitem[{Baid et~al.(2018)Baid, Talbar, Rane, Gupta, Thakur, Moiyadi, Thakur,
  and Mahajan}]{baid2018deep}
Baid U, Talbar S, Rane S, Gupta S, Thakur MH, Moiyadi A, Thakur S, Mahajan A
  (2018) Deep learning radiomics algorithm for gliomas (drag) model: a novel
  approach using 3d unet based deep convolutional neural network for predicting
  survival in gliomas. In: International MICCAI Brainlesion Workshop, Springer,
  pp 369--379

\bibitem[{Bakas et~al.(2015)Bakas, Zeng, Sotiras, Rathore, Akbari, Gaonkar,
  Rozycki, Pati, and Davatzikos}]{bakas2015glistrboost}
Bakas S, Zeng K, Sotiras A, Rathore S, Akbari H, Gaonkar B, Rozycki M, Pati S,
  Davatzikos C (2015) Glistrboost: combining multimodal mri segmentation,
  registration, and biophysical tumor growth modeling with gradient boosting
  machines for glioma segmentation. In: BrainLes 2015, Springer, pp 144--155

\bibitem[{Bakas et~al.(2017{\natexlab{a}})Bakas, Akbari, Sotiras, Bilello,
  Rozycki, Kirby, Freymann, Farahani, and Davatzikos}]{bakas2017segmentation}
Bakas S, Akbari H, Sotiras A, Bilello M, Rozycki M, Kirby J, Freymann J,
  Farahani K, Davatzikos C (2017{\natexlab{a}}) Segmentation labels and
  radiomic features for the pre-operative scans of the tcga-gbm collection. the
  cancer imaging archive (2017)

\bibitem[{Bakas et~al.(2017{\natexlab{b}})Bakas, Akbari, Sotiras, Bilello,
  Rozycki, Kirby, Freymann, Farahani, and Davatzikos}]{bakas2017segmentation1}
Bakas S, Akbari H, Sotiras A, Bilello M, Rozycki M, Kirby J, Freymann J,
  Farahani K, Davatzikos C (2017{\natexlab{b}}) Segmentation labels and
  radiomic features for the pre-operative scans of the tcga-lgg collection. The
  Cancer Imaging Archive 286

\bibitem[{Bakas et~al.(2017{\natexlab{c}})Bakas, Akbari, Sotiras, Bilello,
  Rozycki, Kirby, Freymann, Farahani, and Davatzikos}]{bakas2017advancing}
Bakas S, Akbari H, Sotiras A, Bilello M, Rozycki M, Kirby JS, Freymann JB,
  Farahani K, Davatzikos C (2017{\natexlab{c}}) Advancing the cancer genome
  atlas glioma mri collections with expert segmentation labels and radiomic
  features. Scientific data 4:170117

\bibitem[{Bakas et~al.(2018)Bakas, Reyes, Jakab, Bauer, Rempfler, Crimi,
  Shinohara, Berger, Ha, Rozycki et~al.}]{bakas2018identifying}
Bakas S, Reyes M, Jakab A, Bauer S, Rempfler M, Crimi A, Shinohara RT, Berger
  C, Ha SM, Rozycki M, et~al. (2018) Identifying the best machine learning
  algorithms for brain tumor segmentation, progression assessment, and overall
  survival prediction in the brats challenge. arXiv preprint arXiv:181102629

\bibitem[{Banerjee et~al.(2018)Banerjee, Mitra, and
  Shankar}]{banerjee2018multi}
Banerjee S, Mitra S, Shankar BU (2018) Multi-planar spatial-convnet for
  segmentation and survival prediction in brain cancer. In: International
  MICCAI Brainlesion Workshop, Springer, pp 94--104

\bibitem[{Barron(2019)}]{barron2019general}
Barron JT (2019) A general and adaptive robust loss function. In: Proceedings
  of the IEEE Conference on Computer Vision and Pattern Recognition, pp
  4331--4339

\bibitem[{Bastien et~al.(2012)Bastien, Lamblin, Pascanu, Bergstra, Goodfellow,
  Bergeron, Bouchard, Warde-Farley, and Bengio}]{bastien2012theano}
Bastien F, Lamblin P, Pascanu R, Bergstra J, Goodfellow I, Bergeron A, Bouchard
  N, Warde-Farley D, Bengio Y (2012) Theano: new features and speed
  improvements. arXiv preprint arXiv:12115590

\bibitem[{Bauer et~al.(2012)Bauer, Fejes, Slotboom, Wiest, Nolte, and
  Reyes}]{bauer2012segmentation}
Bauer S, Fejes T, Slotboom J, Wiest R, Nolte LP, Reyes M (2012) Segmentation of
  brain tumor images based on integrated hierarchical classification and
  regularization. In: MICCAI BraTS Workshop. Nice: Miccai Society, p~11

\bibitem[{Bernal et~al.(2019)Bernal, Kushibar, Asfaw, Valverde, Oliver,
  Mart{\'\i}, and Llad{\'o}}]{bernal2019deep}
Bernal J, Kushibar K, Asfaw DS, Valverde S, Oliver A, Mart{\'\i} R, Llad{\'o} X
  (2019) Deep convolutional neural networks for brain image analysis on
  magnetic resonance imaging: a review. Artificial intelligence in medicine
  95:64--81

\bibitem[{Bharath et~al.(2017)Bharath, Colleman, Sima, and
  Van~Huffel}]{bharath2017tumor}
Bharath HN, Colleman S, Sima DM, Van~Huffel S (2017) Tumor segmentation from
  multimodal mri using random forest with superpixel and tensor based feature
  extraction. In: International MICCAI Brainlesion Workshop, Springer, pp
  463--473

\bibitem[{for Biotechnology~Information(2020 (accessed December 30,
  2020))}]{pubmed:2020}
for Biotechnology~Information NC (2020 (accessed December 30, 2020)) National
  Library of Medicine. \urlprefix\url{https://pubmed.ncbi.nlm.nih.gov/}

\bibitem[{Buendia et~al.(2013)Buendia, Taylor, Ryan, and
  John}]{buendia2013grouping}
Buendia P, Taylor T, Ryan M, John N (2013) A grouping artificial immune network
  for segmentation of tumor images. Multimodal Brain Tumor Segmentation 1

\bibitem[{Casamitjana et~al.(2016)Casamitjana, Puch, Aduriz, and
  Vilaplana}]{casamitjana20163d}
Casamitjana A, Puch S, Aduriz A, Vilaplana V (2016) 3d convolutional neural
  networks for brain tumor segmentation: a comparison of multi-resolution
  architectures. In: International Workshop on Brainlesion: Glioma, Multiple
  Sclerosis, Stroke and Traumatic Brain Injuries, Springer, pp 150--161

\bibitem[{Casamitjana et~al.(2017)Casamitjana, Cat{\`a}, S{\'a}nchez, Combalia,
  and Vilaplana}]{casamitjana2017cascaded}
Casamitjana A, Cat{\`a} M, S{\'a}nchez I, Combalia M, Vilaplana V (2017)
  Cascaded v-net using roi masks for brain tumor segmentation. In:
  International MICCAI Brainlesion Workshop, Springer, pp 381--391

\bibitem[{Castillo et~al.(2017)Castillo, Daza, Rivera, and
  Arbel{\'a}ez}]{castillo2017brain}
Castillo LS, Daza LA, Rivera LC, Arbel{\'a}ez P (2017) Brain tumor segmentation
  and parsing on mris using multiresolution neural networks. In: International
  MICCAI Brainlesion Workshop, Springer, pp 332--343

\bibitem[{Center(2019 (accessed April 6, 2020))}]{rochester:2020}
Center RM (2019 (accessed April 6, 2020)) Health Encyclopedia.
  \urlprefix\url{https://www.urmc.rochester.edu/encyclopedia/ content.aspx}

\bibitem[{Chandra et~al.(2019)Chandra, Vakalopoulou, Fidon, Battistella,
  Estienne, Sun, Robert, Deutsch, and Paragios}]{chandra2019context}
Chandra S, Vakalopoulou M, Fidon L, Battistella E, Estienne T, Sun R, Robert C,
  Deutsch E, Paragios N (2019) Context aware 3d cnns for brain tumor
  segmentation. brainles 2018. Springer LNCS 11384:299--310

\bibitem[{Chang(2016)}]{chang2016fully}
Chang PD (2016) Fully convolutional deep residual neural networks for brain
  tumor segmentation. In: International Workshop on Brainlesion: Glioma,
  Multiple Sclerosis, Stroke and Traumatic Brain Injuries, Springer, pp
  108--118

\bibitem[{Chen et~al.(2018{\natexlab{a}})Chen, Bentley, Mori, Misawa, Fujiwara,
  and Rueckert}]{chen2018drinet}
Chen L, Bentley P, Mori K, Misawa K, Fujiwara M, Rueckert D
  (2018{\natexlab{a}}) Drinet for medical image segmentation. IEEE transactions
  on medical imaging 37(11):2453--2462

\bibitem[{Chen et~al.(2017)Chen, Papandreou, Kokkinos, Murphy, and
  Yuille}]{chen2017deeplab}
Chen LC, Papandreou G, Kokkinos I, Murphy K, Yuille AL (2017) Deeplab: Semantic
  image segmentation with deep convolutional nets, atrous convolution, and
  fully connected crfs. IEEE transactions on pattern analysis and machine
  intelligence 40(4):834--848

\bibitem[{Chen et~al.(2018{\natexlab{b}})Chen, Liu, Peng, Sun, and
  Qiao}]{chen2018s3d}
Chen W, Liu B, Peng S, Sun J, Qiao X (2018{\natexlab{b}}) S3d-unet: separable
  3d u-net for brain tumor segmentation. In: International MICCAI Brainlesion
  Workshop, Springer, pp 358--368

\bibitem[{Chollet et~al.(2018)}]{chollet2018keras}
Chollet F, et~al. (2018) Keras: The python deep learning library. Astrophysics
  Source Code Library

\bibitem[{Choudhury et~al.(2018)Choudhury, Vanguri, Jambawalikar, and
  Kumar}]{choudhury2018segmentation}
Choudhury AR, Vanguri R, Jambawalikar SR, Kumar P (2018) Segmentation of brain
  tumors using deeplabv3+. In: International MICCAI Brainlesion Workshop,
  Springer, pp 154--167

\bibitem[{Clark et~al.(1998)Clark, Hall, Goldgof, Velthuizen, Murtagh, and
  Silbiger}]{clark1998automatic}
Clark MC, Hall LO, Goldgof DB, Velthuizen R, Murtagh FR, Silbiger MS (1998)
  Automatic tumor segmentation using knowledge-based techniques. IEEE
  transactions on medical imaging 17(2):187--201

\bibitem[{Colmeiro et~al.(2017)Colmeiro, Verrastro, and
  Grosges}]{colmeiro2017multimodal}
Colmeiro RR, Verrastro C, Grosges T (2017) Multimodal brain tumor segmentation
  using 3d convolutional networks. In: International MICCAI Brainlesion
  Workshop, Springer, pp 226--240

\bibitem[{Cordier et~al.(2013)Cordier, Menze, Delingette, and
  Ayache}]{cordier2013patch}
Cordier N, Menze B, Delingette H, Ayache N (2013) Patch-based segmentation of
  brain tissues

\bibitem[{Corso et~al.(2008)Corso, Sharon, Dube, El-Saden, Sinha, and
  Yuille}]{corso2008efficient}
Corso JJ, Sharon E, Dube S, El-Saden S, Sinha U, Yuille A (2008) Efficient
  multilevel brain tumor segmentation with integrated bayesian model
  classification. IEEE transactions on medical imaging 27(5):629--640

\bibitem[{Cuadra et~al.(2004)Cuadra, Pollo, Bardera, Cuisenaire, Villemure, and
  Thiran}]{cuadra2004atlas}
Cuadra MB, Pollo C, Bardera A, Cuisenaire O, Villemure JG, Thiran JP (2004)
  Atlas-based segmentation of pathological mr brain images using a model of
  lesion growth. IEEE transactions on medical imaging 23(10):1301--1314

\bibitem[{Dai et~al.(2018)Dai, Li, Shu, Zhong, Shen, and
  Zhu}]{dai2018automatic}
Dai L, Li T, Shu H, Zhong L, Shen H, Zhu H (2018) Automatic brain tumor
  segmentation with domain adaptation. In: International MICCAI Brainlesion
  Workshop, Springer, pp 380--392

\bibitem[{Dera et~al.(2016)Dera, Raman, Bouaynaya, and
  Fathallah-Shaykh}]{dera2016interactive}
Dera D, Raman F, Bouaynaya N, Fathallah-Shaykh HM (2016) Interactive
  semi-automated method using non-negative matrix factorization and level set
  segmentation for the brats challenge. In: International Workshop on
  Brainlesion: Glioma, Multiple Sclerosis, Stroke and Traumatic Brain Injuries,
  Springer, pp 195--205

\bibitem[{Dieleman et~al.(2015)Dieleman, Schlüter, Raffel, Olson, Sønderby,
  Nouri et~al.}]{lasagne}
Dieleman S, Schlüter J, Raffel C, Olson E, Sønderby SK, Nouri D, et~al.
  (2015) Lasagne: First release. \doi{10.5281/zenodo.27878},
  \urlprefix\url{http://dx.doi.org/10.5281/zenodo.27878}

\bibitem[{Dong et~al.(2017{\natexlab{a}})Dong, Supratak, Mai, Liu, Oehmichen,
  Yu, and Guo}]{dong2017tensorlayer}
Dong H, Supratak A, Mai L, Liu F, Oehmichen A, Yu S, Guo Y (2017{\natexlab{a}})
  Tensorlayer: a versatile library for efficient deep learning development. In:
  Proceedings of the 25th ACM international conference on Multimedia, pp
  1201--1204

\bibitem[{Dong et~al.(2017{\natexlab{b}})Dong, Yang, Liu, Mo, and
  Guo}]{dong2017automatic}
Dong H, Yang G, Liu F, Mo Y, Guo Y (2017{\natexlab{b}}) Automatic brain tumor
  detection and segmentation using u-net based fully convolutional networks.
  In: annual conference on medical image understanding and analysis, Springer,
  pp 506--517

\bibitem[{Doyle et~al.(2013)Doyle, Vasseur, Dojat, and Forbes}]{doyle2013fully}
Doyle S, Vasseur F, Dojat M, Forbes F (2013) Fully automatic brain tumor
  segmentation from multiple mr sequences using hidden markov fields and
  variational em. Procs NCI-MICCAI BraTS pp 18--22

\bibitem[{Ellwaa et~al.(2016)Ellwaa, Hussein, AlNaggar, Zidan, Zaki, Ismail,
  and Ghanem}]{ellwaa2016brain}
Ellwaa A, Hussein A, AlNaggar E, Zidan M, Zaki M, Ismail MA, Ghanem NM (2016)
  Brain tumor segmantation using random forest trained on iteratively selected
  patients. In: International Workshop on Brainlesion: Glioma, Multiple
  Sclerosis, Stroke and Traumatic Brain Injuries, Springer, pp 129--137

\bibitem[{Feng and Meyer(2017)}]{feng2017patch}
Feng X, Meyer C (2017) Patch-based 3d u-net for brain tumor segmentation. In:
  International Conference on Medical Image Computing and Computer-Assisted
  Intervention (MICCAI)

\bibitem[{Feng et~al.(2018)Feng, Tustison, and Meyer}]{feng2018brain}
Feng X, Tustison N, Meyer C (2018) Brain tumor segmentation using an ensemble
  of 3d u-nets and overall survival prediction using radiomic features. In:
  International MICCAI Brainlesion Workshop, Springer, pp 279--288

\bibitem[{Festa et~al.(2013)Festa, Pereira, Mariz, Sousa, and
  Silva}]{festa2013automatic}
Festa J, Pereira S, Mariz JA, Sousa N, Silva CA (2013) Automatic brain tumor
  segmentation of multi-sequence mr images using random decision forests.
  Proceedings of NCI-MICCAI BRATS 1:23--26

\bibitem[{Geremia et~al.(2012)Geremia, Menze, Ayache
  et~al.}]{geremia2012spatial}
Geremia E, Menze BH, Ayache N, et~al. (2012) Spatial decision forests for
  glioma segmentation in multi-channel mr images. MICCAI Challenge on
  Multimodal Brain Tumor Segmentation 34

\bibitem[{Goetz et~al.(2014)Goetz, Weber, Bloecher, Stieltjes, Meinzer, and
  Maier-Hein}]{goetz2014extremely}
Goetz M, Weber C, Bloecher J, Stieltjes B, Meinzer HP, Maier-Hein K (2014)
  Extremely randomized trees based brain tumor segmentation. Proceeding of
  BRATS challenge-MICCAI pp 006--011

\bibitem[{Goyal et~al.(2018)Goyal, Agrawal, and Sohi}]{goyal2018noise}
Goyal B, Agrawal S, Sohi B (2018) Noise issues prevailing in various types of
  medical images. Biomedical \& Pharmacology Journal 11(3):1227

\bibitem[{Guo et~al.(2013)Guo, Schwartz, and Zhao}]{guo2013semi}
Guo X, Schwartz L, Zhao B (2013) Semi-automatic segmentation of multimodal
  brain tumor using active contours. Multimodal Brain Tumor Segmentation 27

\bibitem[{Hamamci and Unal(2012)}]{hamamci2012multimodal}
Hamamci A, Unal G (2012) Multimodal brain tumor segmentation using the
  tumor-cut method on the brats dataset. Proc MICCAI-BRATS pp 19--23

\bibitem[{Hamamci et~al.(2011)Hamamci, Kucuk, Karaman, Engin, and
  Unal}]{hamamci2011tumor}
Hamamci A, Kucuk N, Karaman K, Engin K, Unal G (2011) Tumor-cut: segmentation
  of brain tumors on contrast enhanced mr images for radiosurgery applications.
  IEEE transactions on medical imaging 31(3):790--804

\bibitem[{Havaei et~al.(2015)Havaei, Dutil, Pal, Larochelle, and
  Jodoin}]{havaei2015convolutional}
Havaei M, Dutil F, Pal C, Larochelle H, Jodoin PM (2015) A convolutional neural
  network approach to brain tumor segmentation. In: BrainLes 2015, Springer, pp
  195--208

\bibitem[{Havaei et~al.(2017)Havaei, Davy, Warde-Farley, Biard, Courville,
  Bengio, Pal, Jodoin, and Larochelle}]{havaei2017brain}
Havaei M, Davy A, Warde-Farley D, Biard A, Courville A, Bengio Y, Pal C, Jodoin
  PM, Larochelle H (2017) Brain tumor segmentation with deep neural networks.
  Medical image analysis 35:18--31

\bibitem[{Healthcareplex(2016 (accessed April 10, 2020))}]{healthcare:2016}
Healthcareplex (2016 (accessed April 10, 2020)) CT Scan vs. MRI.
  \urlprefix\url{https://healthcareplex.com/mri-vs-ct-scan/}

\bibitem[{Hopkins(2019 (accessed April 6, 2020))}]{jhm:2020}
Hopkins J (2019 (accessed April 6, 2020)) Health.
  \urlprefix\url{https://www.hopkinsmedicine.org/health/conditions-and-diseases/basics-of-brain-tumors}

\bibitem[{Hu et~al.(2018{\natexlab{a}})Hu, Li, Zhao, Dong, Menze, and
  Piraud}]{hu2018hierarchical}
Hu X, Li H, Zhao Y, Dong C, Menze BH, Piraud M (2018{\natexlab{a}})
  Hierarchical multi-class segmentation of glioma images using networks with
  multi-level activation function. In: International MICCAI Brainlesion
  Workshop, Springer, pp 116--127

\bibitem[{Hu and Xia(2017)}]{hu20173d}
Hu Y, Xia Y (2017) 3d deep neural network-based brain tumor segmentation using
  multimodality magnetic resonance sequences. In: International MICCAI
  Brainlesion Workshop, Springer, pp 423--434

\bibitem[{Hu et~al.(2018{\natexlab{b}})Hu, Liu, Wen, Niu, and
  Xia}]{hu2018brain}
Hu Y, Liu X, Wen X, Niu C, Xia Y (2018{\natexlab{b}}) Brain tumor segmentation
  on multimodal mr imaging using multi-level upsampling in decoder. In:
  International MICCAI Brainlesion Workshop, Springer, pp 168--177

\bibitem[{Hua et~al.(2018)Hua, Huo, Gao, Sun, and Shi}]{hua2018multimodal}
Hua R, Huo Q, Gao Y, Sun Y, Shi F (2018) Multimodal brain tumor segmentation
  using cascaded v-nets. In: International MICCAI Brainlesion Workshop,
  Springer, pp 49--60

\bibitem[{Huang et~al.(2017)Huang, Liu, Van Der~Maaten, and
  Weinberger}]{huang2017densely}
Huang G, Liu Z, Van Der~Maaten L, Weinberger KQ (2017) Densely connected
  convolutional networks. In: Proceedings of the IEEE conference on computer
  vision and pattern recognition, pp 4700--4708

\bibitem[{Isensee et~al.(2017)Isensee, Kickingereder, Wick, Bendszus, and
  Maier-Hein}]{isensee2017brain}
Isensee F, Kickingereder P, Wick W, Bendszus M, Maier-Hein KH (2017) Brain
  tumor segmentation and radiomics survival prediction: Contribution to the
  brats 2017 challenge. In: International MICCAI Brainlesion Workshop,
  Springer, pp 287--297

\bibitem[{Isensee et~al.(2018)Isensee, Kickingereder, Wick, Bendszus, and
  Maier-Hein}]{isensee2018no}
Isensee F, Kickingereder P, Wick W, Bendszus M, Maier-Hein KH (2018) No
  new-net. In: International MICCAI Brainlesion Workshop, Springer, pp 234--244

\bibitem[{Islam and Ren(2017)}]{islam2017multi}
Islam M, Ren H (2017) Multi-modal pixelnet for brain tumor segmentation. In:
  International MICCAI Brainlesion Workshop, Springer, pp 298--308

\bibitem[{Islam et~al.(2018)Islam, Jose, and Ren}]{islam2018glioma}
Islam M, Jose VJM, Ren H (2018) Glioma prognosis: Segmentation of the tumor and
  survival prediction using shape, geometric and clinical information. In:
  International MICCAI Brainlesion Workshop, Springer, pp 142--153

\bibitem[{Janssen and Hoff(2012)}]{janssen2012teaching}
Janssen PM, Hoff EI (2012) Teaching neuroimages: Subacute intracerebral
  hemorrhage mimicking brain tumor. Neurology 79(21):e183--e183

\bibitem[{Jesson and Arbel(2017)}]{jesson2017brain}
Jesson A, Arbel T (2017) Brain tumor segmentation using a 3d fcn with
  multi-scale loss. In: International MICCAI Brainlesion Workshop, Springer, pp
  392--402

\bibitem[{Jia et~al.(2014)Jia, Shelhamer, Donahue, Karayev, Long, Girshick,
  Guadarrama, and Darrell}]{jia2014caffe}
Jia Y, Shelhamer E, Donahue J, Karayev S, Long J, Girshick R, Guadarrama S,
  Darrell T (2014) Caffe: Convolutional architecture for fast feature
  embedding. In: Proceedings of the 22nd ACM international conference on
  Multimedia, pp 675--678

\bibitem[{Jungo et~al.(2017)Jungo, McKinley, Meier, Knecht, Vera,
  P{\'e}rez-Beteta, Molina-Garc{\'\i}a, P{\'e}rez-Garc{\'\i}a, Wiest, and
  Reyes}]{jungo2017towards}
Jungo A, McKinley R, Meier R, Knecht U, Vera L, P{\'e}rez-Beteta J,
  Molina-Garc{\'\i}a D, P{\'e}rez-Garc{\'\i}a VM, Wiest R, Reyes M (2017)
  Towards uncertainty-assisted brain tumor segmentation and survival
  prediction. In: International MICCAI Brainlesion Workshop, Springer, pp
  474--485

\bibitem[{Kamnitsas et~al.(2016)Kamnitsas, Ferrante, Parisot, Ledig, Nori,
  Criminisi, Rueckert, and Glocker}]{kamnitsas2016deepmedic}
Kamnitsas K, Ferrante E, Parisot S, Ledig C, Nori AV, Criminisi A, Rueckert D,
  Glocker B (2016) Deepmedic for brain tumor segmentation. In: International
  workshop on Brainlesion: Glioma, multiple sclerosis, stroke and traumatic
  brain injuries, Springer, pp 138--149

\bibitem[{Kamnitsas et~al.(2017)Kamnitsas, Bai, Ferrante, McDonagh, Sinclair,
  Pawlowski, Rajchl, Lee, Kainz, Rueckert et~al.}]{kamnitsas2017ensembles}
Kamnitsas K, Bai W, Ferrante E, McDonagh S, Sinclair M, Pawlowski N, Rajchl M,
  Lee M, Kainz B, Rueckert D, et~al. (2017) Ensembles of multiple models and
  architectures for robust brain tumour segmentation. In: International MICCAI
  Brainlesion Workshop, Springer, pp 450--462

\bibitem[{Kao et~al.(2018)Kao, Ngo, Zhang, Chen, and Manjunath}]{kao2018brain}
Kao PY, Ngo T, Zhang A, Chen JW, Manjunath B (2018) Brain tumor segmentation
  and tractographic feature extraction from structural mr images for overall
  survival prediction. In: International MICCAI Brainlesion Workshop, Springer,
  pp 128--141

\bibitem[{Kermi et~al.(2018)Kermi, Mahmoudi, and Khadir}]{kermi2018deep}
Kermi A, Mahmoudi I, Khadir MT (2018) Deep convolutional neural networks using
  u-net for automatic brain tumor segmentation in multimodal mri volumes. In:
  International MICCAI Brainlesion Workshop, Springer, pp 37--48

\bibitem[{Kim(2017)}]{kim2017brain}
Kim G (2017) Brain tumor segmentation using deep fully convolutional neural
  networks. In: International MICCAI Brainlesion Workshop, Springer, pp
  344--357

\bibitem[{Kleesiek et~al.(2014)Kleesiek, Biller, Urban, Kothe, Bendszus, and
  Hamprecht}]{kleesiek2014ilastik}
Kleesiek J, Biller A, Urban G, Kothe U, Bendszus M, Hamprecht F (2014) Ilastik
  for multi-modal brain tumor segmentation. Proceedings MICCAI BraTS (brain
  tumor segmentation challenge) pp 12--17

\bibitem[{Kori et~al.(2018)Kori, Soni, Pranjal, Khened, Alex, and
  Krishnamurthi}]{kori2018ensemble}
Kori A, Soni M, Pranjal B, Khened M, Alex V, Krishnamurthi G (2018) Ensemble of
  fully convolutional neural network for brain tumor segmentation from magnetic
  resonance images. In: International MICCAI Brainlesion Workshop, Springer, pp
  485--496

\bibitem[{Krizhevsky et~al.(2012)Krizhevsky, Sutskever, and
  Hinton}]{krizhevsky2012imagenet}
Krizhevsky A, Sutskever I, Hinton GE (2012) Imagenet classification with deep
  convolutional neural networks. In: Advances in neural information processing
  systems, pp 1097--1105

\bibitem[{Kwon et~al.(2014)Kwon, Akbari, Da, Gaonkar, and
  Davatzikos}]{kwon2014multimodal}
Kwon D, Akbari H, Da X, Gaonkar B, Davatzikos C (2014) Multimodal brain tumor
  image segmentation using glistr. MICCAI brain tumor segmentation (BraTS)
  challenge manuscripts pp 18--19

\bibitem[{Lachinov et~al.(2018)Lachinov, Vasiliev, and
  Turlapov}]{lachinov2018glioma}
Lachinov D, Vasiliev E, Turlapov V (2018) Glioma segmentation with cascaded
  unet. In: International MICCAI Brainlesion Workshop, Springer, pp 189--198

\bibitem[{Le~Folgoc et~al.(2016)Le~Folgoc, Nori, Ancha, and
  Criminisi}]{le2016lifted}
Le~Folgoc L, Nori AV, Ancha S, Criminisi A (2016) Lifted auto-context forests
  for brain tumour segmentation. In: International Workshop on Brainlesion:
  Glioma, Multiple Sclerosis, Stroke and Traumatic Brain Injuries, Springer, pp
  171--183

\bibitem[{LeCun et~al.(1998)LeCun, Bottou, Bengio, and
  Haffner}]{lecun1998gradient}
LeCun Y, Bottou L, Bengio Y, Haffner P (1998) Gradient-based learning applied
  to document recognition. Proceedings of the IEEE 86(11):2278--2324

\bibitem[{Lee et~al.(2019)Lee, Kim, Lee, Lee, and Yoon}]{lee2019ficklenet}
Lee J, Kim E, Lee S, Lee J, Yoon S (2019) Ficklenet: Weakly and semi-supervised
  semantic image segmentation using stochastic inference. In: Proceedings of
  the IEEE conference on computer vision and pattern recognition, pp 5267--5276

\bibitem[{Lefkovits et~al.(2016)Lefkovits, Lefkovits, and
  Szil{\'a}gyi}]{lefkovits2016brain}
Lefkovits L, Lefkovits S, Szil{\'a}gyi L (2016) Brain tumor segmentation with
  optimized random forest. In: International Workshop on Brainlesion: Glioma,
  Multiple Sclerosis, Stroke and Traumatic Brain Injuries, Springer, pp 88--99

\bibitem[{Li and Shen(2017)}]{li2017deep}
Li Y, Shen L (2017) Deep learning based multimodal brain tumor diagnosis. In:
  International MICCAI Brainlesion Workshop, Springer, pp 149--158

\bibitem[{Lin et~al.(2017)Lin, Goyal, Girshick, He, and
  Doll{\'a}r}]{lin2017focal}
Lin TY, Goyal P, Girshick R, He K, Doll{\'a}r P (2017) Focal loss for dense
  object detection. In: Proceedings of the IEEE international conference on
  computer vision, pp 2980--2988

\bibitem[{Litjens et~al.(2017)Litjens, Kooi, Bejnordi, Setio, Ciompi,
  Ghafoorian, Van Der~Laak, Van~Ginneken, and S{\'a}nchez}]{litjens2017survey}
Litjens G, Kooi T, Bejnordi BE, Setio AAA, Ciompi F, Ghafoorian M, Van Der~Laak
  JA, Van~Ginneken B, S{\'a}nchez CI (2017) A survey on deep learning in
  medical image analysis. Medical image analysis 42:60--88

\bibitem[{Long et~al.(2015)Long, Shelhamer, and Darrell}]{long2015fully}
Long J, Shelhamer E, Darrell T (2015) Fully convolutional networks for semantic
  segmentation. In: Proceedings of the IEEE conference on computer vision and
  pattern recognition, pp 3431--3440

\bibitem[{Lopez and Ventura(2017)}]{lopez2017dilated}
Lopez MM, Ventura J (2017) Dilated convolutions for brain tumor segmentation in
  mri scans. In: International MICCAI Brainlesion Workshop, Springer, pp
  253--262

\bibitem[{Ma and Yang(2018)}]{ma2018automatic}
Ma J, Yang X (2018) Automatic brain tumor segmentation by exploring the
  multi-modality complementary information and cascaded 3d lightweight cnns.
  In: International MICCAI Brainlesion Workshop, Springer, pp 25--36

\bibitem[{Maier et~al.(2015)Maier, Wilms, and Handels}]{maier2015image}
Maier O, Wilms M, Handels H (2015) Image features for brain lesion segmentation
  using random forests. In: BrainLes 2015, Springer, pp 119--130

\bibitem[{Marcinkiewicz et~al.(2018)Marcinkiewicz, Nalepa, Lorenzo, Dudzik, and
  Mrukwa}]{marcinkiewicz2018segmenting}
Marcinkiewicz M, Nalepa J, Lorenzo PR, Dudzik W, Mrukwa G (2018) Segmenting
  brain tumors from mri using cascaded multi-modal u-nets. In: International
  MICCAI Brainlesion Workshop, Springer, pp 13--24

\bibitem[{McKinley et~al.(2016)McKinley, Wepfer, Gundersen, Wagner, Chan,
  Wiest, and Reyes}]{mckinley2016nabla}
McKinley R, Wepfer R, Gundersen T, Wagner F, Chan A, Wiest R, Reyes M (2016)
  Nabla-net: A deep dag-like convolutional architecture for biomedical image
  segmentation. In: International Workshop on Brainlesion: Glioma, Multiple
  Sclerosis, Stroke and Traumatic Brain Injuries, Springer, pp 119--128

\bibitem[{McKinley et~al.(2017)McKinley, Jungo, Wiest, and
  Reyes}]{mckinley2017pooling}
McKinley R, Jungo A, Wiest R, Reyes M (2017) Pooling-free fully convolutional
  networks with dense skip connections for semantic segmentation, with
  application to brain tumor segmentation. In: International MICCAI Brainlesion
  Workshop, Springer, pp 169--177

\bibitem[{McKinley et~al.(2018)McKinley, Meier, and
  Wiest}]{mckinley2018ensembles}
McKinley R, Meier R, Wiest R (2018) Ensembles of densely-connected cnns with
  label-uncertainty for brain tumor segmentation. In: International MICCAI
  Brainlesion Workshop, Springer, pp 456--465

\bibitem[{Media(2004 (accessed April 10, 2020))}]{amzreg:2019}
Media H (2004 (accessed April 10, 2020)) CT Scan vs. MRI.
  \urlprefix\url{https://www.healthline.com/health/ct-scan-vs-mri/}

\bibitem[{Mehta and Arbel(2018)}]{mehta20183d}
Mehta R, Arbel T (2018) 3d u-net for brain tumour segmentation. In:
  International MICCAI Brainlesion Workshop, Springer, pp 254--266

\bibitem[{Meier et~al.(2013)Meier, Bauer, Slotboom, Wiest, and
  Reyes}]{meier2013hybrid}
Meier R, Bauer S, Slotboom J, Wiest R, Reyes M (2013) A hybrid model for
  multimodal brain tumor segmentation. Multimodal Brain Tumor Segmentation
  31:31--37

\bibitem[{Meier et~al.(2014)Meier, Bauer, Slotboom, Wiest, and
  Reyes}]{meier2014appearance}
Meier R, Bauer S, Slotboom J, Wiest R, Reyes M (2014) Appearance-and
  context-sensitive features for brain tumor segmentation. Proceedings of
  MICCAI BRATS Challenge pp 020--026

\bibitem[{Meier et~al.(2015)Meier, Karamitsou, Habegger, Wiest, and
  Reyes}]{meier2015parameter}
Meier R, Karamitsou V, Habegger S, Wiest R, Reyes M (2015) Parameter learning
  for crf-based tissue segmentation of brain tumors. In: BrainLes 2015,
  Springer, pp 156--167

\bibitem[{Meier et~al.(2016)Meier, Knecht, Wiest, and Reyes}]{meier2016crf}
Meier R, Knecht U, Wiest R, Reyes M (2016) Crf-based brain tumor segmentation:
  alleviating the shrinking bias. In: International workshop on brainlesion:
  glioma, multiple sclerosis, stroke and traumatic brain injuries, Springer, pp
  100--107

\bibitem[{Menze et~al.(2012)Menze, Geremia, Ayache, and
  Szekely}]{menze2012segmenting}
Menze BH, Geremia E, Ayache N, Szekely G (2012) Segmenting glioma in
  multi-modal images using a generative-discriminative model for brain lesion
  segmentation. Proc MICCAI-BRATS (Multimodal Brain Tumor Segmentation
  Challenge) 8

\bibitem[{Menze et~al.(2014)Menze, Jakab, Bauer, Kalpathy-Cramer, Farahani,
  Kirby, Burren, Porz, Slotboom, Wiest et~al.}]{menze2014multimodal}
Menze BH, Jakab A, Bauer S, Kalpathy-Cramer J, Farahani K, Kirby J, Burren Y,
  Porz N, Slotboom J, Wiest R, et~al. (2014) The multimodal brain tumor image
  segmentation benchmark (brats). IEEE transactions on medical imaging
  34(10):1993--2024

\bibitem[{Myronenko(2018)}]{myronenko20183d}
Myronenko A (2018) 3d mri brain tumor segmentation using autoencoder
  regularization. In: International MICCAI Brainlesion Workshop, Springer, pp
  311--320

\bibitem[{Nuechterlein and Mehta(2019)}]{nuechterlein20193d}
Nuechterlein N, Mehta S (2019) 3d-espnet with pyramidal refinement for
  volumetric brain tumor image segmentation. brainles 2018. Springer LNCS
  11384:245--253

\bibitem[{Nvidia(2020 (accessed April 28, 2020))}]{gpuspec:2019}
Nvidia (2020 (accessed April 28, 2020)) Nvidia.
  \urlprefix\url{https://www.nvidia.com/en-in/}

\bibitem[{Ny{\'u}l and Udupa(1999)}]{nyul1999standardizing}
Ny{\'u}l LG, Udupa JK (1999) On standardizing the mr image intensity scale.
  Magnetic Resonance in Medicine: An Official Journal of the International
  Society for Magnetic Resonance in Medicine 42(6):1072--1081

\bibitem[{Pan and Yang(2009)}]{pan2009survey}
Pan SJ, Yang Q (2009) A survey on transfer learning. IEEE Transactions on
  knowledge and data engineering 22(10):1345--1359

\bibitem[{Paszke et~al.(2019)Paszke, Gross, Massa, Lerer, Bradbury, Chanan,
  Killeen, Lin, Gimelshein, Antiga et~al.}]{paszke2019pytorch}
Paszke A, Gross S, Massa F, Lerer A, Bradbury J, Chanan G, Killeen T, Lin Z,
  Gimelshein N, Antiga L, et~al. (2019) Pytorch: An imperative style,
  high-performance deep learning library. In: Advances in Neural Information
  Processing Systems, pp 8024--8035

\bibitem[{Pawar et~al.(2017)Pawar, Chen, Shah, and Egan}]{pawar2017residual}
Pawar K, Chen Z, Shah NJ, Egan G (2017) Residual encoder and convolutional
  decoder neural network for glioma segmentation. In: International MICCAI
  Brainlesion Workshop, Springer, pp 263--273

\bibitem[{Pereira et~al.(2015)Pereira, Pinto, Alves, and
  Silva}]{pereira2015deep}
Pereira S, Pinto A, Alves V, Silva CA (2015) Deep convolutional neural networks
  for the segmentation of gliomas in multi-sequence mri. In: BrainLes 2015,
  Springer, pp 131--143

\bibitem[{Phophalia and Maji(2017)}]{phophalia2017multimodal}
Phophalia A, Maji P (2017) Multimodal brain tumor segmentation using ensemble
  of forest method. In: International MICCAI Brainlesion Workshop, Springer, pp
  159--168

\bibitem[{Piedra et~al.(2016)Piedra, Ellingson, Taira, El-Saden, Bui, and
  Hsu}]{piedra2016brain}
Piedra EAR, Ellingson BM, Taira RK, El-Saden S, Bui AA, Hsu W (2016) Brain
  tumor segmentation by variability characterization of tumor boundaries. In:
  International Workshop on Brainlesion: Glioma, Multiple Sclerosis, Stroke and
  Traumatic Brain Injuries, Springer, pp 206--216

\bibitem[{Pourreza et~al.(2017)Pourreza, Zhuge, Ning, and
  Miller}]{pourreza2017brain}
Pourreza R, Zhuge Y, Ning H, Miller R (2017) Brain tumor segmentation in mri
  scans using deeply-supervised neural networks. In: International MICCAI
  Brainlesion Workshop, Springer, pp 320--331

\bibitem[{Prastawa et~al.(2004)Prastawa, Bullitt, Ho, and
  Gerig}]{prastawa2004brain}
Prastawa M, Bullitt E, Ho S, Gerig G (2004) A brain tumor segmentation
  framework based on outlier detection. Medical image analysis 8(3):275--283

\bibitem[{Puch et~al.(2019)Puch, S{\'a}nchez, Hern{\'a}ndez, Piella, and
  Pr{\'c}kovska}]{puch2019global}
Puch S, S{\'a}nchez I, Hern{\'a}ndez A, Piella G, Pr{\'c}kovska V (2019) Global
  planar convolutions for improved context aggregation in brain tumor
  segmentation. brainles 2018. Springer LNCS 11384:393--405

\bibitem[{Puybareau et~al.(2018)Puybareau, Tochon, Chazalon, and
  Fabrizio}]{puybareau2018segmentation}
Puybareau E, Tochon G, Chazalon J, Fabrizio J (2018) Segmentation of gliomas
  and prediction of patient overall survival: a simple and fast procedure. In:
  International MICCAI Brainlesion Workshop, Springer, pp 199--209

\bibitem[{of~Radiology(1999 (accessed April 10, 2020))}]{amercr:2020}
of~Radiology AC (1999 (accessed April 10, 2020)) Brain Tumor Treatment.
  \urlprefix\url{https://www.radiologyinfo.org/}

\bibitem[{Rajendran and Dhanasekaran(2012)}]{rajendran2012brain}
Rajendran A, Dhanasekaran R (2012) Brain tumor segmentation on mri brain images
  with fuzzy clustering and gvf snake model. International Journal of Computers
  Communications \& Control 7(3):530--539

\bibitem[{Randhawa et~al.(2016)Randhawa, Modi, Jain, and
  Warier}]{randhawa2016improving}
Randhawa RS, Modi A, Jain P, Warier P (2016) Improving boundary classification
  for brain tumor segmentation and longitudinal disease progression. In:
  International Workshop on Brainlesion: Glioma, Multiple Sclerosis, Stroke and
  Traumatic Brain Injuries, Springer, pp 65--74

\bibitem[{Raval et~al.(2021)Raval, Rajput, Roy, and
  Agravat}]{raval2021glioblastoma}
Raval M, Rajput S, Roy M, Agravat R (2021) Glioblastoma multiforme patient
  survival prediction

\bibitem[{Raviv et~al.(2012)Raviv, Leemput, and Menze}]{raviv2012multi}
Raviv TR, Leemput KV, Menze BH (2012) Multi-modal brain tumor segmentation via
  latent atlases. Proceeding MICCAIBRATS 64

\bibitem[{Reza and Iftekharuddin(2013)}]{reza2013multi}
Reza S, Iftekharuddin K (2013) Multi-class abnormal brain tissue segmentation
  using texture. Multimodal Brain Tumor Segmentation 38

\bibitem[{Reza and Iftekharuddin(2014)}]{reza2014improved}
Reza S, Iftekharuddin K (2014) Improved brain tumor tissue segmentation using
  texture features. Proceedings MICCAI BraTS (brain tumor segmentation
  challenge) pp 27--30

\bibitem[{Ronneberger et~al.(2015)Ronneberger, Fischer, and
  Brox}]{ronneberger2015u}
Ronneberger O, Fischer P, Brox T (2015) U-net: Convolutional networks for
  biomedical image segmentation. In: International Conference on Medical image
  computing and computer-assisted intervention, Springer, pp 234--241

\bibitem[{Saha et~al.(2016)Saha, Phophalia, and Mitra}]{saha2016brain}
Saha R, Phophalia A, Mitra SK (2016) Brain tumor segmentation from multimodal
  mr images using rough sets. In: International Conference on Computer Vision,
  Graphics, and Image processing, Springer, pp 133--144

\bibitem[{Serrano-Rubio and Everson(2019)}]{serrano2019brain}
Serrano-Rubio J, Everson R (2019) Brain tumour segmentation method based on
  supervoxels and sparse dictionaries. brainles 2018. Springer LNCS
  11384:210--221

\bibitem[{Shaikh et~al.(2017)Shaikh, Anand, Acharya, Amrutkar, Alex, and
  Krishnamurthi}]{shaikh2017brain}
Shaikh M, Anand G, Acharya G, Amrutkar A, Alex V, Krishnamurthi G (2017) Brain
  tumor segmentation using dense fully convolutional neural network. In:
  International MICCAI Brainlesion Workshop, Springer, pp 309--319

\bibitem[{Shboul et~al.(2017)Shboul, Vidyaratne, Alam, and
  Iftekharuddin}]{shboul2017glioblastoma}
Shboul ZA, Vidyaratne L, Alam M, Iftekharuddin KM (2017) Glioblastoma and
  survival prediction. In: International MICCAI Brainlesion Workshop, Springer,
  pp 358--368

\bibitem[{Shin(2012)}]{shin2012hybrid}
Shin HC (2012) Hybrid clustering and logistic regression for multi-modal brain
  tumor segmentation. In: Proc. of Workshops and Challanges in Medical Image
  Computing and Computer-Assisted Intervention (MICCAI’12)

\bibitem[{Shridhar et~al.(2019)Shridhar, Laumann, and
  Liwicki}]{shridhar2019comprehensive}
Shridhar K, Laumann F, Liwicki M (2019) A comprehensive guide to bayesian
  convolutional neural network with variational inference. arXiv preprint
  arXiv:190102731

\bibitem[{Soltaninejad et~al.(2017)Soltaninejad, Zhang, Lambrou, Yang,
  Allinson, and Ye}]{soltaninejad2017mri}
Soltaninejad M, Zhang L, Lambrou T, Yang G, Allinson N, Ye X (2017) Mri brain
  tumor segmentation and patient survival prediction using random forests and
  fully convolutional networks. In: International MICCAI Brainlesion Workshop,
  Springer, pp 204--215

\bibitem[{Song et~al.(2016)Song, Chou, Chen, Huang, and Liu}]{song2016anatomy}
Song B, Chou CR, Chen X, Huang A, Liu MC (2016) Anatomy-guided brain tumor
  segmentation and classification. In: International Workshop on Brainlesion:
  Glioma, Multiple Sclerosis, Stroke and Traumatic Brain Injuries, Springer, pp
  162--170

\bibitem[{Stawiaski(2018)}]{stawiaski2018pretrained}
Stawiaski J (2018) A pretrained densenet encoder for brain tumor segmentation.
  In: International MICCAI Brainlesion Workshop, Springer, pp 105--115

\bibitem[{Subbanna and Arbel(2012)}]{subbanna2012probabilistic}
Subbanna N, Arbel T (2012) Probabilistic gabor and markov random fields
  segmentation of brain tumours in mri volumes. Proc MICCAI Brain Tumor
  Segmentation Challenge (BRATS) pp 28--31

\bibitem[{Sudre et~al.(2017)Sudre, Li, Vercauteren, Ourselin, and
  Cardoso}]{sudre2017generalised}
Sudre CH, Li W, Vercauteren T, Ourselin S, Cardoso MJ (2017) Generalised dice
  overlap as a deep learning loss function for highly unbalanced segmentations.
  In: Deep learning in medical image analysis and multimodal learning for
  clinical decision support, Springer, pp 240--248

\bibitem[{Sun et~al.(2018)Sun, Zhang, and Luo}]{sun2018tumor}
Sun L, Zhang S, Luo L (2018) Tumor segmentation and survival prediction in
  glioma with deep learning. In: International MICCAI Brainlesion Workshop,
  Springer, pp 83--93

\bibitem[{Szegedy et~al.(2015)Szegedy, Liu, Jia, Sermanet, Reed, Anguelov,
  Erhan, Vanhoucke, and Rabinovich}]{szegedy2015going}
Szegedy C, Liu W, Jia Y, Sermanet P, Reed S, Anguelov D, Erhan D, Vanhoucke V,
  Rabinovich A (2015) Going deeper with convolutions. In: Proceedings of the
  IEEE conference on computer vision and pattern recognition, pp 1--9

\bibitem[{Szegedy et~al.(2016)Szegedy, Vanhoucke, Ioffe, Shlens, and
  Wojna}]{szegedy2016rethinking}
Szegedy C, Vanhoucke V, Ioffe S, Shlens J, Wojna Z (2016) Rethinking the
  inception architecture for computer vision. In: Proceedings of the IEEE
  conference on computer vision and pattern recognition, pp 2818--2826

\bibitem[{Targ et~al.(2016)Targ, Almeida, and Lyman}]{targ2016resnet}
Targ S, Almeida D, Lyman K (2016) Resnet in resnet: Generalizing residual
  architectures. arXiv preprint arXiv:160308029

\bibitem[{Taylor et~al.(2013)Taylor, John, Buendia, and Ryan}]{taylor2013map}
Taylor T, John N, Buendia P, Ryan M (2013) Map-reduce enabled hidden markov
  models for high throughput multimodal brain tumor segmentation. Multimodal
  Brain Tumor Segmentation 43

\bibitem[{Tomas-Fernandez and Wareld(2012)}]{tomas2012automatic}
Tomas-Fernandez X, Wareld S (2012) Automatic brain tumor segmentation based on
  a coupled global-local intensity bayesian model. MICCAI Challenge on
  Multimodal Brain Tumor Segmentation 34

\bibitem[{Tustison et~al.(2010)Tustison, Avants, Cook, Zheng, Egan, Yushkevich,
  and Gee}]{tustison2010n4itk}
Tustison NJ, Avants BB, Cook PA, Zheng Y, Egan A, Yushkevich PA, Gee JC (2010)
  N4itk: improved n3 bias correction. IEEE transactions on medical imaging
  29(6):1310--1320

\bibitem[{Urban et~al.(2014)Urban, Bendszus, Hamprecht, and
  Kleesiek}]{urban2014multi}
Urban G, Bendszus M, Hamprecht F, Kleesiek J (2014) Multi-modal brain tumor
  segmentation using deep convolutional neural networks. MICCAI BraTS (brain
  tumor segmentation) challenge Proceedings, winning contribution pp 31--35

\bibitem[{Van~Griethuysen et~al.(2017)Van~Griethuysen, Fedorov, Parmar, Hosny,
  Aucoin, Narayan, Beets-Tan, Fillion-Robin, Pieper, and
  Aerts}]{van2017computational}
Van~Griethuysen JJ, Fedorov A, Parmar C, Hosny A, Aucoin N, Narayan V,
  Beets-Tan RG, Fillion-Robin JC, Pieper S, Aerts HJ (2017) Computational
  radiomics system to decode the radiographic phenotype. Cancer research
  77(21):e104--e107

\bibitem[{Wang et~al.(2017)Wang, Li, Ourselin, and
  Vercauteren}]{wang2017automatic}
Wang G, Li W, Ourselin S, Vercauteren T (2017) Automatic brain tumor
  segmentation using cascaded anisotropic convolutional neural networks. In:
  International MICCAI brainlesion workshop, Springer, pp 178--190

\bibitem[{Wang et~al.(2018)Wang, Li, Ourselin, and
  Vercauteren}]{wang2018automatic}
Wang G, Li W, Ourselin S, Vercauteren T (2018) Automatic brain tumor
  segmentation using convolutional neural networks with test-time augmentation.
  In: International MICCAI Brainlesion Workshop, Springer, pp 61--72

\bibitem[{Weninger et~al.(2018)Weninger, Rippel, Koppers, and
  Merhof}]{weninger2018segmentation}
Weninger L, Rippel O, Koppers S, Merhof D (2018) Segmentation of brain tumors
  and patient survival prediction: Methods for the brats 2018 challenge. In:
  International MICCAI Brainlesion Workshop, Springer, pp 3--12

\bibitem[{Xiao and Hu(2012)}]{xiao2012hierarchical}
Xiao Y, Hu J (2012) Hierarchical random walker for multimodal brain tumor
  segmentation. MICCAI Challenge on Multimodal Brain Tumor Segmentation

\bibitem[{Xu et~al.(2018)Xu, Gong, Fu, Tao, Zhang, and
  Batmanghelich}]{xu2018multi}
Xu Y, Gong M, Fu H, Tao D, Zhang K, Batmanghelich K (2018) Multi-scale masked
  3-d u-net for brain tumor segmentation. In: International MICCAI Brainlesion
  Workshop, Springer, pp 222--233

\bibitem[{Yao et~al.(2019)Yao, Zhou, and Zhang}]{yao2019automatic}
Yao H, Zhou X, Zhang X (2019) Automatic segmentation of brain tumor using 3d
  se-inception networks with residual connections. brainles 2018. Springer LNCS
  11384:346--357

\bibitem[{Zeng et~al.(2016)Zeng, Bakas, Sotiras, Akbari, Rozycki, Rathore,
  Pati, and Davatzikos}]{zeng2016segmentation}
Zeng K, Bakas S, Sotiras A, Akbari H, Rozycki M, Rathore S, Pati S, Davatzikos
  C (2016) Segmentation of gliomas in pre-operative and post-operative
  multimodal magnetic resonance imaging volumes based on a hybrid
  generative-discriminative framework. In: International Workshop on
  Brainlesion: Glioma, Multiple Sclerosis, Stroke and Traumatic Brain Injuries,
  Springer, pp 184--194

\bibitem[{Zhao et~al.(2012)Zhao, Wu, and Corso}]{zhao2012brain}
Zhao L, Wu W, Corso JJ (2012) Brain tumor segmentation based on gmm and active
  contour method with a model-aware edge map. BRATS MICCAI pp 19--23

\bibitem[{Zhao et~al.(2013)Zhao, Sarikaya, and Corso}]{zhao2013automatic}
Zhao L, Sarikaya D, Corso JJ (2013) Automatic brain tumor segmentation with mrf
  on supervoxels. Multimodal Brain Tumor Segmentation 51

\bibitem[{Zhao et~al.(2016)Zhao, Wu, Song, Li, Fan, and Zhang}]{zhao2016brain}
Zhao X, Wu Y, Song G, Li Z, Fan Y, Zhang Y (2016) Brain tumor segmentation
  using a fully convolutional neural network with conditional random fields.
  In: International Workshop on Brainlesion: Glioma, Multiple Sclerosis, Stroke
  and Traumatic Brain Injuries, Springer, pp 75--87

\bibitem[{Zhao et~al.(2017)Zhao, Wu, Song, Li, Zhang, and Fan}]{zhao20173d}
Zhao X, Wu Y, Song G, Li Z, Zhang Y, Fan Y (2017) 3d brain tumor segmentation
  through integrating multiple 2d fcnns. In: International MICCAI Brainlesion
  Workshop, Springer, pp 191--203

\bibitem[{Zhou et~al.(2018)Zhou, Chen, Ding, and Tao}]{zhou2018learning}
Zhou C, Chen S, Ding C, Tao D (2018) Learning contextual and attentive
  information for brain tumor segmentation. In: International MICCAI
  Brainlesion Workshop, Springer, pp 497--507

\bibitem[{Zhou et~al.(2017)Zhou, Li, Li, and Zhu}]{zhou2017tpcnn}
Zhou F, Li T, Li H, Zhu H (2017) Tpcnn: two-phase patch-based convolutional
  neural network for automatic brain tumor segmentation and survival
  prediction. In: International MICCAI Brainlesion Workshop, Springer, pp
  274--286

\bibitem[{Zikic et~al.(2012{\natexlab{a}})Zikic, Glocker, Konukoglu, Criminisi,
  Demiralp, Shotton, Thomas, Das, Jena, and Price}]{zikic2012decision}
Zikic D, Glocker B, Konukoglu E, Criminisi A, Demiralp C, Shotton J, Thomas OM,
  Das T, Jena R, Price SJ (2012{\natexlab{a}}) Decision forests for
  tissue-specific segmentation of high-grade gliomas in multi-channel mr. In:
  International Conference on Medical Image Computing and Computer-Assisted
  Intervention, Springer, pp 369--376

\bibitem[{Zikic et~al.(2012{\natexlab{b}})Zikic, Glocker, Konukoglu, Shotton,
  Criminisi, Ye, Demiralp, Thomas, Das, Jena et~al.}]{zikic2012context}
Zikic D, Glocker B, Konukoglu E, Shotton J, Criminisi A, Ye D, Demiralp C,
  Thomas O, Das T, Jena R, et~al. (2012{\natexlab{b}}) Context-sensitive
  classification forests for segmentation of brain tumor tissues. In: Proc.
  MICCAI-BRATS, pp 22--30

\bibitem[{Zikic et~al.(2014)Zikic, Ioannou, Brown, and
  Criminisi}]{zikic2014segmentation}
Zikic D, Ioannou Y, Brown M, Criminisi A (2014) Segmentation of brain tumor
  tissues with convolutional neural networks. Proceedings MICCAI-BRATS pp
  36--39

\end{thebibliography}
% etc
%\end{thebibliography}

\end{document}